\documentclass[journal]{IEEEtran}
\usepackage[noadjust]{cite}
\usepackage{amsmath}
\usepackage{array}
\usepackage{mdwmath}
\usepackage{amssymb}
\usepackage{algorithm}
\usepackage[noend]{algpseudocode}
\usepackage{eqparbox}
\usepackage{stfloats}
\usepackage{array}% http://ctan.org/pkg/array
\usepackage{url}
\usepackage[pdftex]{graphicx}
\graphicspath{{../pdf/}{../jpeg/}}
\DeclareGraphicsExtensions{.pdf,.jpeg,.png}
\usepackage{epstopdf}
\usepackage{amsfonts}
\usepackage[first=0,last=9]{lcg}
\usepackage{color, colortbl}
\definecolor{LightCyan}{rgb}{0.88,1,1}

\usepackage[tight,footnotesize]{subfigure}
\begin{document}

\title{Elevation Beamforming with Full Dimension MIMO Architectures in 5G Systems: A Tutorial}
\author{Qurrat-Ul-Ain~Nadeem,~\IEEEmembership{Student Member,~IEEE,} Abla~Kammoun,~\IEEEmembership{Member,~IEEE,}
         and ~ Mohamed-Slim~Alouini,~\IEEEmembership{Fellow,~IEEE}% <-this % stops a space
\thanks{Manuscript received xx xx, 2018.}
\thanks{Q.-U.-A. Nadeem, A. Kammoun and M.-S. Alouini are with the Computer, Electrical and Mathematical Sciences and Engineering (CEMSE) Division, Building 1, Level 3, King Abdullah University of Science and Technology (KAUST), Thuwal, Makkah Province, Saudi Arabia 23955-6900 (e-mail: \{qurratulain.nadeem,abla.kammoun,slim.alouini\}@kaust.edu.sa)}% <-this % stops a space
}

\markboth{}%
{Shell \MakeLowercase{\textit{et al.}}: Bare Demo of IEEEtran.cls for Journals}

\maketitle

\begin{abstract}

Full dimension (FD) multiple-input multiple-output (MIMO) technology has attracted substantial research attention from both wireless industry and academia in the last few years as a promising technique for next-generation wireless communication networks. FD-MIMO scenarios utilize a planar two-dimensional (2D) active antenna system (AAS) that not only allows a large number of antenna elements to be placed within feasible base station (BS) form factors, but also provides the ability of adaptive electronic beam control over both the elevation and the traditional azimuth dimensions.  This paper presents a tutorial on elevation beamforming analysis for cellular networks utilizing FD Massive MIMO antenna arrays. In contrast to existing works that focus on the standardization of FD-MIMO in the 3rd Generation Partnership Project (3GPP), this tutorial is distinguished by its depth with respect to the theoretical aspects of antenna array and 3D channel modeling. In an attempt to bridge the gap between industry and academia, this preliminary tutorial introduces the relevant array and transceiver architecture designs  proposed in the 3GPP Release 13 that enable elevation beamforming.  Then it presents and compares two different 3D channel modeling approaches that can be utilized for the performance analysis of elevation beamforming techniques. The spatial correlation in FD-MIMO arrays is characterized and compared based on both channel modeling approaches and some insights into the impact of different channel and array parameters on the correlation are drawn. All these aspects are put together to provide a mathematical framework for the design of elevation beamforming schemes in single-cell and multi-cell scenarios.  Simulation examples associated with comparisons and discussions are also presented. To this end, this paper highlights the state-of-the-art research and points out future research directions.

\end{abstract}


\section{Introduction}

Historically, an online presence on the Internet was enough for one-way broadcasting and dissemination of information. Today, social networks such as Facebook and Twitter and video streaming applications like YouTube are driving new forms of social interaction, dialogue, exchange and collaboration, most of which happens over smart phones utilizing the underlying cellular network resources. The evolution of social networks and other Web based applications  has led to a growing number of mobile broadband subscribers, who require real-time connectivity and consume bandwidth-hungry video content. This has resulted in a huge explosion in the wireless data traffic, with the amount of wireless data handled by cellular networks expected to exceed five hundred exabytes by 2020 \cite{5G, 5G1}. 

\begin{figure}
\centering
\includegraphics[scale=.35]{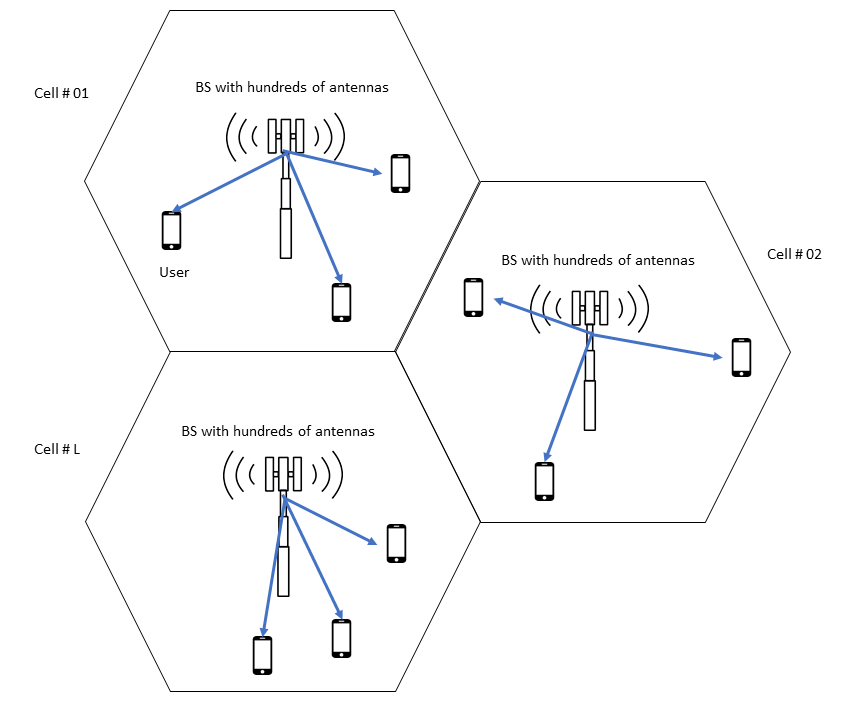}
\caption{Massive MIMO model.}
\label{massivelayout}
\end{figure}

\begin{figure*}
\centering
\includegraphics[scale=.55]{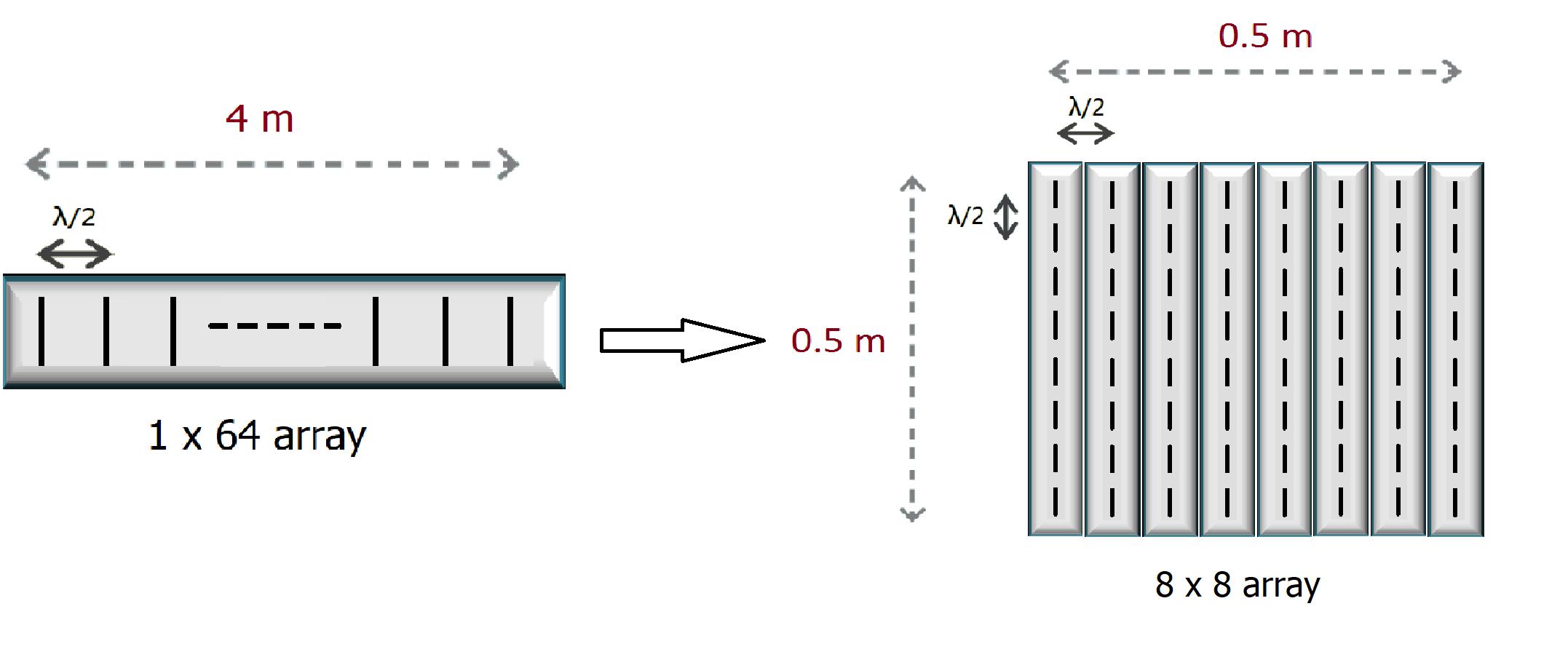}
\caption{Antenna array dimensions at 2.5 GHz operating frequency.}
\label{form_factor}
\end{figure*}

The need to support this wireless data traffic explosion led to the study, development and standardization of several cutting-edge techniques in the 3rd Generation Partnership Project (3GPP) Long Term Evolution (LTE)  Releases 8 through 12.  These include multiple-input multiple-output (MIMO) transmission/reception, coordinated multi-point (CoMP) transmission/reception, carrier aggregation (CA), wireless cooperative networks and heterogeneous networks \cite{CoMP, CA, coop1, coop2, hetro1, hetro}. Although the existing  LTE systems implementing these 4G  technologies have been able to achieve incremental improvements in the network capacity, but these are not enough to meet the demands the networks will face by 2020.  Much of the standardization effort in the 3GPP is now being focused on developing the 5G standard with the vision of meeting unprecedented demands beyond the capacity of previous generation systems.  Some key 5G technologies outlined in the 3GPP Release 13 and Release 14 for evolution towards LTE-Advanced Pro systems include Massive MIMO, millimeter (mm)-Wave MIMO, LTE-Unlicensed, small cell deployments, elevation beamforming with full-dimension (FD)-MIMO and device-to-device (D2D) communications \cite{LTE_unlicensed, 5Gfeasible, small_cells,FD1, FD2, D2D}.  This article will focus on FD Massive MIMO technology.

\subsection{Massive MIMO Model and Practical Challenges}

Most of the wireless communication research in the last two decades has focused on designing schemes to benefit from MIMO techniques. As far as the practical implementation is concerned, the existing MIMO systems consider the deployment of fewer than ten antennas in a uniform linear array (ULA) at the top of the base station (BS) towers. In fact, the 3GPP LTE standard allows for upto eight linearly arranged antennas at the BS \cite{massive, LTE}, which is why the corresponding improvement in the spectral efficiency (SE) of these wireless systems, although important, is still relatively modest and can be vastly improved by scaling up these systems by possibly orders of magnitude as compared to the current state-of-the-art. In fact, recent works that aim at achieving higher SE gains,  have proposed Massive MIMO or large-scale MIMO systems, where each BS is equipped with a very large number of antennas, allowing it to serve many users in the same time-frequency resource using simple linear precoding methods and thereby reaping the benefits of conventional multi-user (MU)-MIMO on a much larger scale \cite{massive, massive1, mass_sur2, mass_sur, massive2,  massive_energy_eff}. A typical multi-cell Massive MIMO system is illustrated in Fig. \ref{massivelayout}. The idea behind the development of Massive MIMO is that when the number of antennas at the BS goes to infinity, the MU interference caused by downlink user co-scheduling and uplink multiple access approaches tends to zero, thereby increasing the throughput by several orders of magnitude. The energy efficiency is increased dramatically too as the radiated energy can be concentrated into smaller regions in space  towards the targeted directions \cite{massive_energy_eff}. Using tools from random matrix theory (RMT), it can be shown that the effects of noise, fading and hardware impairments are also eliminated when a large number of antennas is deployed \cite{massive1, massive2}. All these advantages can be achieved using simple linear MU precoding/detection methods like maximum ratio transmission/combining \cite{massive1, FD2}. 

 One of the main challenges to build Massive MIMO systems in practice is that the number of antennas that can be equipped at the top of the BS towers in the conventional ULA is often limited by the BS form factor and the operating LTE carrier frequency. The BS form factor refers to the physical dimensions of the space available at the top of the BS tower for the deployment of antenna arrays. Some realistic BS form factor indications have been provided in Table II of  \cite{formfactor}, according to which a macro-cell BS has a form factor of $1430 \times 570 \times 550$ mm, while pico-cell and femto-cell BSs have even smaller rooms available for the deployment of antenna arrays. Now consider the deployment of $64$ antennas in a ULA with $0.5 \lambda$ spacing, where $\lambda$ is the carrier wavelength at the typical LTE operating frequency of $2.5$ GHz. This requires a horizontal room of about $4 m$ at the top of the BS tower. Comparing it with the form factor of a macro-cell BS, it is clear that installing 64 antennas linearly at existing BSs is impractical.  

Another limitation of the existing MIMO systems is that they consider the deployment of antennas in the azimuth plane alone. As a consequence, most of the works considering Massive MIMO deployments model the channel and investigate the beamforming methods in the horizontal plane only. However, the performance of real world Massive MIMO deployments is strongly dependent on the channels' characteristics in the elevation domain. Recently, the authors in \cite{elevationcamp} show that 65\% of the energy of the propagation paths incident on a rectangular antenna array has an elevation larger than $10^{o}$. The authors in \cite{elevationcamp1} state that 90\% of the energy of the propagation channel is confined within an elevation range of $0^{o}$ and $40^{o}$. Due to the 3D nature of the real-world wireless channels, beamforming in the horizontal plane cannot fully exploit all the degrees of freedom offered by the channel. Moreover, with the advent of 3D directional antennas that transmit with a certain radiation pattern in both the azimuth and elevation planes, the vertical plane of the antenna pattern can also be used for transmission optimization \cite{portapp}. This implies that the MIMO precoding of data streams and the antenna radiation pattern can be simultaneously optimized in both the azimuth and elevation planes to bring about significant improvements in the system performance. 

The two limitations of the existing Massive MIMO models discussed above motivated the development and study of an advanced Massive MIMO technology that not only allows a large number of antennas to be packed within feasible BS form factors but also exploits the channels' degrees of freedom in both the horizontal and vertical planes to improve the system performance. The next subsection presents FD-MIMO as the key enabling technology that resolves these issues and that is the subject of this tutorial.

\subsection{FD-MIMO Systems}

Although the maximum benefits of Massive MIMO techniques can be realized when a large number of antennas are placed in the horizontal domain with an acceptable inter-antenna spacing, the spatial limitations at the top of the BS towers in real-world deployment scenarios render the use of massive linear antenna arrays infeasible as discussed earlier \cite{FD2, FD1}.

To tackle the deployment challenge associated with massive linear antenna arrays and to enable antenna pattern adaptation in the 3D space,  Full Dimension (FD) MIMO was identified as a promising candidate technology for evolution towards the next generation LTE systems during the 3GPP LTE Release-12 workshop  in 2012 \cite{3GPP3D}. FD-MIMO utilizes a 2D active antenna array (AAA) that integrates a 2D planar passive antenna element array and an active transceiver unit array into an active antenna system (AAS). The 2D array structure allows a large number of antenna elements to be packed within feasible BS form factors. As an example, again consider the deployment of $64$ antennas but now in an $8 \times 8$ 2D planar array with $0.5 \lambda$ inter-antenna spacing. This  would require  an array of dimensions $\sim 50 cm \times 50 cm$, which can be readily installed at existing BSs. This form factor limitation has been illustrated in Fig. \ref{form_factor}.

\begin{figure}
\centering
\includegraphics[scale=.5]{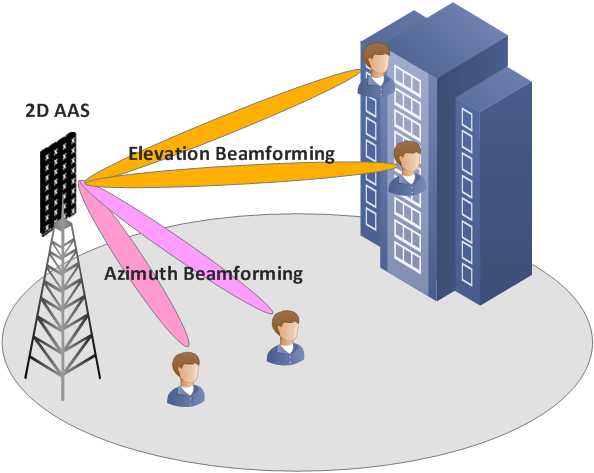}
\caption{FD-MIMO layout.}
\label{layout}
\end{figure}

 FD-MIMO has two distinguishing features as compared to the current LTE systems. First, it benefits from the extra degrees of freedom offered by massive transmit (Tx) antenna arrays installed within feasible BS form factors. Second, the use of an AAS provides the ability of dynamic beam pattern adaptation in the 3D space. 

In traditional cell site architectures, the base transceiver station (BTS) equipment is located remotely from the passive antenna element array and both are connected via long cables. The passive antenna system can not change its radiation pattern dynamically  and can support geographically separated users simultaneously using only MU-MIMO precoding methods in the baseband. The last decade, however, has witnessed an evolution in this traditional cell site architecture  to the one wherein the active transceiver components, including amplifiers and phase shifters, are located in the remote radio unit (RRU) closer to the passive antenna element array. This separation of the digital radio in the base band unit (BBU), from the analog radio in RRU, not only allows for a reduction in the equipment foot-print at the cell site but also enables a more efficient network operation. 

The next stage in the evolution of cell site architecture is the integration of the active transceiver unit array into the passive antenna element array at the top of BS tower, resulting in an AAS \cite{AAS, AAS1, AAS2, AAS3}.  The AAS can change its directional radiation pattern at each transmission, since the amplitude and phase weight applied to each antenna element in the system can be dynamically controlled through the power amplifier (PA) connected directly to that element. Arranging the transceivers in a 2D array and mapping each transceiver to a group of antenna elements arranged in the vertical direction extends this adaptive electronic beamforming capability to the elevation plane in addition to the conventional azimuth plane. This results in an electric downtilt feature \cite{3Dbeamforming, AAS, AAS1, AAS2}. This beam-tilt feature of AAS enables the vertical radiation pattern of the array to be optimized simultaneously along with the MU-MIMO precoding.  Popularly known as 3D/elevation beamforming, this technique can help realize more directed and spatially separated  transmissions to a larger number of users, leading to an additional increase in throughput and coverage as confirmed through several field trials \cite{practicals, practicals1, practicals2, practicals3}. A typical FD-MIMO deployment scenario is illustrated in Fig. \ref{layout}, for one sector of a macro-cell BS equipped with a 2D AAS.

\subsection{Overview -  From 3GPP to Theory}

\begin{figure*}
\centering
\includegraphics[scale=1.2]{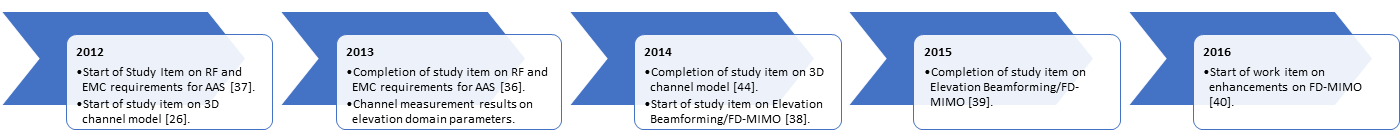}
\caption{3GPP roadmap for the standardization of FD-MIMO.}
\label{roadmap}
\end{figure*}

FD-MIMO transmission/reception was first identified as a promising  technology for the next generation cellular systems in the 3GPP Release 12 workshop in 2012 \cite{3GPP3D}. To realize FD-MIMO techniques, the radio frequency (RF) and electromagnetic compatibility (EMC) requirements for AAAs were outlined in the 3GPP technical report (TR) 37.840 \cite{TR37.84}. This TR was a result of the study item approved at Technical Specification Group (TSG) Radio Access Network (RAN) Meeting number (No.) 53 \cite{TR37.84S}.  The report defined relevant terminology for the study of an AAS BS, studied the  transmitter and receiver characteristics along with their impact on the system performance and determined the appropriate approaches for the standardization, specification implementation and testing of a BS equipped with a 2D AAS. 

FD-MIMO then became a subject of extensive study in the 3GPP Release 13 and Release 14, with focus on identifying key areas in the LTE-Advanced standard that need enhancement to support elevation beamforming with upto $64$ antennas placed in a 2D array \cite{3GPP3D2, TR36.897, 3GPP3DRel14}. In 2015, the 3GPP made great progress on the development of transceiver architectures for FD-MIMO systems through the study item in \cite{3GPP3D2}. The findings were documented in  TR36.897 with the objective to help understand the performance benefits of standard enhancements targeting 2D antenna array operation with 8 or more transceiver units (TXRUs) per transmission point, where each TXRU has its own independent amplitude and phase control and is mapped to a group of physical antenna elements \cite{TR36.897}. A TXRU  is a generic term defined in TR37.840 for signal transmission under identical channel conditions. Note that the term antenna port is often used inter-changeably with antenna TXRU, where an antenna port is generally defined in LTE in conjunction with a reference signal (RS). The TR36.897 presented the structure of the 2D AAA, along with guidelines on the inter-element spacing, number of TXRUs, number of antenna elements per TXRU and their polarization. The virtualization model for the antenna elements per TXRU and the target operating frequency range, considering practical antenna size limitations, were also discussed. Based on the conclusions of this study item, the 3GPP developed specification support for FD-MIMO in Release 13 by enhancing the relevant RSs and channel state information (CSI) reporting mechanism. However, the specification in Release 13 supported upto 16 antenna TXRUs and therefore the benefit from an antenna array with more than 16 TXRUs was limited. There was also no support for providing higher robustness against CSI impairments and no enhancement on CSI reports to enable efficient MU spatial multiplexing. FD-MIMO work item  for Release 14 was proposed in \cite{3GPP3DRel14} to address these Release 13 limitations. 

In order to facilitate the evaluation of FD-MIMO systems, a large effort in developing a 3D spatial channel model (SCM), that not only takes into account both the azimuth and elevation angles of the propagation paths but also incorporates the 3D radiation patterns of the active antenna elements, was also needed. There was some discussion on the 3D SCM in Wireless World Initiative New Radio (WINNER) II project \cite{Winner}, but the complete model was not developed. In 2010, WINNER+ \cite{Winner+} generalized the 2D SCM to the third dimension by including the elevation angles, but many parameters were not determined. The elevation plane was considered in the large scale in the ITU channel to model the radiation pattern of the antenna ports, whereas in the small scale the propagation paths were  modeled only in the azimuth plane \cite{ITU}. The ITU approach is approximate in modeling the vertical antenna radiation pattern because it abstracts the role played by the antenna elements constituting an antenna port in performing the downtilt by approximating the vertical radiation pattern of each port by a narrow beam in the elevation. The ITU model is, therefore, not a full 3D channel model, even though it is widely used in many works on FD-MIMO and 3D beamforming. The first full 3D SCM was completed in 2014 and documented in the 3GPP TR36.873 \cite{TR36.873}, with the objective of facilitating the proper modeling and evaluation of physical layer FD-MIMO techniques. This model captures the characteristics of the AAS at an element level accounting for the weights applied to the individual elements to perform downtilt and now forms the basis of all subsequent 3GPP TRs on FD-MIMO. A detailed literature review on the development of the ITU based and 3GPP TR36.873 based 3D SCMs will appear in the next section. The 3GPP progress on FD-MIMO has been illustrated in Fig. \ref{roadmap}.

In parallel to the standardization efforts being made in the 3GPP, several research papers addressing this subject through theoretical analysis, system-level simulations and measurement campaigns have appeared.  The authors in \cite{FD1} showed using system-level simulations that under the 3D SCM  2-3.6 times cell average throughput gain and 1.5-5 times cell edge throughput gain can be achieved by FD-MIMO arrays when compared to the reference configuration of 2 antenna ports used in the LTE Release 9 and Release 10 \cite{TR36.814}. The authors in \cite{FD2} studied the characteristics of FD-MIMO in terms of deployment scenarios, 2D antenna array design and 3D channel modeling, and outlined the potential enhancements required in the LTE standard to support this technology. The performance gain of FD-MIMO over legacy MIMO systems with a fewer number of antenna ports was confirmed through system-level evaluations. The authors in \cite{3DMIMO} evaluated the performance of FD-MIMO with upto 64 antenna ports using system-level simulations and field trials measurements. Several important insights were drawn such as the performance gain of FD-MIMO is higher in the urban micro (UMi) scenario than in the urban macro (UMa) scenario. The authors in \cite{BS_loc} demonstrated that the performance of FD-MIMO arrays varies in an indoor environment depending on the location of the array. 

The authors in \cite{FD3} provided a very comprehensive overview of FD-MIMO systems in 3GPP LTE Advanced Pro in context of the discussion and studies conducted during the 3GPP Release 13. The article discussed the  standardization of FD-MIMO technology focusing on antenna configurations,  transceiver architectures, 3D channel model, pilot transmission, and CSI measurement and  feedback schemes. A new transceiver architecture proposed in TR36.897 as full/array-connection model, a new RS transmission scheme referred to as beamformed CSI-RS transmission and an enhanced channel feedback scheme referred to as beam index feedback were introduced and discussed. The authors in a recent work \cite{FD4} identified interference among MIMO streams for a large number of users with limited channel feedback and hardware limitations, such as calibration errors, as practical challenges to the successful commercialization of FD-MIMO technology in 5G systems. They developed a  proof of concept (PoC) BS and user prototype to overcome these challenges and demonstrated the practical performance and implementation feasibility of FD-MIMO technology through field trials.  The extension of FD-MIMO to distributed AAAs as a means to satisfy the wireless data demand beyond 5G was discussed in \cite{FD5}.

The authors in \cite{ourwork, ourworkGC, ourworkTCOM, ourwork_access,ourworkSPAWC,ourworkWCNCold} focused on the high spatial correlation experienced by FD-MIMO arrays due to their compact structure. They characterized this correlation and took it into account in the performance analysis of FD-MIMO techniques. In fact, the high spatial correlation in FD-MIMO arrays has been utilized in some works to form the  Rayleigh correlated 3D channel model, which depends on the quasi-static spatial channel covariance matrices of the users. FD-MIMO techniques, like elevation beamforming, can be designed theoretically using these correlation based models with the help of tools from random matrix theory (RMT) that are quite useful in the large antenna regime. The implementation of these schemes will then require the estimation of large-scale parameters only instead of the full channel vectors. More discussion on the importance of spatial correlation in FD-MIMO analysis will appear in Section II and Section V. 

The additional control over the elevation dimension provided by FD-MIMO enables a variety of strategies such as sector-specific elevation beamforming, cell splitting, and user-specific elevation beamforming \cite{beamformingg}, which have been a subject of several works in the recent years. The initial works in this area relied on system-level simulations and field trials to confirm the performance gains achievable through elevation beamforming \cite{practicals, practicals1, practicals2, practicals3, beamforming1, 3Dbeamformingmulticell}. Later, some theoretical studies on downtilt adaptation utilizing the approximate antenna port radiation pattern expression from the ITU  report \cite{ITU} and 3GPP TR36.814 \cite{TR36.814} appeared in \cite{downtilt3, utility, portapp, 3Dbeamforming, beamforming2, beamforming3,7268913}. Only a couple of works have very recently performed elevation beamforming analysis for the FD-MIMO transceiver architectures proposed in the 3GPP TR36.897 utilizing the most recent, theoretically accurate and complete 3D channel model developed in TR36.873 \cite{ourworkTCOM, ourworkWCNC}. Detailed discussion on these works can be found in Section II and Section VI.

\subsection{Motivation and Structure of the Tutorial}

Due to the expanding interest in elevation beamforming with the advent of FD-MIMO technology, it is required to have a unified and deep, yet elementary, tutorial that introduces elevation beamforming analysis for beginners in this field. Although there are some excellent magazine articles that focus on different aspects of FD-MIMO, including the design of AAAs, development of TXRU architectures, extension of the 2D SCM to the third dimension, and enhancements to the CSI-RS and feedback schemes \cite{FD1, FD2, FD3, FD4, FD8, 3Dbeamformingmulticell, 3GPP_survey, drabla}, but these works address these aspects from an industrial point of view, focusing on the ongoing discussion in the 3GPP on the standardization of FD-MIMO technology. While these works provide an excellent background for readers interested in the practical system design and the system-level performance evaluations of FD-MIMO scenarios, they do not provide a theoretical framework for the design of elevation beamforming schemes for these systems. The ability of elevation beamforming allowed by AAAs needs careful study and design to enable an efficient implementation in future 5G systems.    

This tutorial is distinguished in being the preliminary one on FD-MIMO systems and in its objective of equipping the readers with the necessary information on the underlying array structures, the TXRU virtualization models, the 3D  channel modeling approaches and the correlation characterization methods  to allow them to devise, formulate and solve beam adaptation and optimization problems in the 3D space. Although, some useful elevation beamforming strategies have been proposed in existing theoretical works  \cite{downtilt3, utility, portapp, 3Dbeamforming, beamforming2, beamforming3,7268913}, but almost all of them have one of the following limitations.

\begin{itemize}
\item Majority of these works utilize the approximate antenna port radiation pattern expression introduced in 2009 in the ITU report and in 2010 in the 3GPP TR36.814. This approach approximates the main lobe of the vertical radiation pattern of a port  by a narrow beam in the elevation and discards the effect of sidelobes. The practical relevance of these works when considering actual FD-MIMO AASs, where the antenna port radiation pattern depends on the underlying geometry of the port, including the number of elements in it, their patterns, relative positions and applied downtilt weights, is doubtful. 
\item Since the radiation pattern of the array depends on the underlying geometry of the antenna elements, so it is important to model the 3D channel on an element level.  The TR36.873 now documents the generation of this element level 3D channel  in a very comprehensive manner. Works on 3D beamforming should consider this or related models to make the proposed methods more reliable and relevant to realistic 3D propagation environments. However, existing works optimize the downtilt angle using the ITU radiation pattern with channels modeled with respect to antenna ports. In practice, the downtilt weight functions applied to the  physical antenna elements constituting the ports should be optimized under an element level channel model. 
\item Compact structure of large-scale antenna arrays and small values of elevation angular spread in realistic propagation environments drastically increase the spatial correlation in FD-MIMO systems. The adverse impact of correlation on the capacity of MIMO systems has been widely studied for over a decade now in several works \cite{correlation_new, correlation_new1, correlation4, Shafi06polarizedmimo, correlation6,general}. The effects of correlation and mutual coupling are even more pronounced in the compact arrays considered in FD Massive MIMO settings, making it imperative to take these effects into account in the design and performance evaluation of elevation beamforming schemes. This is again missing in existing works. 
\end{itemize}

The authors believe that the absence of a theoretical framework for 3D beamforming that utilizes realistic antenna array designs and 3D channel models is because most of the literature on FD-MIMO architectures and channel modeling is found in 3GPP reports  instead of theoretical works, which creates difficulty in establishing a proper link between the industrial standards for FD-MIMO and the theoretical study of 3D beamforming.   This tutorial aims to bridge this gap between theory and industry. 

In contrast to the existing magazine articles that provide an overview of different aspects of FD-MIMO from the 3GPP's point of view \cite{FD1, FD2, FD3, FD4, FD8, 3Dbeamformingmulticell, 3GPP_survey, drabla}, the authors have written this article in a tutorial style, with the intention of equipping the readers with the necessary mathematical tools, including definitions and basic analytical developments, that will  be highly useful for them to build their theoretical analysis of elevation beamforming using FD-MIMO architectures. The authors have ensured that the mathematical equations are presented at a level comprehensible to readers with preliminary  background in channel modeling and beamforming design. All technical portions of this tutorial contain adequate references that provide detailed  development of the included equations. To facilitate the navigation of the tutorial, we present its organization below. 

We first present the key features of FD-MIMO and summarize the available literature in Section II. Section III introduces the relevant array and TXRU architectures that enable elevation beamforming. This is followed by the presentation and comparison of the ITU based and TR36.873 based 3D ray-tracing channel models  in Section IV. This section requires preliminary background in MIMO channel modeling, which can be acquired using the references provided in it. The spatial correlation in FD-MIMO arrays is characterized and compared based on both channel modeling approaches in Section V. The resulting expressions are used to form the Rayleigh correlated channel model. This section can also be easily followed with some background in  channel modeling and correlation analysis. The authors believe that the equations in Sections IV and V are presented in a very sequential manner and their use and significance can be easily understood without delving into the technical derivations behind them. 

Finally, all the aspects of FD-MIMO are put together to provide a theoretical framework for the design of elevation beamforming schemes in single-cell single-user, single-cell multi-user and multi-cell multi-user multiple-input single-output (MISO) systems in Section VI. This section mostly contains qualitative analysis and comparison between different schemes.  Some related research directions and open issues are outlined in Section VII, followed by concluding remarks in Section VIII.

\section{Key Features of FD-MIMO}

\begin{figure*}
\centering
\includegraphics[scale=.4]{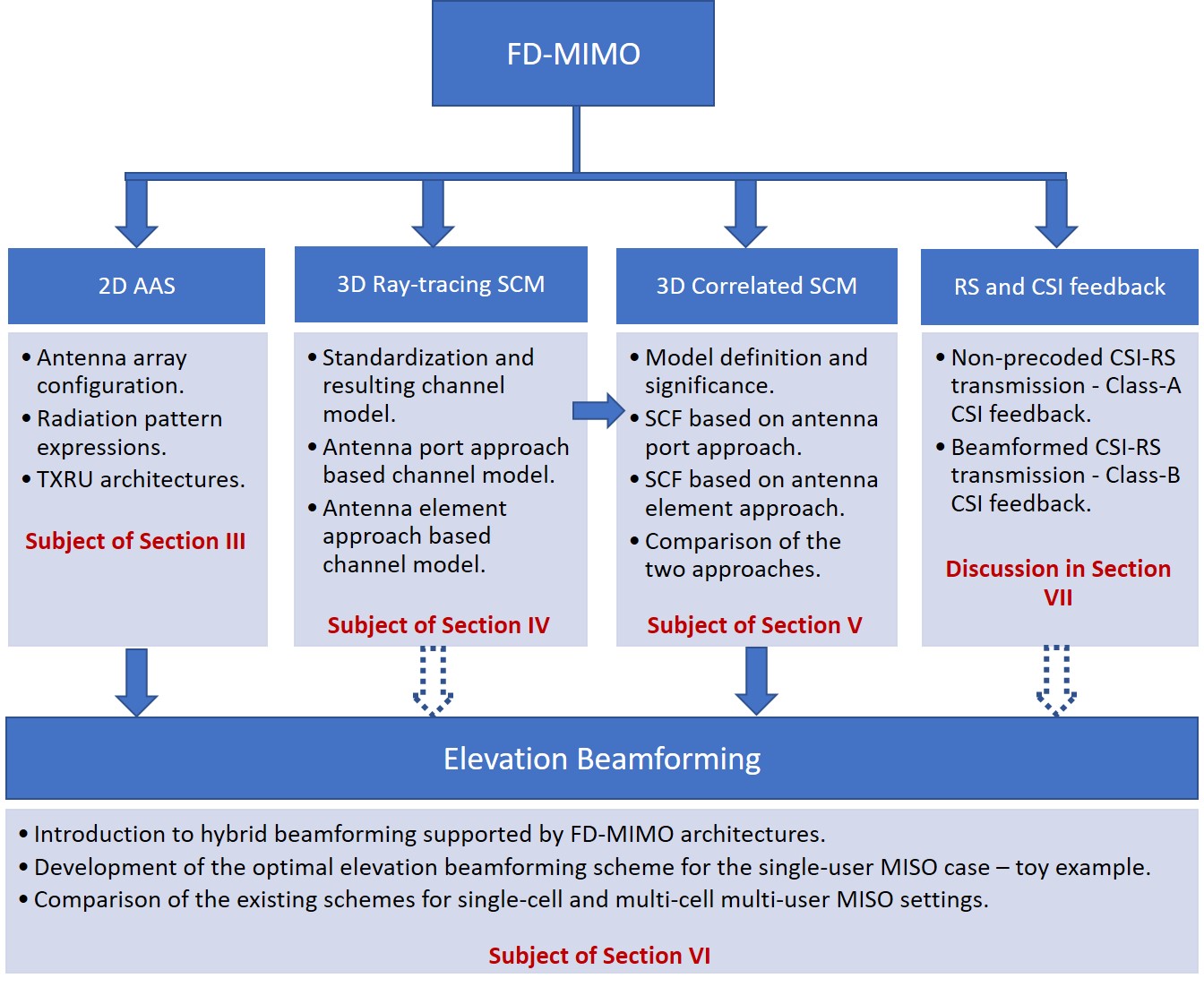}
\caption{Key Features of FD-MIMO. The solid arrows represent the relationships covered in detail in this tutorial while the dotted arrows represent the connections that have been discussed but not dealt with in detail. }
\label{structure}
\end{figure*}

In this section, we introduce the key features of FD-MIMO systems and discuss the related literature in detail. The main aspects of FD-MIMO highlighted here include the design of efficient 2D AAAs, the development of 3D ray-tracing  SCMs, the characterization of spatial correlation and the development of 3D correlated  channel models, the design of RS transmission and CSI feedback schemes, and the development of elevation beamforming strategies. These aspects and their relationships with each other are illustrated in Fig. \ref{structure}. This figure also outlines the concepts covered under each aspect and the related sections in this tutorial.  

\subsection{2D Active Antenna System (AAS)}

In order to realize the benefits of FD-MIMO, an efficient implementation of a 2D AAA is a key requirement. In an active antenna-based system, the gain and phase of the transmitted beam is controlled dynamically by adjusting the excitation current applied to the active components such as PA and low noise amplifier (LNA) attached directly to each antenna element. Arranging these active antenna elements in a 2D array and mapping a group of vertically arranged antenna elements to a TXRU fed with a data symbol, one can control the transmitted radio wave in both the vertical (elevation) and horizontal (azimuth) directions. This type of wave control mechanism is referred to as 3D beamforming.

In addition to providing the ability of 3D beamforming, another benefit of 2D AAA is that it can accommodate a large number of antenna elements without increasing the deployment space. The form factor considerations in the design of 2D AAA have been discussed in \cite{FD1, FD2, FD3, FD_nonunform} and illustrated in Fig. \ref{form_factor} in the introduction. The authors in \cite{FD1} proposed and studied a 2D AAA comprising of 8 antenna ports in the horizontal direction and 4 antenna ports in the vertical directions, resulting in a total of 32 feed ports, where each port consisted of a 4-element vertical sub-array to provide enhanced directional antenna gain in the elevation domain. With a $0.5 \lambda$ horizontal antenna spacing and $2 \lambda$ vertical antenna spacing, they showed that it was possible to construct an FD-MIMO 2D AAA of size $0.5 m \times 1m$ at $2.5$ GHz operating frequency, which could comfortably fit on a macro-cell BS tower. An actual functioning example of this 2D AAA was shown in Fig. 3 of \cite{FD2}, where the patch antenna elements were installed in the $+/- 45^{o}$ directions, resulting in dual-polarization on the two diagonal planes. Further details on the physical construction of the FD-MIMO AAA can be found in \cite{FD2,  AAS1}. 

Another important aspect related to the AAA design in a typical FD-MIMO implementation is that the radio resource is organized on the basis of antenna ports, antenna TXRUs and physical antenna elements. The 3D MIMO precoding of a data stream is therefore implemented sequentially in three stages; port-to-TXRUs precoding, TXRU-to-physical elements precoding, and application of the element radiation pattern. These stages have been discussed in detail in Section III.

\subsection{3D Ray-tracing SCMs}

The 2D SCMs, that capture the characteristics of the propagation paths in the azimuth plane only, have been conventionally used in both academia and industry for the study and evaluation of technologies designed for BSs equipped with ULAs of antennas. The design and study of FD-MIMO systems utilizing 2D AAAs, however, require 3D SCMs that not only take into account both the azimuth and elevation angles of the propagation paths but also incorporate the 3D radiation patterns of the active antenna elements. Note that most existing SCMs proposed by the industry and academia follow a ray-tracing approach where the channel model takes the actual physical wave propagation into account. These ray-tracing models express the channel  between the BS and the user as a sum of the propagation paths characterized by their powers, delays and angles.  This subsection will focus on the development of the 3D ray-tracing SCMs.

Encouraged by the preliminary results on the potential of 3D beamforming to improve the performance of contemporary cellular networks \cite{practicals, practicals1, practicals2, practicals3}, an extensive research activity was carried out in the 3GPP to develop and standardize 3D channel models. The preliminary studies considered the 3D channel between the antenna ports rather than between the individual antenna elements constituting these ports. Both ITU and 3GPP TR36.814 proposed the use of 3D directional antenna port radiation pattern expressions in the modeling of the wireless channel in \cite{ITU} and \cite{TR36.814} respectively while assuming the propagation paths to be characterized using azimuth angles only. These reports considered a three sector macro-cell environment and modeled the horizontal and vertical radiation patterns of each antenna port using a half power beamwidth (HPBW) of $70^{o}$ and $15^{o}$ respectively. The vertical radiation pattern of each antenna port was expressed as a function of the electrical downtilt angle. This downtilt is actually a result of the amplitude and  phase weights applied to the vertical column of physical antenna elements constituting an antenna port. The ITU approach is therefore approximate in the sense that it abstracts the role played by antenna elements constituting an antenna port in performing the downtilt by approximating the vertical radiation pattern of each port by a narrow beam in the elevation.  

Utilizing these expressions, the authors in \cite{drabla, FD1} extended the 3GPP 2D SCM to the third dimension by taking into account the elevation domain parameters. To generate these parameters like the elevation spread at departure (ESD), \cite{FD1} relied on statistics provided by WINNER and COST273 \cite{Winner}, \cite{Winner+}, \cite{COST}.  An overview of 3D channel modeling was provided in \cite{PES1}. The authors used some examples of 3D measurements to discuss the distance dependent nature of elevation parameters, correlation between azimuth and elevation  angles and extension of antenna array modeling to the third dimension. The authors in \cite{drabla} presented the preliminary 3GPP activity on developing the 3D channel model for elevation beamforming and positioned it with respect to previous standardization works. It established the common link between different channel modeling approaches in theory and standards and presented some most used standardized channel models, focusing mostly on the ITU and 3GPP TR36.814 approach to model the 3D radiation pattern.  This work also pointed out the approximate nature of this approach and stressed on the importance of developing a more accurate 3D channel model that takes into the account the characteristics of individual antenna elements.  Since using the ITU approach allows us to model the channel with respect to antenna ports instead of antenna elements, so it is  referred to as the `antenna port approach' towards 3D channel modeling in this work. 

In order to enable the optimization of 3D beamforming techniques in practice, the channels between antenna ports should be expressed as a function of the channels between the antenna elements constituting these ports and the applied weight functions.  A study item was initiated in 2012 \cite{3GPP3D} to finalize the detailed specifications of this element-level 3D SCM. The outcome of this study item resulted in the first complete 3D channel model introduced in the 3GPP TR 36.873 \cite{TR36.873}, which takes into account the geometry of the array and the characteristics of the antenna elements constituting the AAS quite meticulously.  Using the guidelines provided in this TR, the antenna port radiation pattern can be expressed as a function of the number of antenna elements constituting the port, their patterns,  positions and the phase and amplitude weights applied to these elements.  This model has been  presented and utilized in several papers on FD-MIMO \cite{FD2, FD3, 3GPP_survey} and is expected to form the basis of future studies on elevation beamforming. This new and more accurate approach is referred to as the `antenna element approach' towards 3D channel modeling in this work. Very recently, the authors in  \cite{AAS_reviewer} have established a similar full 3D channel model to support the performance evaluation of AAA-based wireless communication systems. The vertical radiation pattern of an antenna port has been expressed as a function of the individual element radiation pattern  and the array factor of the column of elements constituting the port. The authors used the developed 3D SCM to study the effect of mechanical downtilt, electrical downtilt and the combination of two downtilts on the coverage and capacity performance of different AAA configurations and showed that downtilt optimization can introduce significant gains in coverage and capacity, when antenna ports have narrower vertical HPBWs.

There is still ongoing work in both industry and academia on improving the 3D channel model and the estimates of different elevation domain parameters \cite{elevation_camp1, elevation_camp2, elevation_camp3, elevation_camp4}.  An analysis of the elevation domain parameters in the urban microcell scenario with channel measurements at 2.3 GHz center frequency can be found in \cite{elevation_camp1}. A smaller ESD was observed for  higher Tx antenna height. Also, a significant difference in the ESDs between the line-of-sight (LOS) and the non-line-of-sight (NLOS) propagation conditions was reported, with the ESD following a negative exponential model with respect to the distance for the former and a linear model for the latter. The authors in \cite{elevation_camp2} identified the limitations of the existing 3D channel models in describing the cross-correlation coefficients of channel large-scale parameters, the distance dependent properties of elevation domain parameters  and the inter-dependence between azimuth and elevation angles. They used an outfield measurement campaign to propose a reliable 3D stochastic channel model that addressed these limitations. A log-normal distribution was proposed to fit the probability density function (PDF) of ESD, with a mean which decreased with distance. A mixture of Von-Mises Fisher distributions with a log-normally distributed concentration parameter was used to model the interdependency between azimuth and elevation angles. A summary of the 3D channel model development in the 3GPP can be found in \cite{3GPP_survey}.

Both the antenna port and the antenna element based ray-tracing channel modeling approaches are outlined and compared in Section IV.

\subsection{3D Correlated SCMs}

The compact structure of 2D AAAs deployed in FD-MIMO systems to meet the form factor requirements often results in small inter-element spacing between the antenna elements. This increases the spatial correlation in the array. Recently, measurement campaigns have also confirmed the small values of elevation angular spreads in realistic propagation environments, resulting in the elements to be highly correlated in the vertical domain \cite{PES1}. This dramatic increase in spatial correlation makes it imperative to characterize it and take it into account while evaluating the performance gains realizable through elevation beamforming techniques. 

Spatial correlation has been popularly known to deteriorate the system performance. While this is always true for point-to-point MIMO communications \cite{corr_impact, corr_impact1}, spatial correlation can actually be beneficial in multi-user Massive MIMO settings, where each user can experience high spatial correlation within its channel vector, but the correlation matrices are generally almost orthogonal for different users, resulting in each user getting the full array gain proportional to the number of antennas as shown in \cite{massiveMIMObook}. More discussion on this can be found in Section V-A.

The high correlation in FD-MIMO arrays can also reduce the CSI feedback overhead incurred in the implementation of elevation beamforming techniques. This is possible through the design of elevation beamforming schemes using correlated 3D channel models that depend on the quasi-static spatial channel covariance matrices of the users. A popular such channel model for point-to-point MIMO system is the Kronecker model \cite{kronecker_model} and for MU-MISO system is the Rayleigh correlated model \cite{assumptions}. The covariance matrices used to form these models can be estimated using knowledge of the slowly-varying large scale channel parameters instead of the small-scale channel parameters that vary instantaneously. Note that the 3D SCM discussed in the last subsection is a ray-tracing model, which is one way to generate correlated FD-MIMO channels. However, the explicit dependence of this channel model on the number of propagation paths and associated small-scale parameters like angles, powers and delays makes the theoretical analysis of 3D beamforming generally intractable. This has further motivated the characterization of 3D spatial correlation functions (SCF)s for FD-MIMO systems that can be used to form these so-called correlated channel models that depend only on the channel covariance matrices and facilitate the design of 3D beamforming methods using tools from RMT.

The initial SCFs proposed in literature were developed for 2D channels that ignore the elevation parameters in describing the antenna patterns and propagation paths \cite{correlation5, additional_correlation, correlation2, correlation3,correlation7,correlation9, MC5}.  In \cite{correlation3}, approximate closed-form expressions for the spatial correlation matrices were derived for clustered 2D MIMO channel models, assuming a Laplacian azimuth angle of arrival (AoA) distribution.  The Kronecker channel model was shown to provide a good fit to the ray-tracing channel model. This is encouraging for researchers interested in the use of the former for the design of Massive MIMO techniques.  

%In \cite{correlation2}, the authors derived exact closed-form expressions for the spatial correlation between the elements of linear and circular arrays using cosine, Gaussian and Von Mises azimuth AoA distributions. The authors encouraged the use of Von Mises distribution to model the AoAs and showed the correlation expressions to be simpler for it. The impact of mutual coupling (MC) on the correlation was also studied for VM distribution and it was shown that half wavelength inter-element separation is not enough to decorrelate adjacent elements.  The authors in \cite{MC5} obtained closed-form expressions for the spatial correlation and the average bit error rate (BER) for a compact antenna array under Gaussian angular distribution, taking into account the effects of MC between the antenna elements. The effects of the geometry of the antenna array and the propagation conditions, including fading, angular spread and mean AoA on the BER performance were studied. 

 %The authors in \cite{correlation8} derived the spatial correlation functions of  the uniform circular array (UCA) and the uniform linear array (ULA) of antennas for Gaussian and uniform angular distributions, considering only  azimuth angles and omnidirectional antennas.  The proposed analysis  relied on numerical integration methods to compute the correlation coefficients for  the circular array for Gaussian angular distribution.

The notion of spatial correlation in 3D propagation environments has been addressed in some research works. An important contribution in this area appeared in \cite{Shafi06polarizedmimo}. The authors developed closed-form expressions for the spatial correlation and large system ergodic mutual information (MI) for a 3D cross-polarized channel model, assuming the angles to be distributed according to Von Mises distribution. The authors in \cite{correlation1}, showed that elevation plays a crucial role in determining the SCF. The derivation was based on the spherical harmonic expansion (SHE) of plane waves and assumed the distribution of AoAs to be 3D Von Mises-Fisher. 

%Recently, some works on 3D channels have made use of unified 3D non-isotropic scattering models, such as the Von Mises Fisher and the Fisher-Bingham distributions, for  the generation of azimuth and elevation angles in a joint fashion. The angular spread in the azimuth and elevation is quantified as a single parameter in these distributions. Consequently, several useful insights into the impact of the angular parameters, parametrized by the mean and the spread, on the system performance can not be obtained independently for the two dimensions. Also, in the recent 3GPP standardization works on 3D channel modeling, independent distributions are proposed for the azimuth and elevation angles. 

In \cite{correlation4}, closed-form expressions for the SCFs of several omnidirectional antenna arrays utilizing a 3D MIMO channel model were derived. These SCFs then formed the covariance matrices that were used for the evaluation of channel capacity. The derived results were expressed as a function of angular and array parameters and used to study the impact of azimuth and elevation angular spreads on the capacity. However, this work assumed the angular distributions to be uniform. The uniform distribution although widely used due to its simplicity, does not accurately capture the characteristics of non-isotropic wave propagation and can underestimate the correlation in realistic propagation environments. The first attempt to develop a general analytical expression for the 3D SCF was made in \cite{general}, where the authors used the SHE of plane waves to derive  closed-form expressions for the correlation that can be applied to a variety of angular distributions.

 %(Laplacian distribution for the generation of elevation angles and Wrapped Gaussian distribution for the generation of azimuth angles)

Some recent works have dealt with spatial correlation in FD-MIMO architectures \cite{ourwork, ourworkGC, ourworkTCOM, ourwork_access, ourworkWCNCold, ourworkSPAWC}. In \cite{ourwork}, an exact closed-form expression for the SCF of FD-MIMO channels constituted by antenna ports arranged linearly, with each port mapped to a group of physical antenna elements in the vertical direction, was developed using ITU's antenna port radiation pattern expression. The proposed SCF is general in the sense that it can be used for any arbitrary choice of  antenna pattern and distribution of azimuth and elevation angles and can be computed using knowledge of the Fourier Series (FS) coefficients of the Power Azimuth Spectrum (PAS) and the Power Elevation Spectrum (PES) of the propagation scenario under study. A similar analysis was done in \cite{ourwork_access} for a uniform circular array (UCA) of antenna ports, utilizing  the ITU based channel representation. 

%However, here again the ITU approach was utilized and the role of the individual physical antenna elements in determining the radiation pattern and the spatial correlation between the antenna ports was ignored. 

A more recent work \cite{ourworkTCOM} addressed the limitations of existing correlation models and provided the correlation analysis for an FD-MIMO array, taking into account the correlations between all the elements constituting the antenna ports.  The SCF for the antenna elements was first derived using the 3GPP 3D ray-tracing SCM in \cite{TR36.873}. The correlation between the antenna ports was then expressed as a function of the correlation matrix of the elements constituting the ports and the applied downtilt weight vectors. This SCF can be easily utilized to form the Tx and receive (Rx) covariance matrices that constitute the Kronecker channel model in a point-to-point MIMO system and the Rayleigh correlated channel model in a MU-MISO system. Elevation beamforming schemes can be readily developed using these correlated 3D channel models utilizing RMT tools in the regime where the number of BS antennas grows large.  This has been discussed in detail in Section V-A.

 %In this work, the authors derived the exact SCF for the 3D channels constituted by individual antenna elements in the 2D AAS by expanding the array response of each antenna element using the SHE of plane waves.

\subsection{RS Transmission and CSI Feedback}

Reference signals (RS)s are necessary for both the demodulation of downlink data signals and CSI estimation. In the LTE standards, two types of RSs are used to support multi-antenna transmissions: CSI reference signal (CSI-RS) and demodulation reference signal (DM-RS) \cite{FD2}. CSI-RS is a low overhead downlink RS that is common to all users in a cell and allows each user to measure the downlink CSI. On the other hand, DM-RS is a user-specific downlink RS with the same precoding as the data signal and is transmitted on the same frequency/time resource as the data signal, to provide the user with a reference for data demodulation. The CSI feedback mechanism then allows each user to report a recommended set of values, including the rank indicator (RI), the precoding matrix indicator (PMI) and the channel quality indicator (CQI), where the first two are used to assist the BS in performing beamforming  \cite{3GPP3D_channelfeed}.

 If CSI-RSs and DM-RSs increase proportionally with the number of antenna ports as in the traditional LTE networks, it would impose a prohibitively high overhead of downlink RSs in FD Massive MIMO settings. To avoid this large overhead, an alternative is to support FD-MIMO in time-division duplex (TDD) systems only, where BS can exploit the channel reciprocity in determining the downlink CSI. However, the current standards and dominant current cellular systems, e.g. 3GPP LTE and LTE-Advanced, are all based on FDD protocol. Therefore enhancements to the CSI-RS schemes are required to support FD-MIMO and elevation beamforming in existing LTE-Advanced systems. 

In the 3GPP standardization process  of FD-MIMO, two CSI-RS transmission schemes have been proposed - non-precoded CSI-RS transmission and beamformed CSI-RS transmission \cite{FD3,  feedback_FD}. In the first strategy, users observe the non-precoded CSI-RS transmitted from each antenna port. By feeding back the precoder maximizing a specific performance criterion to the BS, the system can adapt to the channel variations. This scheme is referred to as the Class-A CSI feedback in the 3GPP. On the other hand, beamformed RS transmission uses multiple precoding weights in the spatial domain and the user picks the best beam and feeds back its index. This scheme, referred to as the Class-B CSI feedback, provides many benefits over non-precoded CSI-RS transmission that have been discussed in detail in \cite{FD3}. 

It is important to note that the conventional 2D codebooks cannot measure the CSI for FD-MIMO systems.  Kronecker product based codebook (KPC)  using a discrete Fourier transform (DFT) structure is usually considered in studies on FD-MIMO \cite{feedback_FD2, feedback_FD3, feedback_FD4, 3GPP3Dcodebook}. The authors in \cite{feedback_FD2} showed analytically that this codebook is suitable to quantize channels formed by antennas arranged in a 2D AAA and proposed some improvements to the channel quantization quality. Depending on whether the CSI values are computed independently for the azimuth and elevation dimensions or jointly based on the full channel, Kronecker product based CSI feedback schemes can be differently defined as explained in \cite{FD6}. More recently the beam index feedback method has been proposed in the 3GPP Release 13 to support beamformed CSI-RS transmission. Enhancements to the CSI reporting mechanism, to take into account the large number of antennas and the 2D array structure, have been made in 3GPP TR36.897 \cite{TR36.897}. More discussion on the two types of CSI-RS transmission and CSI feedback schemes can be found in Section VII.

%Existing works on elevation beamforming assume the availability of perfect CSI, which is not true in practice. Elevation beamforming schemes should be studied in light of the channel estimation and feedback schemes proposed for FDD FD-MIMO systems in the 3GPP and academia, and the effect of channel estimation errors on the performance of these schemes should be analyzed. 

\subsection{Elevation Beamforming}

%The traditional cell site architecture consists of a distributed baseband unit (BBU), a RRU, and passive antenna elements, which are connected by cables. This architecture not only requires more connection cables which increases the cost  but also allows only static phase shifts to be applied to the antenna elements in each antenna port resulting in a fixed downtilt of the beam transmitted from each port. To enable complete freedom in the design of beamforming in the spatial domain, more transceivers need to be integrated with the passive antenna elements at the top of BS towers.  This allows for adaptive electronic beamforming since the phase shift applied to each element can be controlled through the PA connected directly to each element. Arranging the transceivers in a 2D array and mapping each transceiver to a group of antenna elements arranged in the vertical direction extends this adaptive beamforming capability to the elevation plane in addition to the conventional azimuth plane.

The main distinguishing feature of FD-MIMO is its ability of dynamic beam pattern adaptation in the vertical plane, resulting in 3D beamforming. All the aspects of FD-MIMO presented so far are directly related to the development of efficient elevation beamforming schemes both in theory and practice. This tutorial brings together these aspects to provide a mathematical framework for the design of elevation beamforming schemes.

%The additional control over the elevation dimension in FD-MIMO systems enables a variety of strategies such as sector-specific elevation beamforming, cell splitting, and user-specific elevation beamforming \cite{beamformingg}. Vertical sectorization and cell splitting improve the system performance by using different downtilt angles and HPBWs of the radiation pattern for data transmission in different sectors. User-specific elevation beamforming attempts to increase the signal-to-interference-plus-noise ratio (SINR) at each user by adjusting the downtilt of the vertical radiation pattern of the antenna port in the direction of the targeted user while minimizing the interference leakage to the adjacent users. 

Dynamic tilt adaptation  for performance optimization has attracted a lot of research interest  since the introduction of the FD-MIMO concept. The initial works relied on system-level simulations and field trials to confirm the performance gains achievable through elevation beamforming \cite{practicals, practicals1, practicals2, practicals3, beamforming1, 3Dbeamformingmulticell}. The authors in \cite{practicals} used lab and field trials to show that 3D beamforming can achieve significant performance gains in real indoor and outdoor deployments by adapting the vertical dimension of the antenna pattern at the BS individually for each user according to its location.  Different terminal specific downtilt adaptation methods for interference avoidance with and without requirements on inter-BS coordination were considered in \cite{practicals2} for a multi-cell scenario. The authors in \cite{beamforming1} explored different realizations of vertical beam steering in both noise and interference limited scenarios and studied the impact of the downtilt, the vertical HPBW and the inter-site distance (ISD) on the SE and cell coverage. Since these works are based on pure system-level simulations and/or field trials results, they do not provide any theoretical design guidelines to determine the optimal tilts for the BS ports.

Later, some theoretical studies on downtilt adaptation utilizing the approximate antenna port radiation pattern expression from the ITU  report \cite{ITU} and 3GPP TR36.814 \cite{TR36.814} appeared in \cite{downtilt3, utility, portapp, 3Dbeamforming, beamforming2, beamforming3,7268913}.  The authors in \cite{downtilt3} utilized this expression to provide a vertical plane coordination framework to control inter-cell interference through the joint adaptation of the tilt angles according to  the scheduled users' locations, while maximum ratio transmission (MRT) was used in the horizontal domain to maximize the desired signal at the active users. The authors in \cite{beamforming3} considered a single-cell scenario and proposed to partition the cell into vertical regions and apply one out of a finite number of tilts and HPBW pairs when serving each region. A scheduler was used to schedule transmission to one of the vertical regions in each time slot to maximize a suitable utility function of the users' throughput.  The ITU based radiation pattern utilized in these works does not take into account the individual contributions of the antenna elements in determining the downtilt of each antenna port. More sophisticated elevation beamforming methods need to be devised which focus on optimizing the weights applied to the individual antenna elements in each antenna port to perform downtilt. 

In \cite{downtilt2}, the cell was partitioned into smaller sectors according to the traffic load and an optimal 3D beam pattern design for each sector utilizing a 2D AAS, with downtilt weights applied to the elements in each port, was achieved using convex optimization.  A recent work \cite{ourworkTCOM} made use of an FD-MIMO transceiver architecture proposed in the 3GPP TR36.897 and a simplified 3D channel model inspired from TR36.873 to propose algorithms for weight vector optimization in a single cell multi-user MISO setup, that were shown to outperform existing elevation beamforming methods. Further details on popular elevation beamforming schemes proposed for the single-cell single user, single-cell multi-user and multi-cell multi-user MISO systems can be found in Section VI.  The subsequent three sections will now deal with the AAA and transceiver architecture design, 3D channel modeling based on the ray-tracing method, and spatial correlation characterization and formation of correlated FD-MIMO channel models respectively. The results of these sections will be utilized in the design of elevation beamforming schemes in Section VI.

%This tutorial delves into the underlying array and TXRU architectures and 3D channel modeling principles to present and compare different methods for spatial correlation characterization and beam adaptation in FD-MIMO systems. Most of the literature on FD-MIMO architectures and channel modeling is found in the 3GPP reports  instead of theoretical works, which is the main reason researchers have difficulty establishing a proper link between 3D beamforming and the underlying practical array structures and 3D channel models.  This tutorials aims to bridge this gap between theory and industry, by presenting all the relevant information to the design of 3D beamforming, while focusing on practical transceiver architectures, 3D standardized channel models and spatial correlation between antenna elements. 

 \begin{figure*}
\centering
\includegraphics[scale=.425]{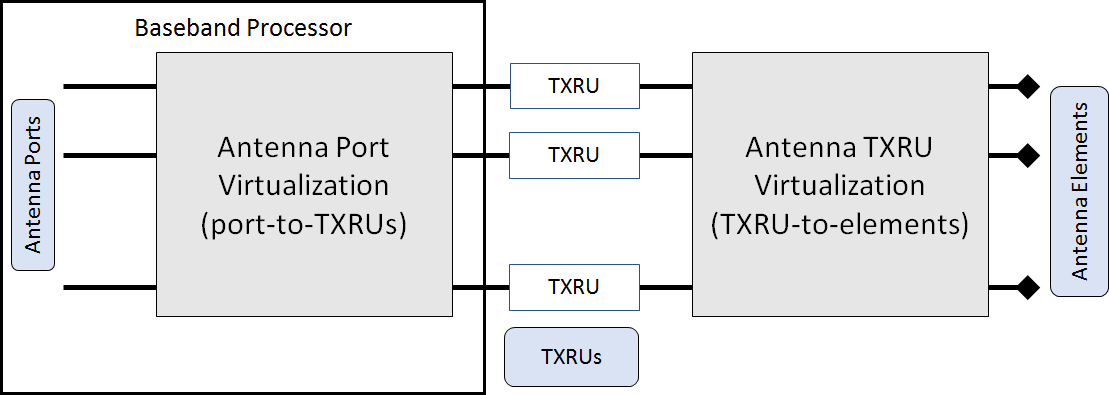}
\caption{Signal processing model for FD-MIMO precoding.}
\label{virtualization}
\end{figure*}

\section{Antenna Array Design}

In an FD-MIMO architecture reflecting a typical implementation of a 2D AAS, the radio resource is organized on the basis of antenna ports,  antenna TXRUs and physical antenna elements. The MIMO precoding of a data stream is implemented in three stages. 

First, a data stream on an antenna port is precoded on $Q$ TXRUs in the digital domain. This stage is referred to as antenna port virtualization. In LTE standard, an antenna port is defined together with a pilot, or a RS and is often referred to as a CSI-RS resource \cite{FD6}. Traditionally, a static one-to-one mapping is assumed between the antenna ports and the TXRUs and  both terms are often interchangeably used in literature. This tutorial will also assume this one-to-one mapping and therefore digital precoding is performed across the TXRUs.

Second, the signal on a TXRU is precoded on a group of physical antenna elements in the analog domain. This is the second stage referred to as antenna TXRU virtualization. The 3GPP has proposed some TXRU virtualization models that define the mapping of each TXRU to a group of co-polarized antenna elements through analog phase shifters or variable gain amplifiers. The signal from each TXRU is, therefore, fed to the underlying physical elements with corresponding virtualization weights (interchangeably referred to as downtilt weights in this article) to focus the transmitted wavefront in the direction of the targeted user.  

%
%
%Massive MIMO techniques rely on increasing the number of TXRUs to improve the spatial multiplexing gains, but considering that the cost of TXRUs is much higher than that of the antenna elements, each transceiver is connected to a group of  antenna elements through a well-designed mapping function. This is the second stage referred to as TXRU virtualization.  By balancing the tradeoff between cost and performance, the 3GPP has proposed some TXRU virtualization models for the future commercialization of FD-MIMO technology in 5G systems, where each TXRU is mapped to a group of co-polarized physical antenna elements through analog phase shifters or variable gain amplifiers. Consequently, this stage is also referred to as analog  precoding, where the signal from each TXRU is fed to the underlying physical elements with corresponding virtualization weights (interchangeably referred to as downtilt weights in this article) to focus the wavefront in the direction of the targeted user.  

The third stage is the application of the antenna element radiation pattern. A signal transmitted from each active antenna element will have a directional radiation pattern. This signal processing model is illustrated in Fig. \ref{virtualization}.

%In FD-MIMO systems, MIMO precoding of the data stream is done in three stages. First, the data stream is precoded on $M$ TXRUs; second, the data symbol on every TXRU is precoded onto $K$ antenna elements with corresponding weights; and third, the signal emitted from every antenna element is given the directional radiation pattern given in (\ref{minpattern}). The first stage is referred to as antenna port virtualization \cite{FD3}, while the second stage is referred to as TXRU virtualization. Antenna port virtualization is implemented in the digital domain generally using linear precoding methods like zero-forcing and maximum ratio transmission.  
%Since these weights are not extracted from a codebook, so this technique is referred to as non-codebook based precoding.
 %A FD-MIMO architecture utilizing a 2D active antenna systems, therefore, implements MIMO precoding of a data stream in three stages: (1) antenna-port virtualization: Each stream on an antenna port is mapped to  $Q$ transceiver units (TXRUs); (2) TXRU virtualization: a signal on each TXRU is mapped to a group of antenna elements; and (3) antenna element pattern: a signal transmitted by each active antenna element will have a directional radiation pattern. 

In order to enable practical design of elevation beamforming techniques through the optimization of applied virtualization weight functions, it is important to understand the antenna array configuration and the TXRU architectures proposed in the 3GPP for FD-MIMO technology. This is the subject of this section. The important terms used in this section have been defined in Table \ref{def}.

\begin{table*}[!t]
\caption{Important definitions}
\begin{tabular}{ l l}
\hline
\rowcolor{LightCyan}
 \textbf{Term} & \textbf{Definition}   \\
\hline
& \\
 Active antenna system (AAS) & A system which combines a passive antenna element array with an active transceiver unit array. \\
 &  \\
 Antenna element   & A single physical radiating element with a fixed radiation pattern.  \\
 &  \\
Radiation pattern & The angular distribution of the radiated electromagnetic field in the far field region. \\
& \\
Antenna array & A group of antenna elements characterized by the geometry and the properties of the \\
& individual elements. \\
& \\
Transceiver unit array & An array of transceiver units that generate/accept radio signals in the Tx/Rx directions.  \\
&  \\
Transceiver unit (TXRU) &  A transmitter/receiver mapped to a group of antenna elements that Tx/Rx \\
& the same data symbol.  \\
& \\
TXRU virtualization model & Model that defines the relation between the signals at the TXRUs and the signals \\
& at the antenna elements. \\
& \\
Array factor & The radiation pattern of an array of antenna elements when each element is considered to \\
& radiate isotropically. \\
& \\
Downtilt angle, $\theta_{tilt}$ & The elevation angle between the direction of the maximum antenna gain and the $\hat{\textbf{e}}_{z}$ direction. \\
& \\
Horizontal scan angle, $\phi_{scan}$ & The azimuth angle between the direction of the maximum antenna gain and the $\hat{\textbf{e}}_{x}$ direction. \\
& \\
Vertical virtualization (downtilt) weights, $w$  & The weights applied to vertically arranged elements in a TXRU to steer the beam \\
& at a particular downtilt value given by $\theta_{tilt}$. \\
& \\
Horizontal virtualization weights, $v$  & The weights applied to horizontally arranged elements in a TXRU to steer the beam \\
& at a particular value of $\phi_{scan}$. \\
& \\
Antenna gain (in a given direction), $G$ & The ratio of the radiation intensity, in a given direction, to the radiation intensity if  \\
& antenna radiated isotropically. \\
& \\
Horizontal half power beam-width (HPBW), $\phi_{3dB}$ & The azimuth angular separation in which the magnitude of the radiation pattern \\
& decreases by one-half (3\rm{dB}). \\
& \\
Vertical HPBW, $\theta_{3dB}$ & The elevation angular separation in which the magnitude of the radiation pattern \\
& decreases by one-half (3\rm{dB}). \\
& \\
Front-to-back ratio, $A_{m}$ & The ratio of maximum gain of an antenna to its gain in a specified rearward direction. \\
& \\
Side-lobe attenuation, $SLA$ & The maximum value of the side-lobes in the radiation pattern (away from the main lobe). \\
& \\
\hline
\end{tabular}
\label{def}
\end{table*}

\subsection{Antenna Array Configuration}

Unlike conventional MIMO systems utilizing passive antenna elements with omnidirectional radiation of energy, systems utilizing active antenna elements can control the gain, downtilt and HPBW of the beam transmitted from each TXRU dynamically by  adjusting the amplitude and phase weights applied to the elements within it. Arranging these active antenna elements in a 2D array allows for the dynamic adaptation of the radiation pattern in both azimuth and elevation planes, making it possible to control the radio wave in the 3D space. This type of wave control mechanism allowed by FD-MIMO is referred to as 3D beamforming, and FD-MIMO is also interchangeably known as 3D-MIMO. Since the radiation pattern of each TXRU depends on the number of antenna elements within it, the inter-element spacing and the applied weights, the AAS should be modeled at an element-level.

\begin{figure}
\centering
\includegraphics[width=2.5in]{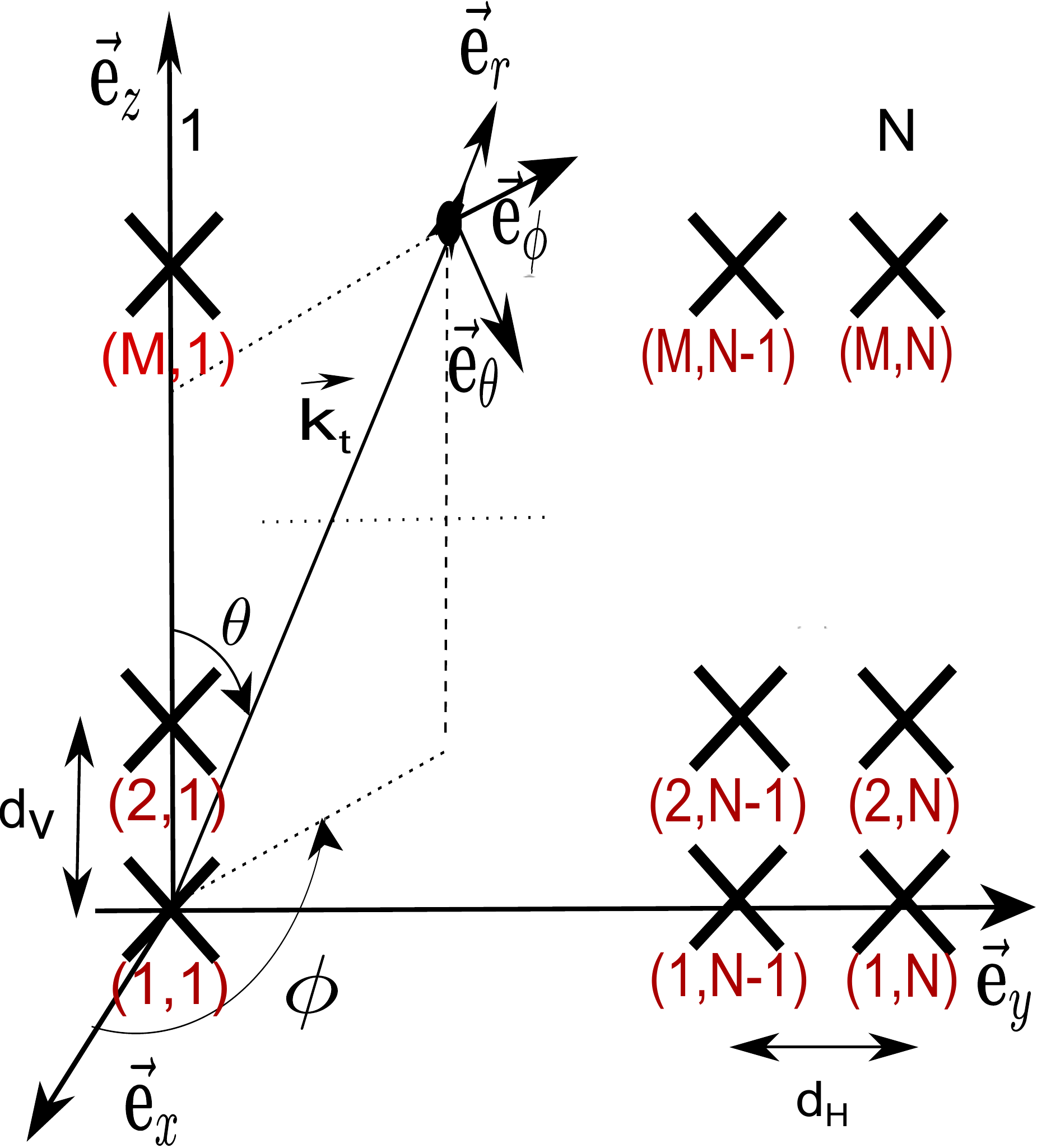}
\caption{2D antenna array system.}
\label{antenna1}
\end{figure}

FD-MIMO systems utilize a 2D planar uniformly spaced antenna element array model. The configuration is represented by $(M, N, P)$, where $M$ is the number of antenna elements with the same polarization in each column along the $\hat{\textbf{e}}_{z}$ direction, $N$ is the number of columns placed at equidistant positions in the $\hat{\textbf{e}}_{y}$ direction and $P$ is the number of polarization dimensions, with $P = 1$ for co-polarized system and $P = 2$ for dual-polarized system. The resulting configuration for cross-polarized antenna elements including indices for co-polarized antenna elements  is shown in Fig. \ref{antenna1}.  The inter-element spacing is represented by $d_{H}$ in the horizontal direction and $d_{V}$ in the vertical direction.

Each antenna element has a directional radiation pattern in the vertical and horizontal directions. The combined 3D radiation pattern of an individual antenna element is given in the 3GPP as  \cite{TR36.897}, \cite{TR36.873},
\begin{align}
\label{port}
A_{E}(\phi,\theta)&= G_{max,E}-\text{min}\{-(A_{E,H}(\phi)+A_{E,V}(\theta)),A_{m}\},
\end{align}
where,
\begin{align}
\label{minpattern}
A_{E,H}(\phi)&= -\text{min}\left[ 12 \left(\frac{\phi}{\phi_{3dB}}\right)^2, A_{m} \right] \rm{dB}, \\
A_{E,V}(\theta)&= -\text{min} \left[12 \left(\frac{\theta - 90^{o}}{\theta_{3dB}}\right)^2, SLA_{v} \right] \rm{dB},
\end{align}
where $\phi$ and $\theta$ are the azimuth and elevation angles respectively, $G_{max,E}$=$8 \rm{dBi}$ is the maximum directional antenna element gain, $\phi_{3dB}=65^{o}$ and $\theta_{3dB}=65^{o}$ are the horizontal and vertical HPBWs respectively, $A_{m}=30\rm{dB}$ is the front-to-back ratio and $SLA_{v}=30\rm{dB}$ is the side lobe attenuation in the vertical direction. The azimuth angles are defined from $0$ to $2\pi$ from the $\hat{\textbf{e}}_{x}$ direction and the elevation angles are defined from $0$ to $\pi$, from the $\hat{\textbf{e}}_{z}$ direction as shown in Fig. \ref{antenna1}. 

\subsection{Transceiver Unit}

\begin{figure}[!t]
\centering
    \includegraphics[width=3 in]{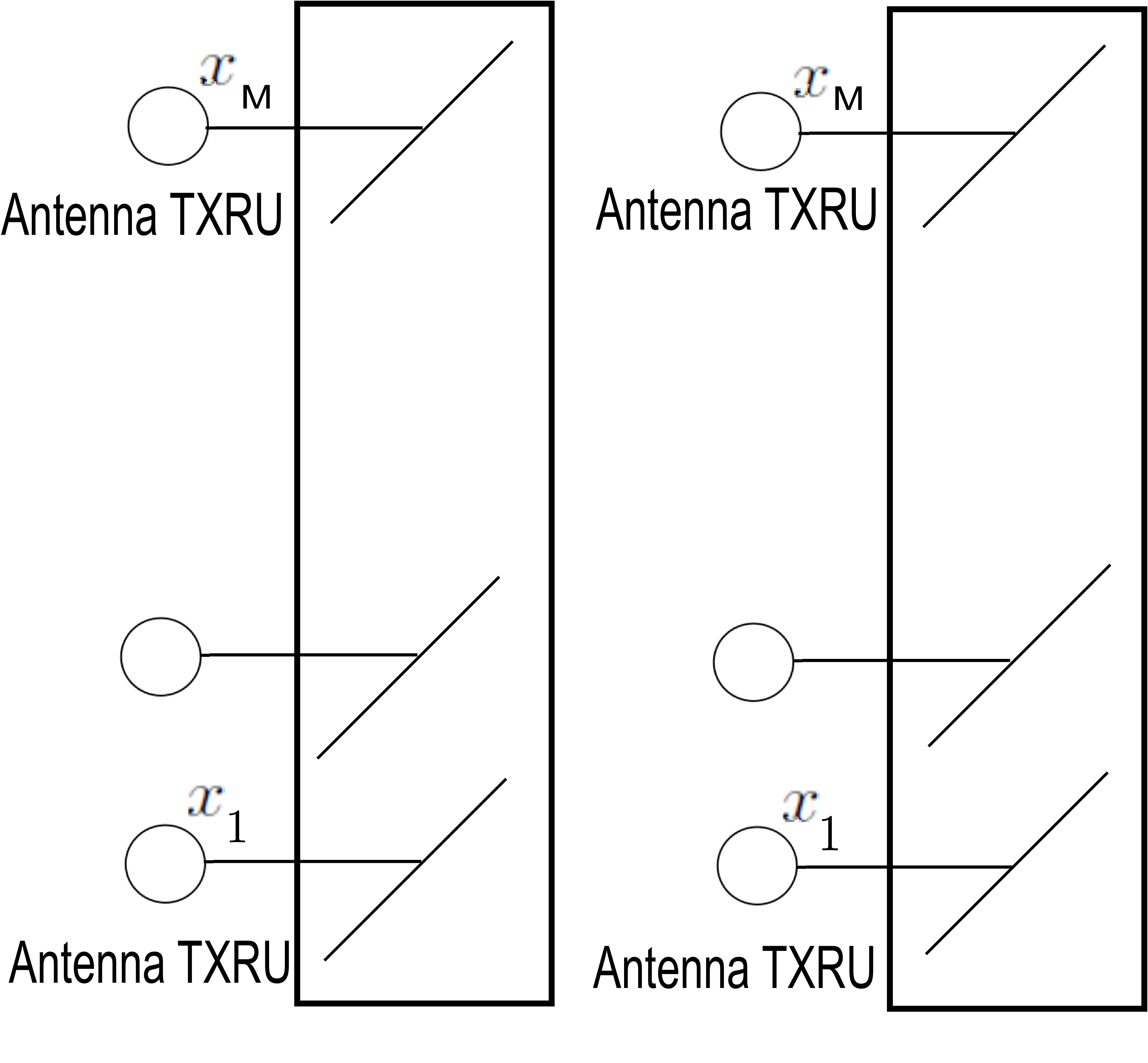}
		\caption{TXRU structure with $K=1$.}
\label{K=1}
\end{figure}

\begin{figure}[!t]
\centering
    \includegraphics[width=3.3 in]{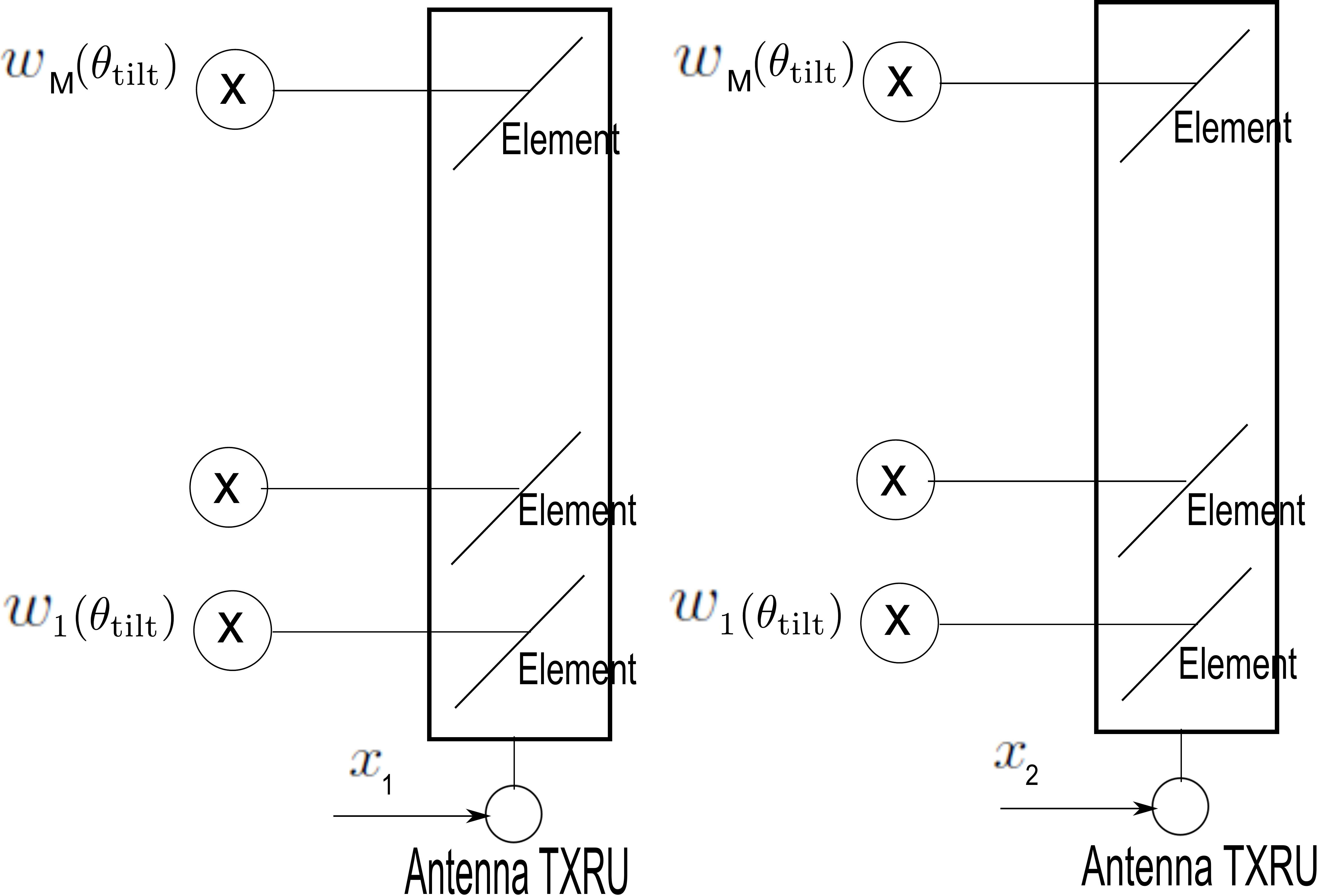}
		\caption{TXRU structure with $K=M$.}
\label{K=M}
\end{figure}

A transceiver unit (TXRU) refers to a group of antenna elements that Tx/Rxs the same data symbol.  To realize the electronic downtilt feature, it is desirable to map a TXRU to $K$ vertically arranged co-polarized antenna elements, with the same signal fed to these elements with corresponding weights to tilt the wavefront transmitted from that TXRU in the targeted direction.   Two values of $K$ were initially considered by the 3GPP TSG-RAN Work Group (WG) 1 in its studies on FD-MIMO; first was to choose $K = 1$ in which case the number of TXRUs per column will be equal to $M$ and second was to choose $K=M$, in which case each column of elements in Fig. \ref{antenna1} will correspond to one TXRU. The two cases have been illustrated in Fig. \ref{K=1} and Fig. \ref{K=M} respectively. Note that in the former, each element acts as a TXRU and carries a different data symbol, $x$, but does not support any beam-tilt feature. The MU-MIMO precoding is performed across all the elements resulting in high spatial multiplexing gains but also requiring high-dimension CSI acquisition. In the latter, the whole column of elements act as a TXRU, with the same symbol fed to the elements within a TXRU with corresponding weights $w_{k}(\theta_{tilt})$, $k=1,\dots, K$, to focus the radiation pattern in the targeted users' direction. The narrow beamwidth of the transmitted beams results in better spatial separation of the users. The MU-MIMO precoding is performed across a reduced number of TXRUs in this case.

The overall radiation pattern of a TXRU is essentially a superposition of the individual element radiation pattern in (\ref{port}) and the array factor for the whole TXRU, wherein the array factor takes into account the downtilt weights and the array responses of the elements in that TXRU. The exact antenna radiation pattern in \rm{dB} for a TXRU/port comprising of $K$ vertically arranged elements, denoted as $A_{P}^{E}(\phi, \theta, \theta_{tilt})$, is given by \cite{TR37.84},
\begin{align}
\label{element_pattern}
&A_{P}^{E}(\phi, \theta, \theta_{tilt})=A_{E}(\phi, \theta) + 20 \log_{10} |A_{F}(\theta, \theta_{tilt})|,
\end{align}
\normalsize
where $A_{E}(\phi, \theta)$ is given by (\ref{port}) and $A_{F}(\theta, \theta_{tilt})$ is the array factor for the column of elements in a TXRU computed as,
\begin{align}
&A_{F}(\theta, \theta_{tilt})= \sum_{k=1}^{K}{w}_{k}(\theta_{tilt}) \exp \left(i \textbf{k}_{t} . \textbf{x}_{t,k} \right),
\end{align}
where \textbf{.} is the scalar dot product, $\textbf{x}_{t,k}$ is the location vector of the $k^{th}$ antenna element and $\textbf{k}_{t}$ is the Tx wave vector.

It is also important to highlight here that the relationship of the HPBW of an antenna port with $K$ and $d_{V}$ is given as \cite{antenna_book},
\begin{align}
\label{BW}
\theta_{3dB,P} \approx 2 \left[\frac{\pi}{2}- \cos^{-1} \left(\frac{1.391 \lambda}{\pi K d_{V}}  \right)  \right].
\end{align}
Also  at $\theta=0, \phi =0$ and $\theta_{tilt}=90^{o}$, the maximum directional antenna port gain is given as follows \cite{antenna_book},
\begin{align}
\label{gain}
G_{max,P}=G_{max,E}+20 \log_{10} \sqrt{K}.
\end{align}
This implies that the vertical HPBW is inversely related to the number of antenna elements in a TXRU and the inter-element separation. Therefore in FD-MIMO systems, it is preferred to work with the $K = M$ setting shown in Fig. \ref{K=M}, since higher $K$ can realize narrower vertical beams resulting in better spatial separation of the users, which is the main objective of elevation beamforming. 

The following subsection will introduce more details on  the different ways of mapping TXRUs to antenna elements, as proposed in the 3GPP TR36.897.

\subsection{Transceiver Architectures}

In FD-MIMO systems, each TXRU is mapped to a column of antenna elements arranged in the vertical direction using a well-designed mapping function. Balancing the tradeoff between cost and performance, the 3GPP TR36.897 has proposed some TXRU architectures and defined the corresponding TXRU virtualization weight functions to realize different elevation beamforming scenarios. 

A TXRU model configuration corresponding to an antenna array model configuration $(M,N, P)$ is represented by $(M_{TXRU}, N, P)$, where $M_{TXRU}$ is the number of TXRUs per column per polarization dimension such that $M_{TXRU} \leq M$. A TXRU is only associated with antenna elements having the same polarization. The total number of TXRUs $Q$ is therefore equal to $M_{TXRU} \times N \times P$.
\\ \\
\textbf{TXRU virtualization model:}
\\ \\
A TXRU virtualization model defines the relation between the signals at the TXRUs and the signals at the antenna elements. Two virtualization methods have been introduced in TR36.897: 1D TXRU virtualization and 2D TXRU virtualization. Both are summarized below. 

\subsubsection{1D TXRU Virtualization}

The following notations will be used in describing this method. $\textbf{q}=[q_{1}, q_{2}, \dots, q_{M}]^{T}$ is a Tx signal vector at the $M$ co-polarized antenna elements within a column, $\textbf{w}=[w_{1}, w_{2}, \dots, w_{K}]^{T}$ is the TXRU virtualization weight vector and $\textbf{x}=[x_{1}, x_{2}, \dots, x_{M_{TXRU}}]^{T}$ is a TXRU signal vector at $M_{TXRU}$ TXRUs.

In this method, $M_{TXRU}$ TXRUs are associated with only the $M$ co-polarized antenna elements that comprise a column in the array. Accounting for $N$ such columns and dual-polarized configuration, the total number of TXRUs is given as $Q=M_{TXRU} \times N \times P$. The main architecture proposed for this method is the sub-array partition model.  

\begin{figure}
\centering
\includegraphics[width=2.35in]{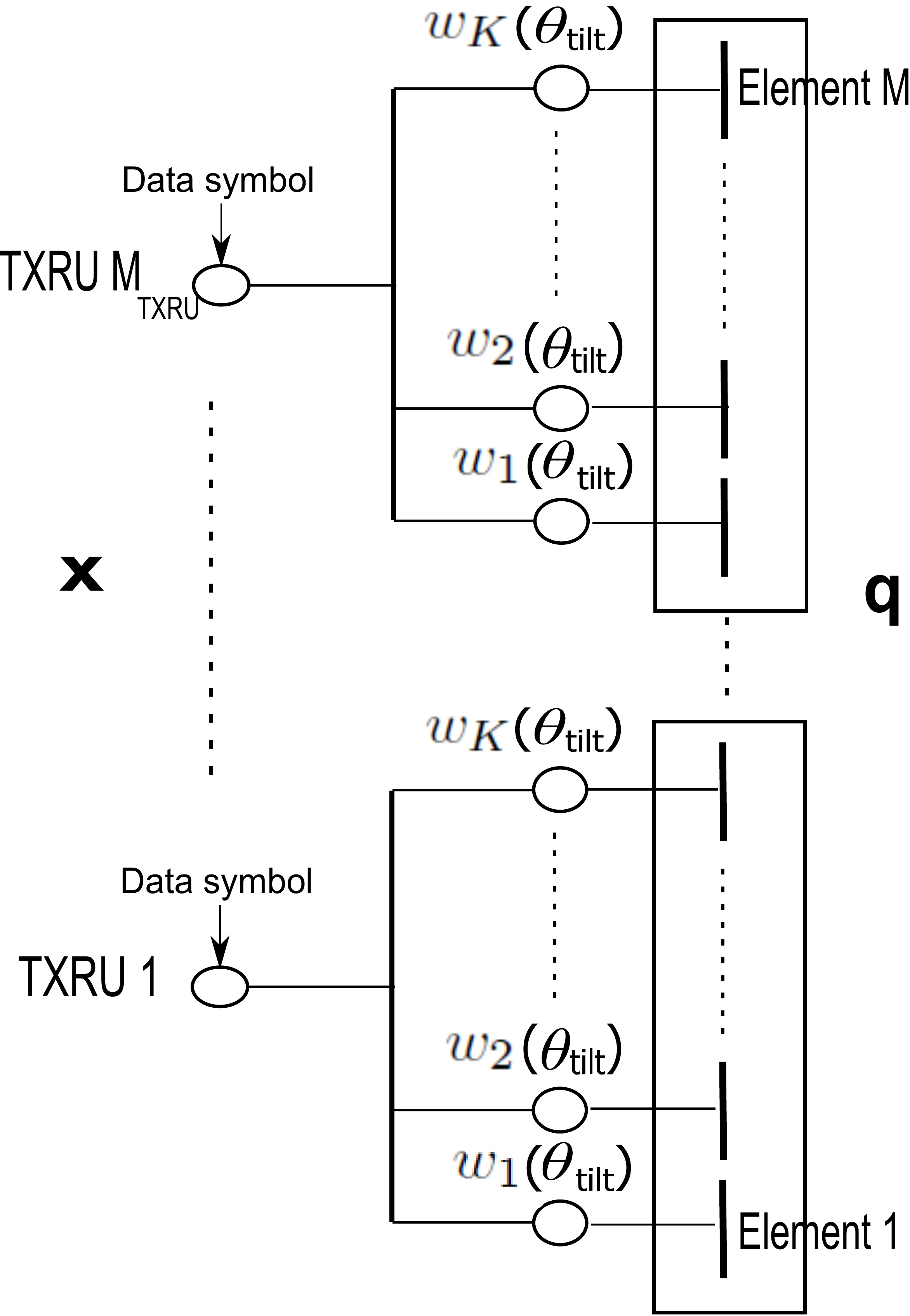}
\caption{Sub-array partition model for 1D TXRU virtualization. There will be $N$ such columns.}
\label{subarray1D}
\end{figure}

%\begin{figure}
%\centering
%\includegraphics[width=2.35in]{fullcon1D.eps}
%\caption{Full connection model for 1D TXRU virtualization. There will be $N$ such columns.}
%\label{fullcon1D}
%\end{figure}

\begin{figure*}[b]
\centering
\includegraphics[width=5in]{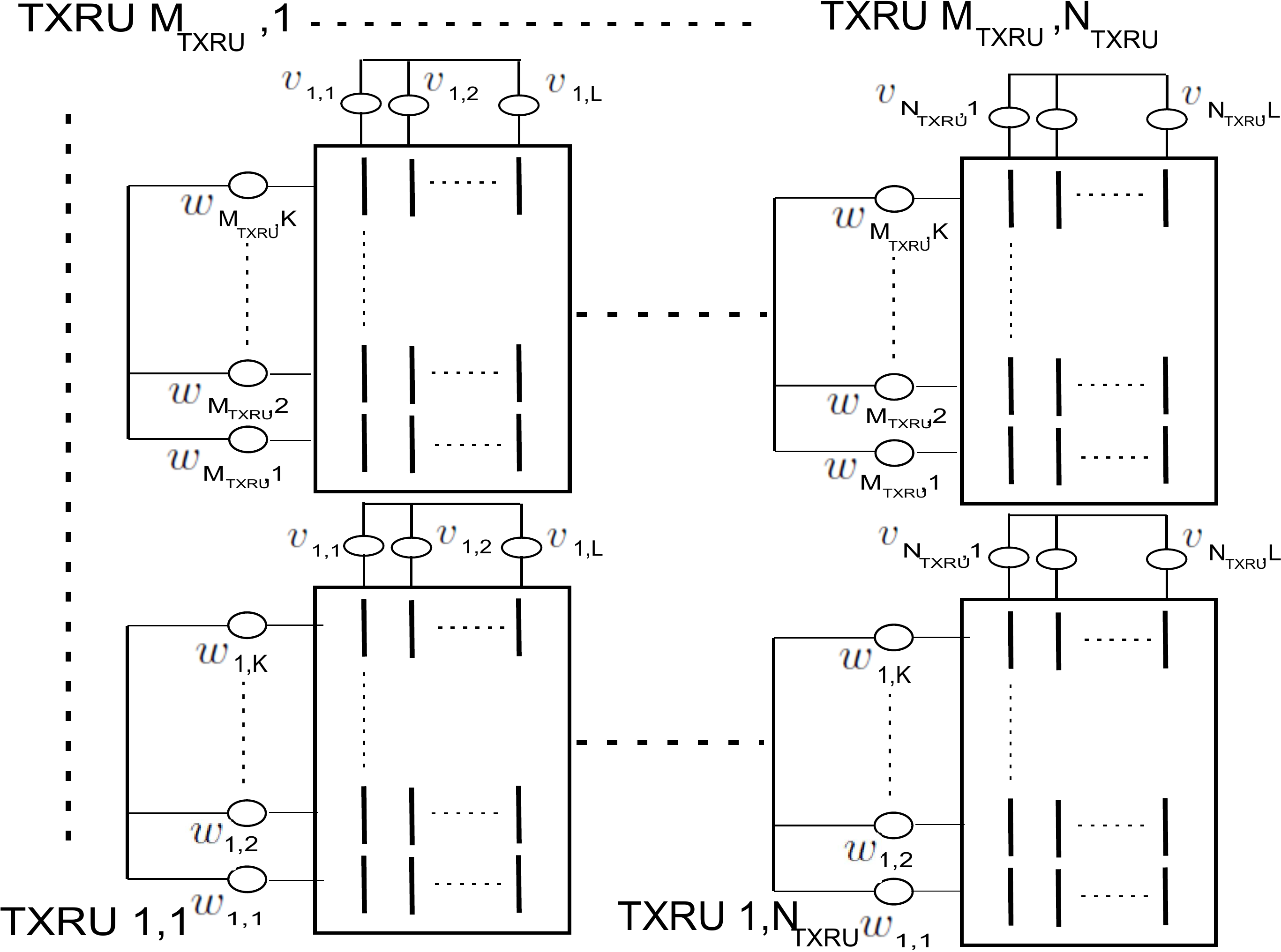}
\caption{Sub-array partition model for 2D TXRU virtualization.}
\label{subarray2D}
\end{figure*}

In this model, $M$ co-polarized antenna elements comprising a column  are partitioned into groups of $K$ elements. Therefore, $M_{TXRU}=M/K$ as shown in Fig. \ref{subarray1D}, for one column of vertically polarized antenna elements. Note that there will be $N$ such columns. The model is defined as,
\begin{align}
& \textbf{q}=\textbf{x} \otimes \textbf{w}, 
\end{align}
where the same TXRU virtualization weight vector $\textbf{w}=[w_{1}, w_{2}, \dots, w_{K}]^{T}$ is applied to all TXRUs. The 3GPP allows the use of any unit norm weight vector that can realize the desired elevation beamforming scenario, with one proposed expression given as \cite{TR36.897},
\begin{align}
\label{weight}
& w_{k}=\frac{1}{\sqrt{K}} \exp\Big(-i \frac{2\pi}{\lambda}(k-1)d_{V}\cos(\theta_{tilt}) \Big),
\end{align}
for $k=1,\dots, K,$ where $\theta_{tilt}$ is the electric downtilt angle. 

Another architecture for 1D virtualization, referred to as the full-connection model, has also been introduced in TR36.897, where the signal output of each TXRU associated with a column of co-polarized antenna elements is split into $M$ signals, which are then precoded by a group of $M$ phase shifters. The $M_{TXRU}$ weighted signals are combined at each antenna element. However, this architecture has not been studied yet in context of elevation beamforming in both industry and academia. Some discussion on it with reference to CSI transmission and feedback can be found in \cite{FD3}.

\subsubsection{2D TXRU Virtualization}

The following notations will be used in describing this method.  $\textbf{q}$ is a Tx signal vector at the $MN$  elements associated with same polarization, $\textbf{w}=[w_{1}, w_{2}, \dots, w_{K}]^{T}$ is the TXRU virtualization weight vector in the vertical direction, $\textbf{v}=[v_{1}, v_{2}, \dots, v_{L}]^{T}$ is the  TXRU virtualization weight vector in the horizontal direction and $\textbf{x}=[x_{1,1}, x_{2,1}, \dots, x_{{M_{TXRU}},1}, x_{1,2}, \dots, x_{M_{TXRU},N_{TXRU}}]^{T}$ is a TXRU signal vector at $M_{TXRU}N_{TXRU}$ TXRUs.

A 2D TXRU virtualization model considers $M_{TXRU}$ TXRUs in the vertical direction and $N_{TXRU}$ TXRUs in the horizontal direction associated with the antenna elements of same polarization. The total $M_{TXRU}N_{TXRU}$ TXRUs can be associated with any of the $MN$ co-polarized antenna elements. If dual-polarized antenna elements are virtualized, the total number of TXRUs is $Q=M_{TXRU}N_{TXRU}P$.   Again, the main architecture proposed for this method is the sub-array partition model.

\begin{table*}
\centering
\caption{Comparison of 1D and 2D virtualization methods.}
\normalsize
\begin{tabular}{p{0.4\textwidth} p{0.4\textwidth}}
\hline
\rowcolor{LightCyan}
\normalsize{\textbf{1D Virtualization}} &  \normalsize{\textbf{2D Virtualization}}  \\
\hline
\begin{itemize}
  \item Each TXRU mapped to co-polarized antenna elements in the vertical direction only.
  \item Control over the downtilt angle only.
  \end{itemize} & \begin{itemize}
  \item Each TXRU mapped to co-polarized antenna elements in vertical and horizontal directions.
  \item Control over both the downtilt and horizontal scan angles.
  \end{itemize} \\
\end{tabular}
\begin{tabular}{p{0.8\textwidth}}
\hline
\hline
\rowcolor{LightCyan}
\normalsize{\textbf{Architectures}}   \\
\hline
\begin{itemize}
  \item Sub-Array Partition
	\begin{itemize}
	\item Signal from each TXRU fed to an independent group of co-polarized antenna elements.
	\item Mapping defined by weight vectors.
	\end{itemize}
  \item Full Connection
		\begin{itemize}
	\item Signal from each TXRU fed to all co-polarized antenna elements.
	\item Mapping defined by a weight matrix.
	\end{itemize}
  \end{itemize} \\
	\hline
\end{tabular}
\label{array_comp}
\end{table*}

In this model, $MN$ antenna elements associated with each polarization are partitioned into rectangular arrays of $K\times L$ elements, where $K$=$M/M_{TXRU}$ and $L$=$N/N_{TXRU}$. The resulting architecture for vertically polarized antenna elements is shown in Fig. \ref{subarray2D}. The length of the vertical virtualization weight vectors denoted as $\textbf{w}_{m'}$, $m'=1,\dots, M_{TXRU}$, is $K$ and the length of the horizontal virtualization weight vectors denoted as $\textbf{v}_{n'}$,  $n'=1,\dots, N_{TXRU}$, is $L$. Note that these virtualization weight vectors can be different for different TXRUs. The 2D subarray partition model is defined as,
\begin{align}
&\textbf{q}_{m',n'}=x_{m',n'} \times (\textbf{v}_{n'} \otimes \textbf{w}_{m'}),
\end{align}
where $\textbf{q}_{m',n'}$ is the $KL\times 1$ vector of the signals at the antenna elements constituting the $(m',n')$ TXRU, $\textbf{w}_{m'}$ is the $K\times 1$ weight vector for the $m'^{th}$ TXRU in the vertical direction with each entry given as,
\begin{align}
\label{w_2Dsa}
&w_{m',k}=\frac{1}{\sqrt{K}} \exp\Big(-i \frac{2\pi}{\lambda}(k-1)d_{V}\cos(\theta_{tilt,m'}) \Big), 
\end{align}
for $k=1,\dots, K,$ and $\textbf{v}_{n'}$ is the $L\times 1$ weight vector for the $n'^{th}$ TXRU in the horizontal direction with each entry given as,
\begin{align}
&v_{n',l}=\frac{1}{\sqrt{L}} \exp\Big(-i \frac{2\pi}{\lambda}(l-1)d_{H}\sin(\phi_{scan,n'}) \Big), 
\end{align}
for $l=1,\dots, L$, where $\phi_{scan,n'}$ is the horizontal steering angle for the $n'^{th}$ TXRU in the azimuth plane. The 2D virtualization method therefore allows control over the radio wave in both the vertical and horizontal directions through $\theta_{tilt}$ and $\phi_{scan}$ respectively. Note that the 3GPP also gives the option of using other unit norm TXRU virtualization weight vectors instead of the expressions presented above.

The full connection model has also been introduced for 2D virtualization in TR36.897, where the signal output of each TXRU associated with co-polarized antenna elements, is split into $MN$ signals, which are then precoded by a group of $MN$ phase shifters. The $M_{TXRU}N_{TXRU}$ weighted signals are combined at each antenna element. However, this architecture has not been studied yet in context of elevation beamforming, so it is hard to comment on its practical performance benefits as compared to the sub-array partition model.

\begin{figure*}
\centering
\includegraphics[scale=.55]{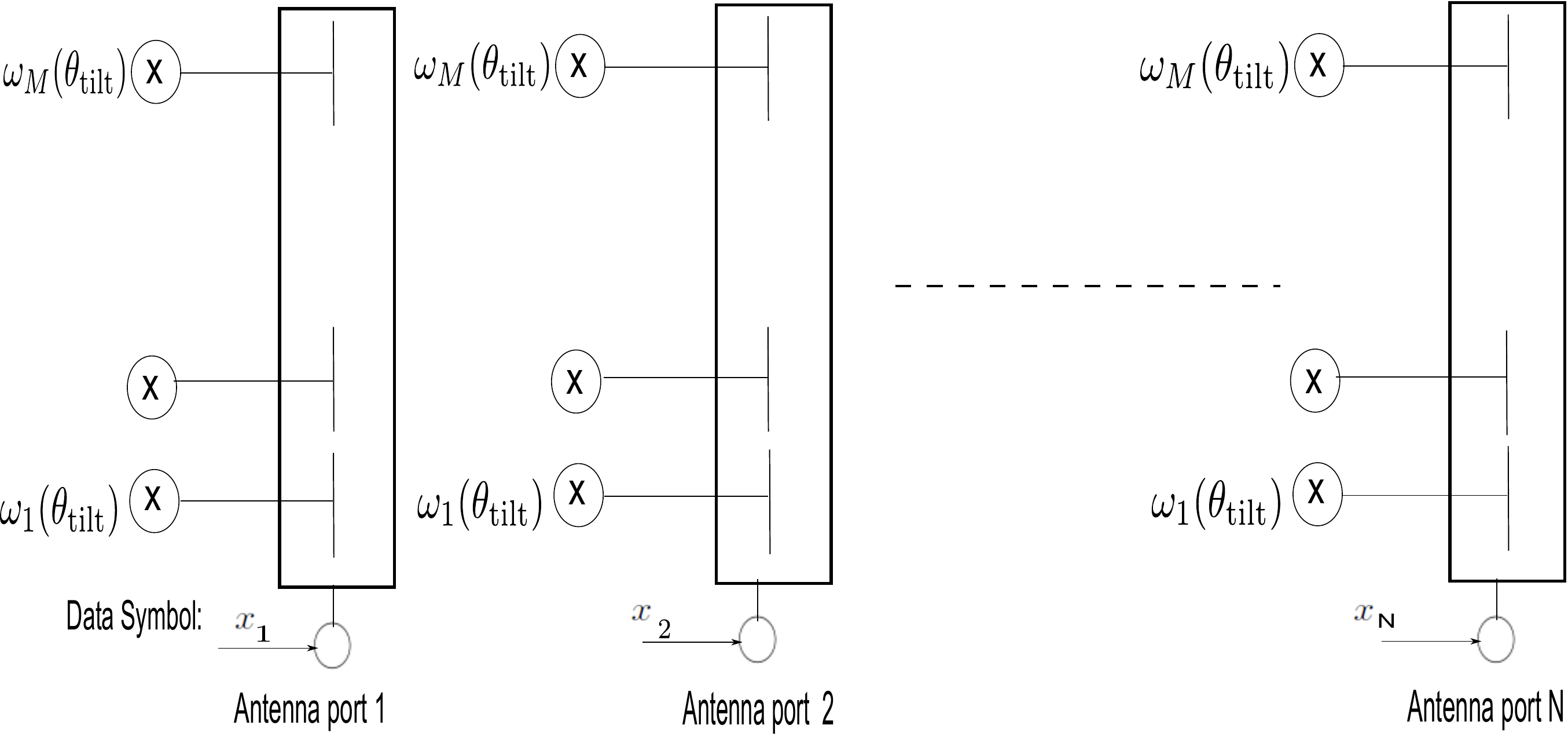}
\caption{1D TXRU virtualization with sub-array partition for $M=K$.}
\label{antenna}
\end{figure*}

The key differences between the 1D and 2D virtualization methods have been summarized in Table \ref{array_comp}.  It is expected that these two virtualization methods and the corresponding architectures will  have different tradeoffs, in terms of hardware complexity and cost, power efficiency and performance. Detailed discussion on these tradeoffs will take place in the future 3GPP meetings.

The TXRU virtualization weight vectors need to be adapted dynamically to realize different elevation beamforming scenarios. Due to the lack of literature available on TXRU virtualization models, the optimization of the vertical and horizontal weight functions has not been a subject of theoretical works on elevation beamforming so far, with the only notable exceptions being \cite{ourworkTCOM, tilt_letter, ourworkWCNC}. This section made an attempt to bridge this gap between the 3GPP and academia by introducing different TXRU architectures and corresponding example weight functions that can be optimized in future works to realize desired elevation beamforming scenarios. 

Over the course of this tutorial, we will develop elevation beamforming schemes for the 1D TXRU virtualization model with sub-array partition. This configuration for $M = K$ vertically polarized antenna elements is shown in Fig. \ref{antenna}. Note that we consider one-to-one mapping between ports and TXRUs, so both terms will be used inter-changeably. The digital precoding is performed across the antenna ports and analog beamforming is performed across the elements in each antenna port. The analysis and discussions provided in subsequent sections can be extended to other TXRU architectures as well. 

In order to support the performance evaluation of the AAAs, we need a 3D SCM that not only takes into account both the azimuth and elevation angles of the propagation paths but also incorporates the 3D
radiation patterns of the active antenna elements. The next section will present two approaches that have been proposed in the standards and academia to model the 3D channel based on the ray-tracing method.

\section{3D Channel Modeling}

The SCM developed over the years as a result of the standardization efforts in the 3GPP \cite{SCM} and WINNER \cite{Winner} initiatives is a 2D SCM, which ignores the elevation angles of the propagation paths for simplicity. It also does not account for the directional antenna radiation patterns, which are a characteristic feature of the active antenna elements. The 2D SCM still remains widely used in academia as well as industry for the performance analysis of MIMO technologies. The evaluation of FD-MIMO techniques, however, requires a 3D channel model. Over the recent years, significant efforts have been made in the 3GPP to get accurate 3D channel models that support the elevation dimension and account for directional radiation patterns of the active antenna elements. The resulting standardized models follow a ray-tracing approach where the channel between the BS and each user is expressed explicitly as a function of the propagation paths and the associated physical parameters. The efforts made in the 3GPP resulted in two different 3D channel modeling approaches that have been studied and utilized in literature on FD-MIMO -  the antenna port approach and the antenna element approach. The main literature related to their development has already been discussed in Section II-B.  In this section, we will provide a brief introduction on the method of 3D channel coefficient generation in the 3GPP standard, which is common to both approaches and then present and compare the two 3D ray-tracing SCMs.

\subsection{Standardized Channel Model}

Most standardized channels like the 3GPP SCM \cite{SCM}, ITU \cite{ITU} and WINNER \cite{Winner} follow a system-level, stochastic channel modeling approach, based on the generation of large scale and small scale parameters defined as: \\
\textit{Large scale parameters:} Random correlated variables drawn from given statistical distributions and specific for each user, that describe the propagation paths at a macroscopic level. These statistical parameters include delay spread (DS), angular spread (AS), and shadow fading (SF). \\
\textit{Small scale parameters:} Physical parameters, including powers, delays and angles of the propagation paths, that describe the channel at a microscopic level.

In the standardized models, the propagation paths are described using large-scale parameters without being physically positioned. These large scale parameters serve to generate the power, delay and angles of each path, referred to as small-scale parameters. The 3GPP 2D SCM uses three large scale parameters - the DS, the azimuth spread at departure (ASD) and the SF. To these large scale parameters, WINNER II and ITU added the azimuth spread at arrival (ASA) and the Rician factor.

In the 3D SCM, each propagation path has to be modeled using both azimuth and elevation angles. For this purpose, the elevation domain large-scale parameters as well as the statistical distribution of the elevation angles need to be introduced to extend the existing 2D SCM to the third dimension.  A double exponential (or Laplace) distribution is proposed in WINNER+ for the generation of elevation angles. The ESD for the generation of elevation angles follows a log normal distribution. Height and distance dependent expressions have been introduced for the mean and variance of the ESD in \cite{TR36.873}. However, measurement campaigns are still going on to finalize the modeling of these parameters. At present, the WINNER+ and the 3GPP TR36.873 account for $7$ large scale parameters by including the ESD and the elevation spread at arrival (ESA). 

\begin{figure}
\centering
\includegraphics[width=3.5 in]{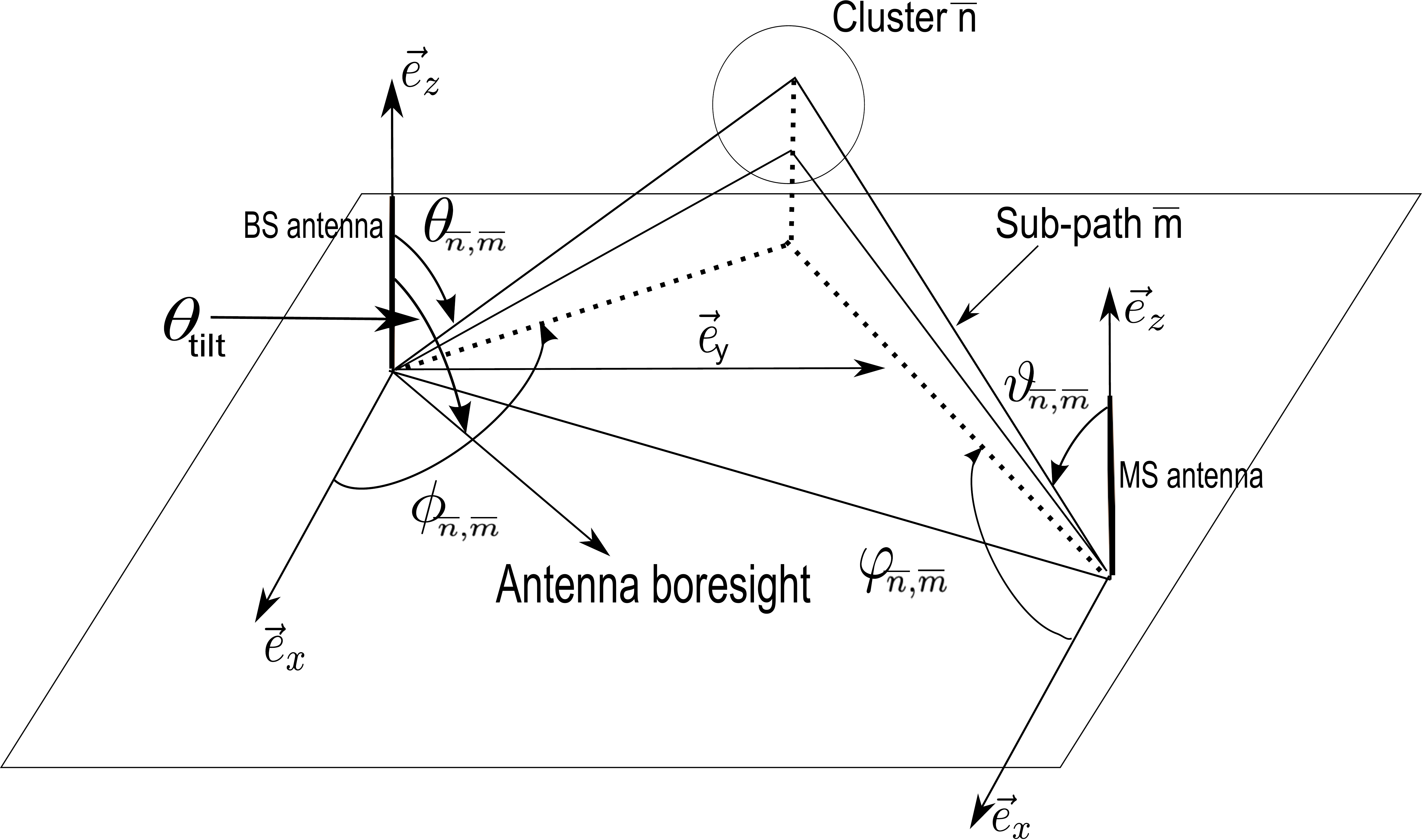}
\caption{3D propagation environment.}
\label{3D_channel}
\end{figure}

The 3D SCM introduced in the 3GPP TR36.873, like its 2D counterpart, is a composite of $\bar{N}$ propagation paths, referred to as clusters. The $\bar{n}^{th}$ cluster is characterized by the delay, angle of departure (AoD) $(\phi_{\bar{n}},\theta_{\bar{n}})$, AoA $(\varphi_{\bar{n}},\vartheta_{\bar{n}})$ and power. Each cluster gives rise to $\bar{M}$ unresolvable sub-paths, which have the same delay as the original cluster and are characterized by the spatial angles $(\phi_{\bar{n},\bar{m}},\theta_{\bar{n},\bar{m}})$ and $(\varphi_{\bar{n},\bar{m}},\vartheta_{\bar{n},\bar{m}})$, $\bar{m}=1, \dots \bar{M}_{\bar{n}}$, where,
\begin{align}
&\theta_{\bar{n},\bar{m}}=\theta_{\bar{n}}+c_{\theta}\alpha_{\bar{m}}, \\
&\phi_{\bar{n},\bar{m}}=\phi_{\bar{n}}+c_{\phi}\alpha_{\bar{m}}, \\
&\vartheta_{\bar{n},\bar{m}}=\vartheta_{\bar{n}}+c_{\vartheta}\alpha_{\bar{m}}, \\
&\varphi_{\bar{n},\bar{m}}=\varphi_{\bar{n}}+c_{\varphi}\alpha_{\bar{m}}, 
\end{align}
where $\alpha_{\bar{m}}$ is a set of symmetric fixed values and $c_{\theta}$, $c_{\phi}$, $c_{\vartheta}$ and $c_{\varphi}$ control the dispersion inside the cluster $\bar{n}$. The 3D propagation environment is illustrated in Fig. \ref{3D_channel}.

According to the 3GPP TR36.873  \cite{TR36.873}, the channel between BS antenna $s$ and mobile station (MS) antenna $u$ corresponding to the $\bar{n}^{th}$ cluster is given by,
\begin{align}
\label{channel1}
& \text{[\textbf{H}]}_{su,\bar{n}}(t)=\sqrt{10^{-(PL+\sigma_{SF})/10}}  \sqrt{P_{\bar{n}}/\bar{M}_{\bar{n}}} \sum\limits_{\bar{m}=1}^{\bar{M}_{\bar{n}}} \textbf{g}_{r}(\varphi_{\bar{n},\bar{m}},\vartheta_{\bar{n},\bar{m}})^{T}  \nonumber \\
& \times \boldsymbol{\alpha}_{\bar{n},\bar{m}}  \textbf{g}_{t}(\phi_{\bar{n},\bar{m}},\theta_{\bar{n},\bar{m}}) [\textbf{a}_{r}(\varphi_{\bar{n},\bar{m}}, \vartheta_{\bar{n},\bar{m}})]_{u} [\textbf{a}_{t}(\phi_{\bar{n},\bar{m}}, \theta_{\bar{n},\bar{m}})]_{s}\nonumber \\
&\times \exp(i 2\pi v_{\bar{n},\bar{m}}t),
\end{align}
where $P_{\bar{n}}$ is the power of the $\bar{n}^{th}$ cluster  and $v_{\bar{n},\bar{m}}$ is the Doppler frequency component for the user corresponding to the $\bar{m}^{th}$ subpath in $\bar{n}^{th}$ cluster. Also, $PL$ and $\sigma_{SF}$ denote the loss incurred in path loss and shadow fading respectively in dB. If polarization is taken into account, $\boldsymbol{\alpha}$ is a $2\times 2$ matrix, describing the coupling between the vertical and horizontal polarizations as,
\begin{align}
&\boldsymbol{\alpha}_{\bar{n},\bar{m}}=\begin{bmatrix} \ \exp(i\Phi_{\bar{n},\bar{m}}^{\theta,\theta}) & \sqrt{\kappa_{\bar{n},\bar{m}}^{-1}} \exp(i\Phi_{\bar{n},\bar{m}}^{\theta,\phi}) \\
\ \sqrt{\kappa_{\bar{n},\bar{m}}^{-1}}\exp(i\Phi_{\bar{n},\bar{m}}^{\phi,\theta}) & \exp(i\Phi_{\bar{n},\bar{m}}^{\phi,\phi}), \end{bmatrix}
\end{align}
where $\kappa_{\bar{n},\bar{m}}$ is the cross-polarization power ratio (XPR) and $\Phi_{\bar{n},\bar{m}}^{\theta,\theta}$, $\Phi_{\bar{n},\bar{m}}^{\theta,\phi}$, $\Phi_{\bar{n},\bar{m}}^{\phi,\theta}$ and $\Phi_{\bar{n},\bar{m}}^{\phi,\phi}$ are the random initial phases for subpath $\bar{m}$ of cluster $\bar{n}$. The diagonal elements of this matrix characterize the co-polarized phase response while the off-diagonal elements characterize the cross-polarized phase response. If polarization is not considered, the matrix is replaced by a scalar $\exp(i\Phi_{\bar{n},\bar{m}})$.

Also $\textbf{g}_{t}(\phi_{\bar{n},\bar{m}},\theta_{\bar{n},\bar{m}})$ and $\textbf{g}_{r}(\varphi_{\bar{n},\bar{m}},\vartheta_{\bar{n},\bar{m}})$ are the radiation patterns of the Tx and Rx antennas respectively. When polarization is considered, these are $2\times 1$ vectors, whose entries represent vertical and horizontal field patterns. Moreover, vectors $\textbf{a}_{t}(\phi_{\bar{n},\bar{m}}, \theta_{\bar{n},\bar{m}})$ and $\textbf{a}_{r}(\varphi_{\bar{n},\bar{m}}, \vartheta_{\bar{n},\bar{m}})$ are the array responses of the Tx and Rx antennas respectively whose entries are given by,
\begin{align}
\label{array_response}
[\textbf{a}_{t}(\phi_{\bar{n},\bar{m}}, \theta_{\bar{n},\bar{m}})]_{s}&=\exp(i \textbf{k}_{t, \bar{n},\bar{m}}\textbf{.}\textbf{x}_{t,s}), \\
\label{array_response1}
[\textbf{a}_{r}(\varphi_{\bar{n},\bar{m}}, \vartheta_{\bar{n},\bar{m}})]_{u}&=\exp(i \textbf{k}_{r,\bar{n},\bar{m}}\textbf{.}\textbf{x}_{r,u}),
\end{align}
where \textbf{.} is the scalar product, $\textbf{x}_{t,s}$ and $\textbf{x}_{r,u}$ are the location vectors of the $s^{th}$ Tx antenna and the $u^{th}$ Rx antenna respectively, $\textbf{k}_{t, \bar{n},\bar{m}}$ and $\textbf{k}_{r, \bar{n},\bar{m}}$ are the Tx and Rx wave vectors corresponding to the $\bar{m}^{th}$ subpath in $\bar{n}^{th}$ cluster respectively, where $\textbf{k}_{\bar{n},\bar{m}}=\frac{2\pi}{\lambda}\hat{\textbf{v}}_{\bar{n},\bar{m}}$, with $\lambda$ being the carrier wavelength and $\hat{\textbf{v}}_{\bar{n},\bar{m}}$ being the direction of wave propagation.

\textit{Remark:} The channel model in (\ref{channel1}) is for the NLoS case. If the user is in the LoS of the BS, then the first cluster contains the LoS contribution of the channel. In this case, the Rician factor is introduced and the channel response corresponding to the first cluster is the scaled sum of the NLoS channels constituted by the $\bar{M}_{1}$ sub-paths and the LoS channel. More details on this can be found in \cite{drabla}.

\begin{figure}
\centering
\includegraphics[scale=.55]{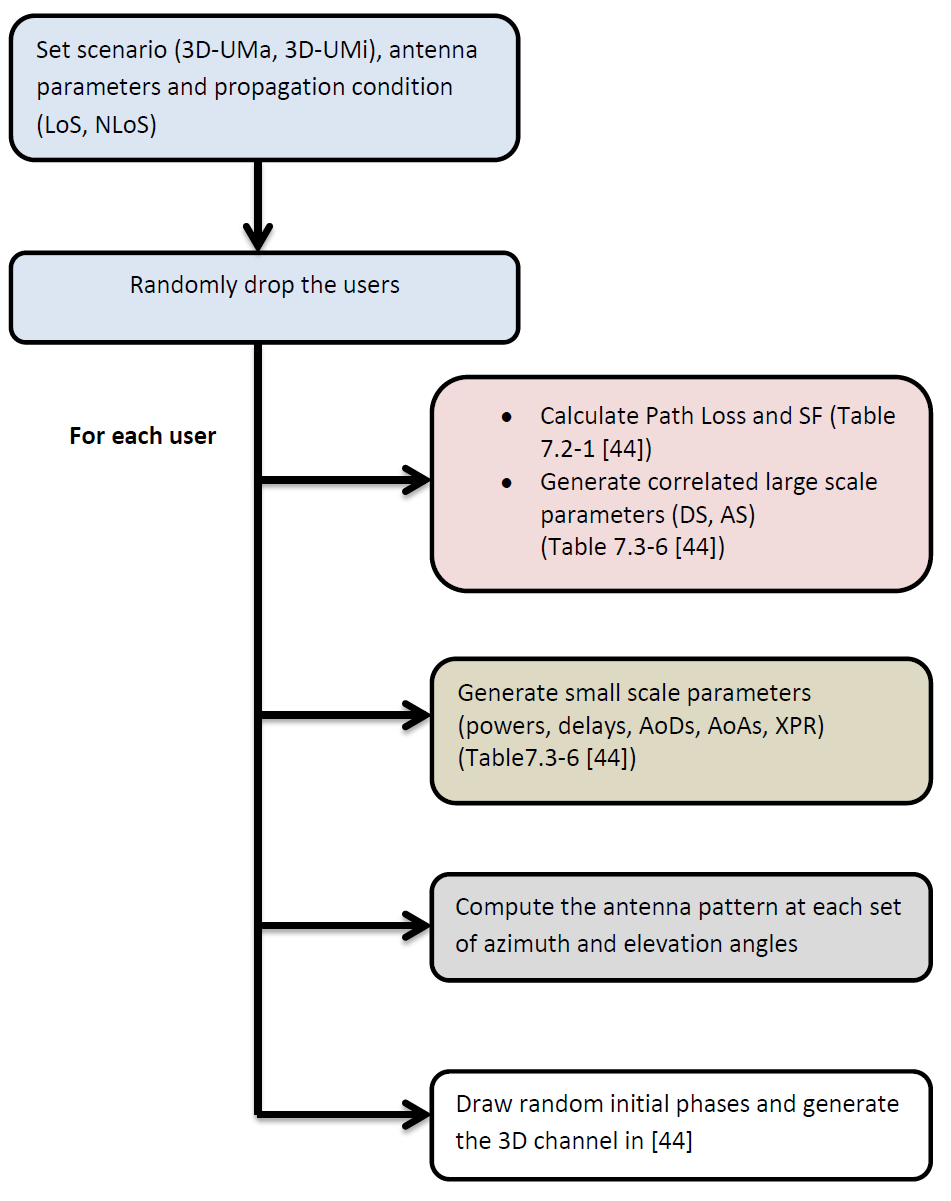}
\caption{Generation of the 3GPP TR36.873 3D channel.}
\label{flowchart}
\end{figure}

The generation of the channel in the 3GPP TR36.873 follows broadly the steps illustrated in Fig. \ref{flowchart}. For each user, the path loss and large scale parameters including DS, ESD, ESA, ASD, ASA, SF and Rician factor are generated according to the statistical distributions outlined in TR36.873. These large scale parameters are then used to generate small scale parameters for each propagation path including powers, delays, AoDs, AoAs and XPRs. The antenna pattern is generated for each set of angles. Finally, the 3D channel coefficient is generated using (\ref{channel1}). The complete outline of the standardized channel coefficient generation can be found in [Fig. 7.3-1 \cite{TR36.873}]. 

The next sections outline the antenna port approach and the antenna element approach towards 3D channel modeling for the 1D sub-array TXRU architecture  shown in Fig. \ref{antenna}. Both follow the standardized model introduced in this section and differ only in the generation of the antenna port radiation pattern.

\subsection{Antenna Port Approach towards 3D Channel Modeling}

The pioneer works on 3D channel modeling consider the channel directly between the antenna ports  instead of between the physical antenna elements constituting these ports \cite{drabla, PES1, FD1, ourwork, ourwork1}, utilizing the approximate antenna port radiation pattern expressions from 3GPP TR36.814 and ITU \cite{TR36.814, ITU}. The motivation behind this approach is that the antenna elements constituting a port carry the same signal with corresponding weights to achieve the desired downtilt angle, so every port appears as a single antenna at the MS. The channel of interest is therefore between the transmitting BS antenna port and the MS. This approach does not  consider the contributions of the individual antenna elements and the downtilt weights applied to these elements in determining the 3D antenna port radiation pattern. In fact, it abstracts the role played by the elements constituting a port in performing the downtilt by approximating the vertical dimension of the antenna radiation pattern of each port by a narrow beam in the elevation plane. 

The combined 3D antenna port radiation pattern approximated in ITU in $\rm{dB}$ is as follows \cite{ITU},
\begin{align}
\label{port_p}
A_{P}(\phi,\theta,\theta_{tilt})&= G_{max,P}-\text{min}\{-(A_{P,H}(\phi)+A_{P,V}(\theta,\theta_{tilt})), \nonumber \\
& A_{m}\},
\end{align}
where,
\begin{align}
\label{minpattern_p}
A_{P,H}(\phi)&= -\text{min}\left[ 12 \left(\frac{\phi}{\phi_{3dB, P}}\right)^2, A_{m} \right] \rm{dB}, \nonumber \\
A_{P,V}(\theta,\theta_{tilt})&= -\text{min} \left[12 \left(\frac{\theta-\theta_{tilt}}{\theta_{3dB, P}}\right)^2,A_{m} \right] \rm{dB},
\end{align}
where $G_{max,P}=17$$\rm{dBi}$ is the maximum directional antenna port gain, $A_{m}=20 \rm{dB}$ is the front-to-back ratio, $\phi_{3dB, P}$ is the horizontal 3$\rm{dB}$ beamwidth that equals $70^{o}$, $\theta_{3dB, P}$ is the vertical 3$\rm{dB}$ beamwidth that equals $15^{o}$ and $\theta_{tilt}$ is the downtilt angle of the antenna boresight. This approach, despite its simplicity, poses a challenge when cross-polarized antenna elements are considered, because the standards do not provide a method for deducing the field patterns along the  horizontal and vertical polarizations from the global antenna port pattern. To circumvent this problem, 3GPP TR36.814 proposes to decompose the global pattern of a port, composed of antenna elements that are slanted perpendicular to the boresight direction by an angle $\beta$, along the horizontal and vertical polarizations as \cite{TR36.814},
\begin{align}
\label{decom}
\textbf{g}_{P}(\phi,\theta, \theta_{tilt})&=[\sqrt{A_{P}(\phi,\theta,\theta_{tilt})|_{lin}}\cos\beta, \nonumber \\
& \sqrt{A_{P}(\phi,\theta,\theta_{tilt})|_{lin}}\sin\beta].
\end{align}

This decomposition is used to compute the vectors $\textbf{g}_{t}(\phi_{\bar{n},\bar{m}},\theta_{\bar{n},\bar{m}})$ and $\textbf{g}_{r}(\varphi_{\bar{n},\bar{m}},\vartheta_{\bar{n},\bar{m}})$, which are then plugged in (\ref{channel1}) to return the 3D channel coefficient between BS antenna port $s$ and MS antenna $u$, under the ITU approach.

It is important to note that this approximation does not hold in practice and to determine the correct decomposition of the global field pattern of an antenna port along horizontal and vertical polarizations, one has to know the exact architecture of the antenna port, i.e. the number of elements constituting a port, the inter-element separation and the applied downtilt weights. This information is missing in the antenna port approach based models.  However, this approach remains widely used in existing works on elevation beamforming due to its simplicity arising from the fact that the channel is directly a function of the downtilt angle of each port through (\ref{minpattern_p}) \cite{downtilt3, utility, portapp, 3Dbeamforming, beamforming2, beamforming3,7268913}.

\subsection{Antenna Element Approach towards 3D Channel Modeling}

In order to allow for realistic and accurate design of 3D beamforming techniques, the channel between individual antenna elements should be considered. In theory, the global radiation pattern of an antenna port depends on the positions and number of the elements within it, the individual patterns of these elements and the applied weights. Mathematically, it is represented as a superposition of the antenna element radiation pattern and the array factor for that port, where the individual element radiation pattern is given by  (\ref{port}) and the array factor will depend on the TXRU architecture considered. 

In order to highlight the difference between the two channel modeling approaches, consider the 1D TXRU virtualization model with sub-array partition, as shown in Fig. \ref{antenna}. Here, the $M\times N$  array response matrix $\textbf{A}_{t}(\phi, \theta)$ is considered instead of the $N\times 1$ array response vector in (\ref{array_response}). This array response matrix is given as \cite{TR37.84},
\begin{equation}
\label{arrayfactor}
\textbf{A}_{t}(\phi, \theta)=[\textbf{a}_{t,1}(\phi, \theta)  \textbf{a}_{t,2}(\phi, \theta) \dots \textbf{a}_{t,N}(\phi, \theta)],
\end{equation}
where,
\begin{align}
\label{xxxx}
[\textbf{a}_{t,s}(\phi, \theta)]_{m}&=\exp(i \textbf{k}_{t} \textbf{.}\textbf{x}_{t,m,s}), \hspace{.03 in} m=1, \dots, M, 
\end{align}
where \textbf{.} is the scalar dot product, $\textbf{x}_{t,m,s}$ is the location vector of the $m^{th}$ antenna element in the $s^{th}$ Tx antenna port, and $\textbf{k}_{t}$ is the Tx wave vector. 

For the antenna configuration shown in Fig. \ref{antenna}, (\ref{xxxx}) will be given by,
\begin{align}
\label{V}
&[\textbf{a}_{t,s}(\phi, \theta)]_{m}=\exp \Big(i 2 \pi \Big((s-1)\frac{d_{H}}{\lambda} \sin\phi \sin\theta \nonumber \\
&+(m-1)\frac{d_{V}}{\lambda} \cos\theta   \Big) \Big), 
\end{align}
where $d_{H}$ is the horizontal separation between the antenna ports and $d_{V}$ is the vertical separation between the antenna elements, with the phase reference at the origin.

Consider a ULA of antenna elements utilized at the MS. The effective 3D channel between the BS antenna port $s$ and the MS antenna element $u$ corresponding to cluster $\bar{n}$ is a weighted sum of the channels constituted by the $M$ elements inside port $s$ to MS antenna element $u$ as,
\begin{align}
&[\textbf{H}]_{su,\bar{n}}(t)=\sum_{m \in \text{port } s}^{M} w_{m}(\theta_{tilt}) [\textbf{H}]_{mu,\bar{n}}(t),
\end{align}
where $w_{m}(\theta_{tilt})$ is the downtilt weight applied to element $m$ in port $s$ calculated using (\ref{weight}) and $[\textbf{H}]_{mu,\bar{n}}(t)$ is the 3D standardized channel coefficient between antenna element $m$ in port $s$ at the BS and antenna element $u$ at the MS, given by (\ref{channel1}).

Utilizing (\ref{channel1}) and the array response matrix just developed, the channel coefficient between the BS antenna port $s$ and the MS antenna element $u$ corresponding to the $\bar{n}^{th}$ cluster can be written as,
%\begin{align}
%\label{channel2}
%& [\textbf{H}]_{su,\bar{n}}(t) = \sqrt{10^{-(PL+\sigma_{SF})/10}}  \sqrt{P_{\bar{n}}/\bar{M}_{\bar{n}}} \sum \limits_{m=1}^{M} {w}_{m}(\theta_{tilt}) \sum \limits_{\bar{m}=1}^{\bar{M}_{\bar{n}}} \nonumber \\
%& \textbf{g}_{E,r}(\varphi_{\bar{n},\bar{m}},\vartheta_{\bar{n},\bar{m}})^{T} \boldsymbol{\alpha_{\bar{n},\bar{m}}} \textbf{g}_{E,t}(\phi_{\bar{n},\bar{m}},\theta_{\bar{n},\bar{m}})  [\textbf{a}_{t,s}(\phi_{\bar{n},\bar{m}}, \theta_{\bar{n},\bar{m}}) ]_{m}  \nonumber \\
%& \times [\textbf{a}_{r}(\varphi_{\bar{n},\bar{m}}, \vartheta_{\bar{n},\bar{m}})]_{u} \exp(i 2\pi v_{\bar{n},\bar{m}}t), 
%\end{align}
\begin{align}
\label{channel3}
& [\textbf{H}]_{su,\bar{n}}(t)  = \sqrt{10^{-(PL+\sigma_{SF})/10}}  \sqrt{P_{\bar{n}}/\bar{M}} \textbf{w}(\theta_{tilt})^{T} \sum \limits_{\bar{m}=1}^{\bar{M}_{\bar{n}}} \nonumber \\
& \textbf{g}_{E,r}(\varphi_{\bar{n},\bar{m}},\vartheta_{\bar{n},\bar{m}})^{T} \boldsymbol{\alpha_{\bar{n},\bar{m}}}  \textbf{g}_{E,t}(\phi_{\bar{n},\bar{m}},\theta_{\bar{n},\bar{m}})  \textbf{a}_{t,s}(\phi_{\bar{n},\bar{m}}, \theta_{\bar{n},\bar{m}})  \nonumber \\
&\times  [\textbf{a}_{r}(\varphi_{\bar{n},\bar{m}}, \vartheta_{\bar{n},\bar{m}})]_{u} \exp(i 2\pi v_{\bar{n},\bar{m}}t), 
\end{align}
where $\textbf{w}(\theta_{tilt})=[w_{1}(\theta_{tilt}),  w_{2}(\theta_{tilt}),  \dots, w_{M}(\theta_{tilt})]^{T}$. For cross-polarized antenna elements, $\textbf{g}_{E,r}$ and $\textbf{g}_{E,t}$ are obtained by decomposing the global antenna element radiation pattern in (\ref{port}) along the vertical and horizontal polarizations, using a similar decomposition as (\ref{decom}). 

This new channel representation is obtained by performing a sum over the channels constituted by individual elements in antenna port $s$ and is different from the `antenna port approach' based standardized models, where the channel is directly characterized between the ports.

\subsection{Comparison of the Two Approaches}

The overall radiation pattern of a port in Fig. \ref{antenna} using the  antenna element approach is essentially a superposition of the individual element radiation pattern and the array factor for the whole port, where the array factor takes into account the downtilt weights and the array responses of the elements in that port as discussed in Section III-B. The antenna port approach, on the other hand,  abstracts the role played by the antenna elements to perform downtilt by approximating  the pattern of the whole port with a narrow beam in the elevation through equations (\ref{port_p}) and (\ref{minpattern_p}). In reality, the global antenna port radiation pattern approximated by the antenna port approach might vary significantly from the actual pattern obtained using the antenna element approach. 

To illustrate this, we compare the antenna port radiation pattern using the two approaches for vertically polarized antenna elements, i.e. $\beta=\pi/2$.  For the antenna port approach, $g_{P,t}(\phi,\theta,\theta_{tilt})=\sqrt{A_{P}(\phi,\theta, \theta_{tilt})|_{\text{lin}}}$, where $A_{P}(\phi,\theta, \theta_{tilt})$ is given by (\ref{port_p}).

For the antenna element approach, the exact antenna port radiation pattern in \rm{dB}, denoted as $A_{P}^{E}(\phi, \theta, \theta_{tilt})$, is given by (\ref{element_pattern}) where  $A_{F}(\theta, \theta_{tilt})$ is computed as,
\begin{align}
&A_{F}(\theta, \theta_{tilt})= \sum_{m=1}^{M}{w}_{m}(\theta_{tilt}) \exp \left(i 2 \pi  (m-1) \frac{d_{V}}{\lambda} \cos \theta  \right).
\end{align}

\begin{figure}
\centering
\includegraphics[width= 2.2in]{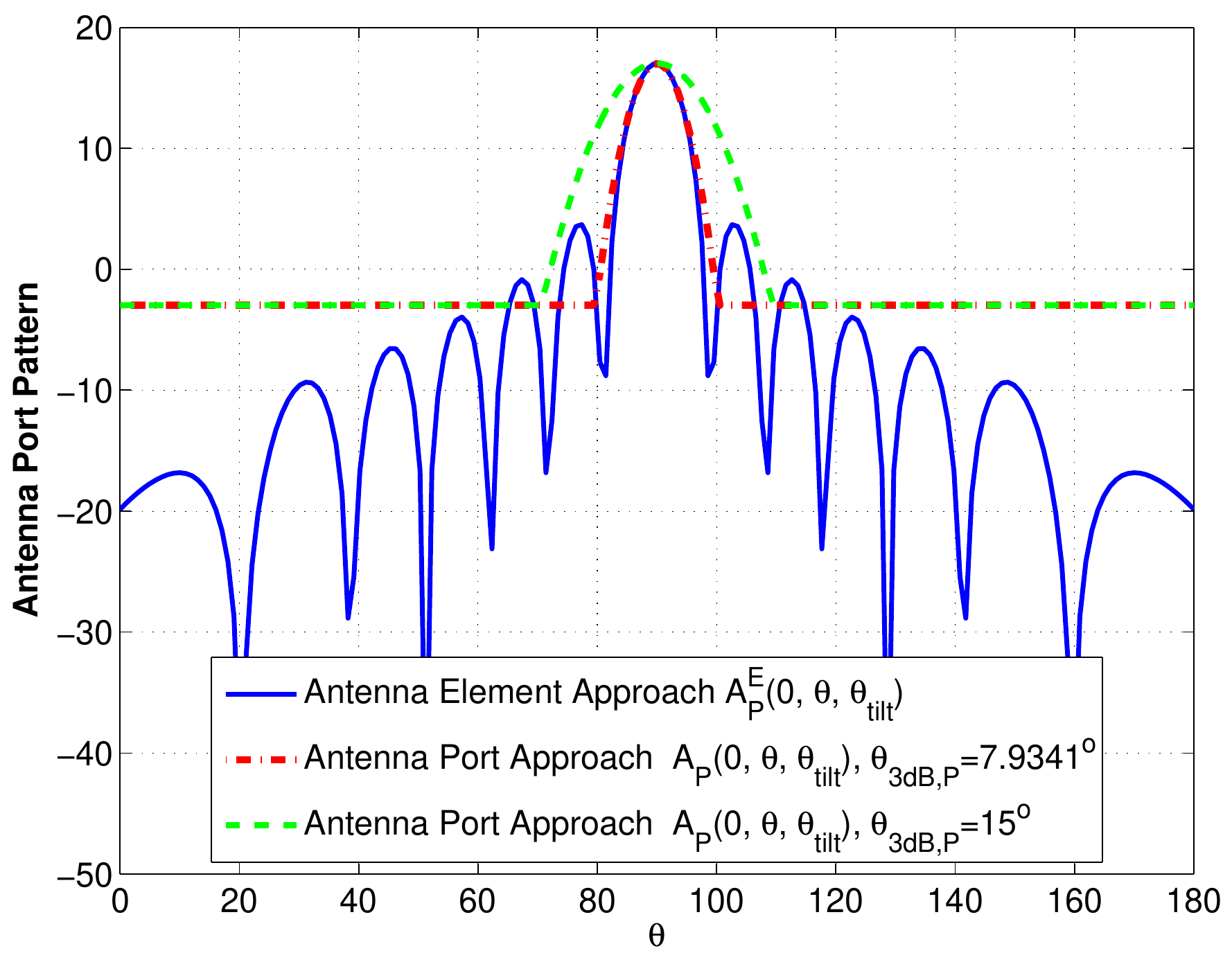}
\caption{Comparison of the antenna port radiation pattern for the element approach and the port approach}
\label{pattern_compere1}
\end{figure}

We now compare the antenna port radiation pattern for both approaches through simulations using the values  from the 3GPP report \cite{TR36.897} for the element approach and the ITU report \cite{ITU} for the port approach.  $A_{P}^{E}(\phi, \theta, \theta_{tilt})$ is plotted at $\phi=0^{o}$ for $N_{E}=8$ and $d_{V}/\lambda=0.8$ in Fig. \ref{pattern_compere1}. The weights are calculated using (\ref{weight}) for $\theta_{tilt}=90^{o}$. We also plot on the same figure, the approximate antenna port radiation pattern $A_{P}(\phi, \theta, \theta_{tilt})$ in (\ref{port_p}) at $\phi=0^{o}$ using the $3$\rm{dB} beamwidth proposed in ITU as $\theta_{3dB,P}=15^{o}$. It can be seen that the antenna port approach discards the side lobes in the radiation pattern. Also, the 3dB beamwidth proposed in ITU is not applicable to any generic port.  For example, for a port with $8$ elements placed at $0.8\lambda$ spacing, $15^{o}$ is too large.

Despite of its approximate nature, the port approach is still very popular in theoretical works on 3D beamforming due to its simplicity. However, some attention must be paid to the value of the $3 \rm{dB}$ beamwidth chosen to approximate the main lobe of the vertical radiation pattern in order to minimize the errors incurred as a result of this approximation.  In other words, the port approach can closely match the element approach if the values of $\theta_{3dB,P}$ and $G_{max,P}$ in (\ref{port_p}) are calculated utilizing the actual values of the number of elements $M$ and the inter-element separation $d_{V}$ used in the construction of the port.

 %The relationship of the $3 \rm{dB}$ beamwidth of an antenna port with $M$ and $d_{V}$ is given as \cite{antenna_book},
%\begin{align}
%\label{BW}
%\theta_{3dB,P} \approx 2 \left[\frac{\pi}{2}- \cos^{-1} \left(\frac{1.391 \lambda}{\pi M d_{V}}  \right)  \right].
%\end{align}
%Also  at $\theta=0, \phi =0$ and $\theta_{tilt}=90^{o}$, the maximum directional antenna port gain is given as follows,
%\begin{align}
%\label{gain}
%G_{max,P}=G_{max,E}+20 \log_{10} \sqrt{M}.
%\end{align}
For an antenna port constructed using $M=8$ antenna elements with $d_{V}=0.8 \lambda$ as proposed in 3GPP TR36.897 \cite{TR36.897}, the values of $G_{max,P}$  and $\theta_{3dB,P}$ are calculated using (\ref{BW}) and (\ref{gain}) to be $17 \rm{dBi}$ and $7.9341^{o}$ respectively.  Using these values, the antenna port radiation pattern approximated by the port approach in (\ref{port_p}) is again plotted  in Fig. \ref{pattern_compere1} at $\phi=0^{o}$ and $\theta_{tilt}=90^{o}$. Both approaches now have the same $\theta_{3dB}=7.9341^{o}$. However, the antenna port approach still ignores the sidelobes in the antenna port radiation pattern, which can result in misleading insights especially in scenarios where the elevation angular spread is high.

If simplicity in analysis is preferred, then the researchers can resort to the antenna port approach as long as the propagation scenario considered does not have rich scattering conditions and the  correct values of HPBW and maximum directional gain are utilized using (\ref{BW}) and (\ref{gain}). In order to see some examples of how this approach is utilized in devising elevation beamforming schemes, the readers are referred to \cite{downtilt3, utility, portapp, 3Dbeamforming, beamforming2, beamforming3,7268913}.  

In practice elevation beamforming is performed by controlling the downtilt weights applied to the elements in the AAS at the BS. The proponents of antenna port approach may argue that the optimal tilt angle can be found utilizing  (\ref{port_p}) and this tilt can then be used to compute the weight function  in (\ref{weight}) to implement elevation beamforming in practice.  However, optimizing the weight functions directly provides more degrees of freedom in controlling the vertical radiation pattern of a port. In fact, the expression in (\ref{weight})  is just an example downtilt weight vector provided by the 3GPP to form narrow beams in the elevation. The 3GPP TR36.897 gives the option of using other unit norm weight vectors that can be directly optimized to increase the energy at desired users and create nulls at interfering users in the same or other cells.  To perform such optimization tasks, it is important to resort to the antenna element approach that shows dependence on the downtilt weights.

This section basically summarized the 3D ray-tracing SCMs developed for the evaluation of FD-MIMO techniques. However, the high correlation in FD-MIMO arrays gives the alternate option of utilizing the correlation based 3D channel models for the better and easier design and analysis of elevation beamforming schemes. This will be the subject of the next section.

\section{Spatial Correlation and Associated Models}

 While a compact 2D planar array benefits from the arrangement of a large number of antenna elements within feasible BS form factors, it also reduces the effective spacing between the antenna elements. Moreover, the reported values of elevation angular spread are much smaller than those for the azimuth angular spread in realistic propagation environments. Both these factors result in the elements of FD-MIMO array to be highly correlated.  It is imperative to characterize and take into account this spatial correlation to realistically evaluate the performance gains realizable through elevation beamforming techniques.

\subsection{Significance in Massive MIMO Analysis}

 It is well known that spatial correlation is detrimental to the performance of MIMO systems and large capacity gains can only be realized when the sub-channels constituted by the elements of the antenna array are potentially decorrelated \cite{Telatar, Foschini}. While this is always true for point-to-point MIMO communications \cite{corr_impact, corr_impact1}, spatial correlation can actually be beneficial in multi-user MIMO settings, where it is the collection of the correlation matrices of all the users  that determines the system performance. The users are generally separated by multiple wavelengths resulting in uncorrelated channels across the users. While each user can experience high spatial correlation within its channel vector, the correlation matrices are generally quite different for different users. As a consequence, it has been shown in \cite{massiveMIMObook} that for mutually orthogonal channel vectors,  each user gets the full array gain proportional to $N$. Spatial  correlation can therefore be beneficial in Massive MIMO scenarios if the users have sufficiently different spatial correlation matrices. This has also been demonstrated for small-scale multi-user MIMO systems in \cite{corr_impact2, corr_impact3, corr_impact4}.

The high correlation in FD-MIMO arrays can not only improve the performance of multi-user Massive MIMO systems but also reduce the CSI feedback overhead incurred in the implementation of elevation beamforming techniques. This is possible through the design of elevation beamforming schemes using these so-called 3D correlated channel models that depend on the quasi-static spatial channel covariance matrices of the users, instead of the small-scale parameters that vary instantaneously. 

The ray-tracing channel model discussed in Section IV is one way to generate correlated FD-MIMO channels. However, the explicit dependence of this channel on the number of paths and associated small-scale parameters (AoDs, AoAs, powers) makes the theoretical analysis of this model generally intractable. Using the developed SCFs from existing works, the Kronecker channel model can be formed and utilized instead, which is defined as \cite{kronecker}, \cite{kronecker_model}, \cite{kronecker_model2},
\begin{equation}
\label{Kronecker1}
\begin{aligned}
\textbf{H} = \sqrt{10^{-(PL+\sigma_{SF})/10}} \textbf{R}_{MS}^{\frac{1}{2}} \textbf{X} \textbf{R}_{BS}^{\frac{1}{2}} ,
\end{aligned}
\end{equation}
where $\textbf{X}$ is a $O$x$N$ matrix with i.i.d. zero mean, unit variance complex Gaussian entries, $N$ is the number of Tx antenna ports, $O$ is the number of Rx antenna ports, and $\textbf{R}_{MS}$ and  $\textbf{R}_{BS}$ are $O \times O$ and $N\times N$ channel covariance matrices for the antenna ports at the MS and the BS respectively.

In multi-user MISO settings, where the user is equipped with a single isotropic Rx antenna element, the Kronecker model is represented by the Rayleigh correlated channel model given as \cite{assumptions},
\begin{equation}
\begin{aligned}
\label{corrchannel}
\textbf{h}=\sqrt{10^{-(PL+\sigma_{SF})/10}} \textbf{R}_{BS}^{\frac{1}{2}}\textbf{z},
\end{aligned}
\end{equation} 
where $\textbf{z}$ has i.i.d. zero mean, unit variance complex Gaussian entries and $\textbf{R}_{BS}$ is the user's channel covariance matrix.

%The covariance matrices should satisfy the following  conditions to allow for the application of random matrix theory (RMT) tools for the analysis of these theoretical channel models:
%\begin{align}
%\label{assumption1}
%\lim \sup_{N_{BS}} || \textbf{R}_{BS} || < +\infty, \\
%\label{assumption2}
%\lim \sup_{N_{MS}} || \textbf{R}_{MS} || < +\infty.
%\end{align}

 The main limitation of these models as compared to the 3GPP based ray-tracing channel model in Section IV is the loss of information on the number of propagation paths. The rank structure of MIMO channel matrix not only depends on the correlation within its elements but also on the structure of scattering in the propagation environment. It is possible to have a rank deficient channel matrix even if the fading is decorrelated at both ends due to mild scattering conditions. This phenomenon is known as the pinhole or keyhole effect \cite{gesbert}, \cite{ourworkTWCnew} and causes discrepancies between the results obtained using (\ref{Kronecker1}) and (\ref{channel1}). However, it was shown in \cite{ourwork} that the effect of this phenomenon diminishes as $\bar{N} \rightarrow \infty$.

Elevation beamforming schemes can be designed theoretically using these correlation based models with the help of tools from RMT that are quite useful in the large antenna regime. The implementation of these schemes will require the estimation of large-scale parameters only instead of the full channel vectors since beamforming is performed using information of the quasi-static channel covariance matrices of the users.  The digital beamforming stage still requires the estimation of the instantaneous channels but these channels will now have a reduced dimension, thanks to the elevation beamforming stage that groups elements into a reduced number of antenna ports using downtilt weight vectors. There is obviously a tradeoff between the system performance and the CSI feedback overhead involved. A similar idea was utilized in \cite{SC_BF} to devise a  spatial-correlation-based partial-channel-aware beamforming scheme. The BSs choose to attain CSI with respect to only the selected antenna elements that transmit the training RS. A good balance is achieved between the system performance and the CSI feedback overhead.

Motivated by the important role played by spatial correlation in the analysis of FD Massive MIMO systems, this section provide guidelines to compute the correlation coefficients based on the two channel representations introduced in the last section for the 2D AAA shown in Fig. \ref{antenna}. The developed SCFs can then be used to form the Kronecker channel model in (\ref{Kronecker1}) and the Rayleigh correlated channel model in (\ref{corrchannel}). Note that the SCF for other TXRU architectures presented in Section III-C can be developed similarly.

\subsection{SCF based on Antenna Port Approach}

Consider the 3D channel representation in (\ref{channel1}) between a BS equipped with an AAA of vertically polarized antenna elements shown in  Fig. \ref{antenna} and a stationary user. The antenna port patterns $\textbf{g}_{P,t}$ and $\textbf{g}_{P,r}$ are computed using (\ref{port_p}) and (\ref{decom}) and the array responses are computed  using (\ref{array_response}) and (\ref{array_response1}). The spatial correlation between Tx antenna ports $s$ and $s'$ corresponding to cluster $\bar{n}$ and $\bar{n}'$ respectively is defined and can be written as \cite{ourwork},
\begin{align}
\label{corr_portap}
&\rho_{P,\bar{n},\bar{n}'}(s,s')=\mathbb{E}[[\textbf{h}]_{s,\bar{n}} [\textbf{h}]_{s',\bar{n}'}^{H}] , \\
\label{corr_portap1}
&=\sum_{\bar{m}=1}^{\bar{M}_{\bar{n}}} \sum_{\bar{m}'=1}^{\bar{M}_{\bar{n}'}} \sqrt{P_{\bar{n}}/\bar{M}_{\bar{n}}} \sqrt{P_{\bar{n}'}/\bar{M}_{\bar{n}'}}  \mathbb{E}\Big[\alpha_{\bar{n},\bar{m}} \alpha_{\bar{n}',\bar{m}'} \nonumber \\
&\times  g_{P,t}(\phi_{\bar{n},\bar{m}},\theta_{\bar{n},\bar{m}},\theta_{tilt}) g_{P,t}(\phi_{\bar{n}',\bar{m}'},\theta_{\bar{n}',\bar{m}'},\theta_{tilt})  \nonumber \\
&\times \exp\left(i\frac{2\pi}{\lambda}d_{H}(s-1)\sin\phi_{\bar{n},\bar{m}}\sin\theta_{\bar{n},\bar{m}}\right) \nonumber \\
& \times \exp\left(-i\frac{2\pi}{\lambda}d_{H}(s'-1)\sin\phi_{\bar{n}',\bar{m}'}\sin\theta_{\bar{n}',\bar{m}'}\right)\Big].
\end{align}
Note that for ports comprising of vertically polarized antenna elements as considered in Fig. \ref{antenna}, $g_{P}$ is a 1x1 entry given by $\sqrt{A_{P}(\phi,\theta,\theta_{tilt})}$ and $\alpha$ is also a scalar. The path loss and shadow-fading just appear as a scaling factor so they are not included in (\ref{corr_portap1}). 

The correlation between the ports $s$ and $s'$ is then given by,
\begin{align}
\label{corr_portap2}
&\rho_{P}(s,s')= \sum_{\bar{n}=1}^{\bar{N}} \sum_{\bar{n}'=1}^{\bar{N}} \rho_{t,n,n'}(s,s').
\end{align}

Since the AoDs and AoAs for the sub-paths in each cluster are correlated, so the expression in (\ref{corr_portap2}) can only be numerically evaluated. In order to enable a tractable closed form formulation of this expression, the authors in \cite{ourwork} drop the assumption made in the standards that every cluster gives rise to $\bar{M}$ unresolvable sub-paths. Since these sub-paths are assumed to be unresolvable in the standards and are centered around the AoD/AoA of the original cluster, so their spatial properties are quite similar and are well-captured by the spatial parameters defined for the overall cluster. The authors assume uniform distribution of power across the clusters and combine it with $\alpha_{\bar{n}}$, so that $\mathbb{E}[|\alpha_{\bar{n}}|^{2}]=\frac{1}{\bar{N}}$. Under these assumptions, the channel between BS antenna port $s$ and MS antenna $u$ is generated by summing the contributions of $\bar{N}$ i.i.d. clusters as follows \cite{ourwork},
\begin{align}
\label{channel_TSP}
& [\textbf{H}]_{su}(t)= \sqrt{10^{-(PL+\sigma_{SF})/10}} \sum\limits_{\bar{n}=1}^{\bar{N}}   \alpha_{\bar{n}} g_{P,t}(\phi_{\bar{n}},\theta_{\bar{n}},\theta_{tilt}) \nonumber \\
&\times g_{P,r}(\varphi_{\bar{n}},\vartheta_{\bar{n}})  \exp \left(i k (s-1) d_{H} \sin\phi_{\bar{n}}\sin\theta_{\bar{n}}\right)  \nonumber \\
&\times \exp\left(i k  (u-1) d_{H}\sin\varphi_{\bar{n}}\sin\vartheta_{\bar{n}}\right) \exp(i 2\pi v_{\bar{n}}t).
\end{align} 

Since the parameters describing the clusters are i.i.d., the double sum over $\bar{n}$ and $\bar{n}'$ in the expression of the correlation between the Tx antenna ports $s$ and $s'$, $\rho_{P}(s,s')=\mathbb{E}[[\textbf{h}]_{s}[\textbf{h}_{s'}]^{H}]$, can be simplified as, 
\begin{align}
\label{Tx_p}
\rho_{P}(s,s')&=\mathbb{E}\Big[|g_{P,t}(\phi,\theta,\theta_{tilt})|^{2}\exp\Big(i\frac{2\pi}{\lambda}d_{H}(s-s') \nonumber \\
&\times \sin\phi\sin\theta\Big)\Big],
\end{align}
for $\mathbb{E}[|\alpha_{\bar{n}}|^{2}]=\frac{1}{\bar{N}}$, $\bar{n}=1,\dots \bar{N}$. 

A closed form expression for (\ref{Tx_p}) was derived and presented in \textbf{Theorem 1} of \cite{ourwork}. The spatial correlation coefficients were calculated by providing the derived Theorem with the FS coefficients of the PAS and PES of the 3D propagation scenario under study. These power spectra are important statistical properties of wireless channels that provide a measure of the power distribution in the azimuth and elevation dimensions respectively. 

%They are defined as,
%\begin{equation}
%\label{PAS}
%\begin{aligned}
%\text{PAS}_{p}(\phi)= p_{\phi}(\phi) A_{P,H}(\phi)|_{lin} ,
%\end{aligned}
%\end{equation}
%\begin{equation}
%\label{PES}
%\begin{aligned}
%\text{PES}_{P}(\theta)= p_{\theta}(\theta) A_{P,V}(\theta,\theta_{tilt})|_{lin} ,
%\end{aligned}
%\end{equation}
%where the angular power density functions, $p_{\phi}(\phi)$ and $p_{\theta}(\theta)$ equal $f_{\phi}(\phi)$ and $\frac{f_{\theta}(\theta)}{\sin(\theta)}$ respectively, with $f_{\phi}(\phi)$ and $f_{\theta}(\theta)$ being the PDFs of the azimuth and elevation angles \cite{Kalliola02angularpower}. 

The resulting expression of $\rho_{P}(s,s')$, after performing the expectation as given in \textbf{Theorem 1} of \cite{ourwork}, depends only on the large-scale parameters of the channel, including the angular spreads and the mean AoDs/AoAs. The proposed expression can be used to form the 3D correlated SCMs outlined in Section V-A. For this, we need to form the covariance matrix for the BS antennas denoted as $\textbf{R}_{BS}$. This can be done using the relationship $[\textbf{R}_{BS}]_{s,s'}=\rho_{P}(s,s')$. The covariance matrix for the MS can be formed similarly. 

%Also $g_{P,H}(\phi)$ and $g_{P,V}(\theta)$ are the field pattern of vertically polarized antenna elements in the horizontal and vertical directions respectively. Note that for $A_{P,H}(\phi)<A_{m}$ and $A_{P,V}(\theta,\theta_{tilt})< A_{m}$ in (\ref{port_p}), $g_{P,t}(\phi,\theta) \approx g_{P,H}(\phi)g_{P,V}(\theta)$.

%The proposed method is unique and is in contrast to most of the previous works that assume 2D channel models and derive the SCF for specific underlying angular distribution and form of antenna patterns. 

It is important to note that the expressions in (\ref{corr_portap2}) and (\ref{Tx_p}) consider the spatial correlation between any two antenna ports without accounting for the correlation between the elements constituting these ports. The role played by the elements is only captured approximately in the antenna port radiation pattern expression in (\ref{port_p}) utilized in (\ref{corr_portap2}) and (\ref{Tx_p}).

\subsection{SCF based on Antenna Element Approach}

The spatial correlation between  antenna ports is actually governed by the correlation between the elements constituting these ports. Here, we give some guidelines to express the SCF between antenna ports as a function of the inter-element correlation and the downtilt weights, utilizing the antenna element based channel representation in (\ref{channel3}). 

\subsubsection{Derivation of the SCF for Antenna Elements}
Consider the 3D channel expression in (\ref{channel3}) and a stationary user. The correlation between any two Tx antenna elements $m$ and $m'$ in antenna ports $s$ and $s'$ respectively in the 2D AAA shown in Fig. \ref{antenna}, corresponding to cluster $\bar{n}$ and $\bar{n}'$, can be written as \cite{ourworkTCOM},
\begin{align}
\label{corr_elementsap}
& \rho_{E,\bar{n},\bar{n}'}((m,s),(m',s'))=\sum \limits_{\bar{m}=1}^{\bar{M}_{\bar{n}}} \sum \limits_{\bar{m}'=1}^{\bar{M}_{\bar{n}'}} \sqrt{P_{\bar{n}}/\bar{M}_{\bar{n}}} \sqrt{P_{\bar{n}'}/\bar{M}_{\bar{n}'}}   \nonumber \\
&\times \mathbb{E}\Big[ \alpha_{\bar{n},\bar{m}} \alpha_{\bar{n}',\bar{m}'} g_{E,t}(\phi_{\bar{n},\bar{m}},\theta_{\bar{n},\bar{m}})g_{E,t}(\phi_{\bar{n}',\bar{m}'},\theta_{\bar{n}',\bar{m}'})   \nonumber \\
&\times [\textbf{a}_{t,s}(\phi_{\bar{n},\bar{m}}, \theta_{\bar{n},\bar{m}}) ]_{m}  [\textbf{a}_{t,s'}(\phi_{\bar{n},\bar{m}}, \theta_{\bar{n},\bar{m}}) ]_{m'}^{*} \Big].
\end{align}
Note that for vertically polarized antenna elements considered in Fig. \ref{antenna}. $g_{E}$ is a 1x1 entry given by $\sqrt{A_{E}(\phi,\theta)}$ and $\alpha$ is also a scalar. 

The correlation between antenna elements $(m,s)$ and $(m',s')$ is written as,
\begin{align}
\label{corr_elementsap1}
& \rho_{E}((m,s),(m',s'))=\sum \limits_{\bar{n}=1}^{\bar{N}} \sum \limits_{\bar{n}'=1}^{\bar{N}}\rho_{E,\bar{n},\bar{n}'}((m,s),(m',s')).
\end{align}

\begin{figure*}
  \centering
  \begin{tabular}{  c  c  }
    \hline
\rowcolor{LightCyan}
    \textbf{Array} & \textbf{Mathematical Formulation} \\ \hline
    \begin{minipage}{.3\textwidth}
		\includegraphics[width=2.1 in]{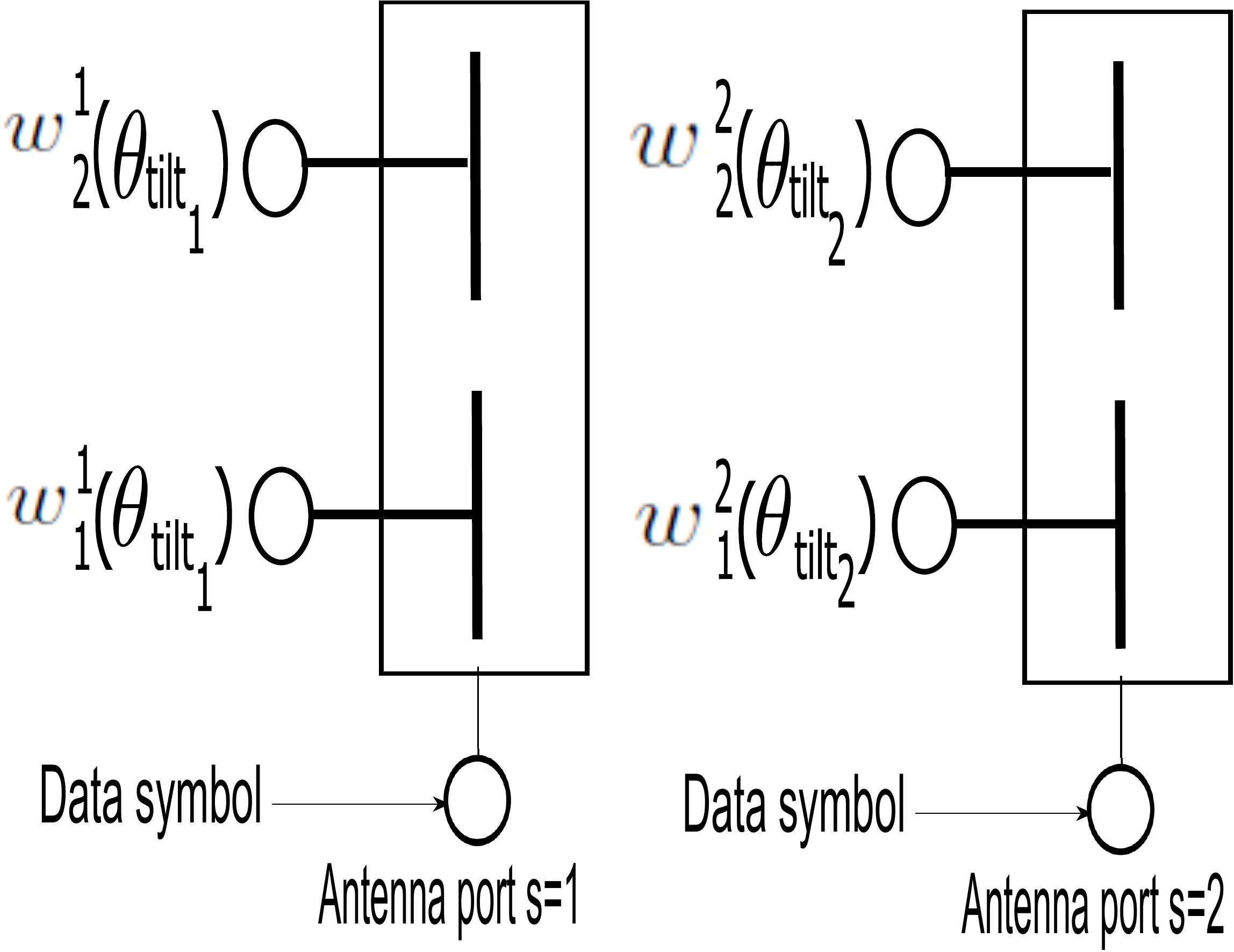}
    \end{minipage}
    &
 $\begin{aligned} 
& \textbf{R}^{BS}=\begin{bmatrix}
\ \rho^{E}_{P}(1,1) & \rho^{E}_{P}(2,1)   \\
\ \rho^{E}_{P}(1,2) & \rho^{E}_{P}(2,2)  \end{bmatrix} \nonumber \\
&= \begin{bmatrix} 
\ w^{*1}_{1} & w^{1*}_{2} & 0 & 0   \\
\ 0 & 0 &  w^{2*}_{1} & w^{2*}_{2} \end{bmatrix} \nonumber \\
&\times \begin{bmatrix} 
\ \rho_{E}((1,1),(1,1)) & \rho_{E}((2,1),(1,1)) & \rho_{E}((1,2),(1,1)) & \rho_{E}((2,2),(1,1))   \\
\ \rho_{E}((1,1),(2,1)) & \rho_{E}((2,1),(2,1)) & \rho_{E}((1,2),(2,1)) & \rho_{E}((2,2),(2,1))   \\
\ \rho_{E}((1,1),(1,2)) & \rho_{E}((2,1),(1,2)) & \rho_{E}((1,2),(1,2)) & \rho_{E}((2,2),(1,2))   \\
\ \rho_{E}((1,1),(2,2)) & \rho_{E}((2,1),(2,2)) &  \rho_{E}((1,2),(2,2)) & \rho_{E}((2,2),(2,2))  \end{bmatrix} \nonumber \\
& \times \begin{bmatrix} 
\ w^{1}_{1}& 0   \\
\ w^{1}_{2}& 0   \\
\ 0 & w^{2}_{1} \\
\ 0 & w^{2}_{2} 
 \end{bmatrix}. \nonumber
\end{aligned}$
    \\ \hline
  \end{tabular}
  \caption{Correlation matrix for a $2\times 2$ AAA.}
	\label{2port}
\end{figure*}

The AoDs and AOAs for the sub-paths in each cluster are correlated, which renders the theoretical analysis of the expression in (\ref{corr_elementsap1}) very hard. The authors in \cite{ourworkTCOM} make an important preliminary contribution of analyzing the expression in (\ref{corr_elementsap1}) and developing a closed-form expression for the spatial correlation between the antenna elements using a simplified channel model, where the assumption that each cluster gives rise to $\bar{M}$ unresolvable sub-paths is again dropped as done in \cite{ourwork}. The authors also consider uniform distribution of power over clusters and combine it with $\alpha_{\bar{n}}$, which now has a variance $\frac{1}{\bar{N}}$. The channel between BS antenna port $s$ and MS antenna element $u$ is generated by summing contributions of $\bar{N}$ i.i.d. clusters as follows \cite{ourworkTCOM},
\begin{align}
\label{channel_sim}
& [\textbf{H}]_{su}(t) =\sqrt{10^{-(PL+\sigma_{SF})/10}} \sum \limits_{m\in \text{port }s}^{M} {w}_{m}(\theta_{tilt}) \sum \limits_{\bar{n}=1}^{\bar{N}} \alpha_{\bar{n}}   \nonumber \\
&\times g_{E,t}(\phi_{\bar{n}},\theta_{\bar{n}}) [\textbf{a}_{t,s}(\phi_{\bar{n}}, \theta_{\bar{n}}) ]_{m} g_{E,r}(\varphi_{n},\vartheta_{n}) [\textbf{a}_{r}(\varphi_{\bar{n}}, \vartheta_{\bar{n}})]_{u} \nonumber \\
&\times  \exp(i 2\pi v_{\bar{n}}t).
\end{align}

Using the  array response expression of an individual element in (\ref{V}), and for $\mathbb{E}[|\alpha_{\bar{n}}|^{2}]=\frac{1}{\bar{N}}$, the SCF for the channels constituted by the $(m, s)$ and $(m', s')$ antenna elements at the Tx side can now be expressed as,
\begin{align}
\label{Tx}
& \rho_{E}((m,s),(m',s'))= \mathbb{E}[|g_{E,t}(\phi,\theta)|^{2} [\textbf{a}_{t,s}(\phi, \theta) ]_{m}  \nonumber \\
&\times [\textbf{a}_{t,s'}(\phi, \theta) ]_{m'}^{*}], \nonumber \\
&=\mathbb{E}\Big[|g_{E,t}(\phi,\theta)|^{2} \exp\Big(i 2 \pi \Big[\frac{d_{H}}{\lambda}(s-s')\sin\phi\sin\theta \nonumber \\
&+ \frac{d_{V}}{\lambda}(m-m') \cos\theta \Big] \Big)\Big], 
\end{align}
where $m,m'=1 \dots M$,  $s,s'=1, \dots N$. The weights are used to group the antenna elements into ports and will play a role when the correlation between two ports will be derived.  

After some reformulations using the SHE of plane waves \cite{jacobi} and properties of Legendre polynomials \cite{legendre}, (\ref{Tx}) is expressed analytically as a linear combination of the FS coefficients of PAS and PES in \textbf{Theorem 1} of \cite{ourworkTCOM}. This SCF has been made available online at \cite{code} to facilitate the interested researchers and industrials in computing the correlation coefficients by providing this function with only the FS coefficients of the PAS and PES for the propagation environment under study.

As discussed in Section III, the radio resource is organized on the basis of antenna TXRUs/ports, where each port is used to transmit a data symbol at a particular value of the downtilt angle  $\theta_{tilt}$, determined using the downtilt weights applied to the elements in that port. Since the spatial multiplexing gains are determined by the number of ports and the correlation between them, so it is important to characterize this correlation.

\subsubsection{Spatial Correlation Function for Ports}

From (\ref{channel3}) it is evident that the SCF for the channels constituted by antenna ports $s$ and $s'$ will be a function of the correlations between all the elements constituting these ports and the weight functions applied to these elements as,
\begin{align}
\label{corr_ports}
&\rho^{E}_{P}(s,s')= \sum_{m=1}^{M}\sum_{m'=1}^{M} w_{m}(\theta_{tilt}) w_{m'}(\theta_{tilt})^{*} \rho_{E}((m,s),(m',s')), 
\end{align}
 where $s,s'=1, \dots N$ and $\rho_{E}((m,s),(m',s'))$ is given by (\ref{corr_elementsap1}) if the exact 3GPP channel model in (\ref{channel3}) is utilized and by (\ref{Tx}) if the simplified channel model in (\ref{channel_sim}) is utilized.

In matrix form, the $N \times N$ correlation matrix for the antenna ports constituting the AAS in Fig. \ref{antenna} can be written as,
\begin{align}
\label{corr_BS}
\textbf{R}_{BS} = \widetilde{\textbf{W}}^{H} \textbf{R}^{E} \widetilde{\textbf{W}},
\end{align}
where $\widetilde{\textbf{W}}$ is $NM \times N$ block diagonal matrix of the weight vectors applied to the $N$ antenna ports given as,
\begin{align}
\label{tilW}
&\widetilde{\textbf{W}}^{H}=
\begin{bmatrix}
\ {\textbf{w}^{s}}^{H}  & \textbf{0}^{1 \times M}  &\textbf{0}^{1 \times M(N-2)}    \\
\ \textbf{0}^{1 \times M} & {\textbf{w}^{s}}^{H} & \textbf{0}^{1 \times M(N-2)}\\
\  & \hspace{.8in} \ddots  & \\
\ \textbf{0}^{1 \times M}   & \textbf{0}^{1 \times M(N-2)}   & {\textbf{w}^{s}}^{H} \end{bmatrix},
\end{align}
where $\textbf{w}^{s}$  is the $M\times 1$ weight vector for antenna port $s$ given by (\ref{weight}). These weight vectors can be different for different ports allowing them to transmit at different downtilt angles, depending on the TXRU architecture being considered. However, for the 1D TXRU virtualization model with sub-array partition shown in Fig. \ref{antenna}, these vectors are equal, i.e. $\textbf{w}^{s}=\textbf{w}$. $\forall s$. Also, $\textbf{R}^{E}$ is the $NM \times NM$ correlation matrix for all the elements constituting the AAS defined as,
\begin{align}
\label{corr_elements}
& [\textbf{R}^{E}]_{(s'-1)M+m',(s-1)M+m}=\rho_{E}((m,s),(m',s')).
\end{align}

Note that with this formulation, $[\textbf{R}_{BS}]_{s',s}=\rho^{E}_{P}(s,s')$. An example formulation of the correlation matrix for a $2 \times 2$ AAA is shown in Fig. \ref{2port}. Also, note that $\textbf{R}_{MS}$ can be formed similarly and the two covariance matrices can be used to form the correlated 3D SCMs given by (\ref{Kronecker1}) and (\ref{corrchannel}).

\normalsize

\subsection{Comparison of the Two Approaches}

\begin{table*}[!b]
\centering
\normalsize
\caption{Simulation parameters.}
\begin{tabular}{|l|l|}
\hline
  \textbf{Parameter} & \textbf{Value} \\ 
\hline
\textbf{AAA parameters for Fig. \ref{antenna}:} & \\
\hline
 Number of antenna ports in the horizontal direction (N) & 4 \\
 Number of antenna elements in the vertical direction (M) & 8 \\
 Horizontal inter-element spacing, $d_{H}$ & 0.5$\lambda$ \\
 Vertical inter-element spacing, $d_{V}$ & 0.8$\lambda$ \\
 Antenna element gain, $G_{max, E}$ & $8\rm{dBi}$ \\
 Vertical HPBW, $\theta_{3dB}$ & $65^{o}$ \\
 Horizontal HPBW, $\phi_{3dB}$ & $65^{o}$ \\
\hline
\textbf{ITU antenna parameters:} & \\
\hline
 Number of antenna ports in the horizontal direction (N) & 4 \\
 Antenna port gain, $G_{max,P}$ & $17\rm{dBi}$ \\
 Vertical HPBW of antenna port, $\theta_{3dB,P}$ & $15^{o}$ \\
 Horizontal HPBW of antenna port, $\phi_{3dB,P}$ & $70^{o}$ \\
\hline
\textbf{Elevation domain parameters:} & \\
Angular distribution, $f_{\theta}(\theta)$ & Laplacian \\
 Mean AoD in the elevation, $\theta_{0}$ & $90^{o}$ \\
 Downtilt angle for each port, $\theta_{tilt}$ & $90^{o}$ \\
 Elevation angular spread at departure, $\sigma_{l,t}$ & $8^{o}$ \\
\hline
\textbf{Azimuth domain parameters:} & \\
Angular distribution, $f_{\phi}(\phi)$ & Von Mises \\
 Mean AoD in the azimuth, $\mu$ & $0$ \\
 Azimuth spread measure at departure, $\kappa_{t}$ & 6 \\
\hline
\textbf{General parameters:} & \\
 Channel estimation & ideal \\
 Number of terms for expressions in \textbf{Theorem 1} of \cite{ourwork} and \textbf{Theorem 1} of \cite{ourworkTCOM},  $N_{0}$ & 15 \cite{ourwork}, 30 \cite{ourworkTCOM}   \\
 \hline
\end{tabular}
\label{parameters}
\end{table*}

Here we compare the correlation between the antenna ports constituting the AAS using both the approximate `antenna port approach' and the exact `antenna element approach'. The spatial correlation between any two antenna ports $s$ and $s'$ $s,s'=1, \dots N$, using the ITU approach has been studied and derived in \cite{ourwork} and summarized in Section V-B. The exact spatial correlation between antenna ports $s$ and $s'$, accounting for the individual contributions of all the elements constituting these ports, is derived in \cite{ourworkTCOM} and summarized in Section V-C. 

For the simulations, we assume a single-user and all antenna ports transmitting at a downtilt angle of $\theta_{tilt}=90^{o}$. In the standards, elevation AoDs and AoAs are drawn from the Laplacian elevation density spectrum with parameters $\theta_{0}$ and $\sigma_{l}$ that represent the mean elevation AoD/AoA and the elevation angular spread respectively.  The characteristics of azimuth angles are well captured by Wrapped Gaussian (WG) density spectrum with mean $\mu$ and spread $\sigma_{WG}^{2}$ \cite{TR36.873,ITU}. However in the recent years, the Von Mises (VM) distribution has received great attention in modeling non-isotropic propagation due to its close association with the WG spectrum \cite{correlation2}, \cite{vonmises1}. The use of VM distribution is also proposed in \cite{ourwork} due to its close match with WG distribution as well as the simplicity in the calculation of its FS coefficients. This distribution is characterized by the mean azimuth AoD/AoA ($\mu$) and the azimuth angular spread (parametrized as $\frac{1}{\kappa}$). We assume $\theta_{0}$ and $\mu$ to be the LoS angles of the user. The values of the parameters at the Tx side as used in the simulations are summarized in Table \ref{parameters}. The values approximately represent the 3D-UMa non-line of sight (NLoS) scenario outlined in TR36.873 \cite{TR36.873}. $N_{0}$ represents the number of terms used in the computation of the correlation using the  analytical expressions in \textbf{Theorem 1} of \cite{ourwork} and \textbf{Theorem 1} of \cite{ourworkTCOM}.

\begin{figure}[!t]
\centering
\includegraphics[width=2.75 in]{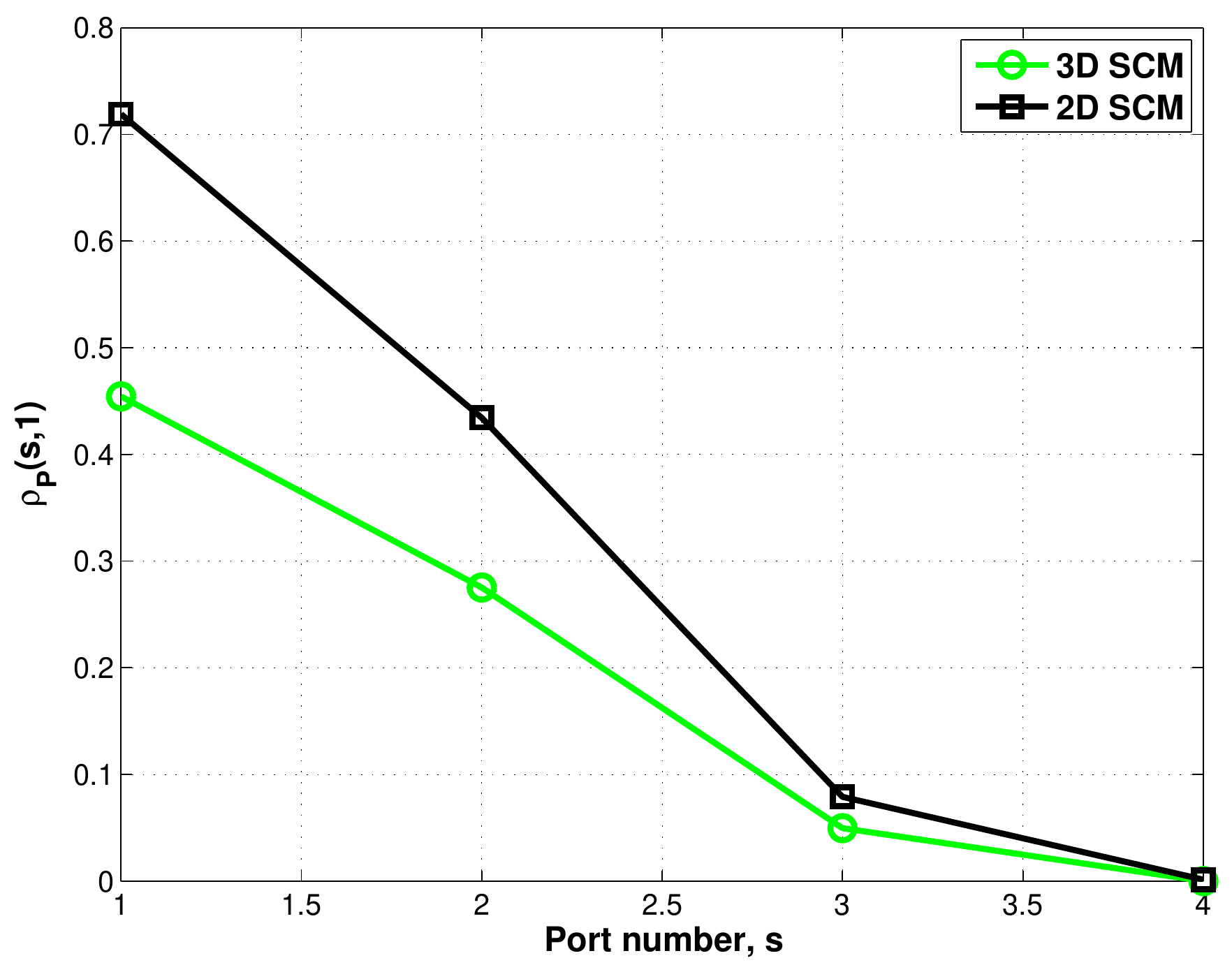}
\caption{Comparison of the correlation between Tx antenna ports for 2D and 3D channels.}
\label{compare_2D}
\end{figure}

We first use Monte-Carlo simulations to compare the correlation between antenna ports for the 2D SCM and the antenna port approach based 3D SCM. The simulations are done for $10000$ realization of the simplified channel model in (\ref{channel_TSP}), dropping the assumption that each cluster gives rise to a group of unresolvable sub-paths.  The results normalized by the antenna gain are plotted in Fig. \ref{compare_2D}, where the correlation for the 3D SCM is computed using the expression in (\ref{Tx_p}) and the correlation for the 2D SCM is computed using the same expression but assuming $\theta=\theta_{tilt}=\frac{\pi}{2}$. The 2D model clearly overestimates the radiated power as well as correlation, which is a consequence of ignoring the directivity of antennas in the elevation. 

\begin{figure}[!t]
\centering
\includegraphics[width=2.75 in]{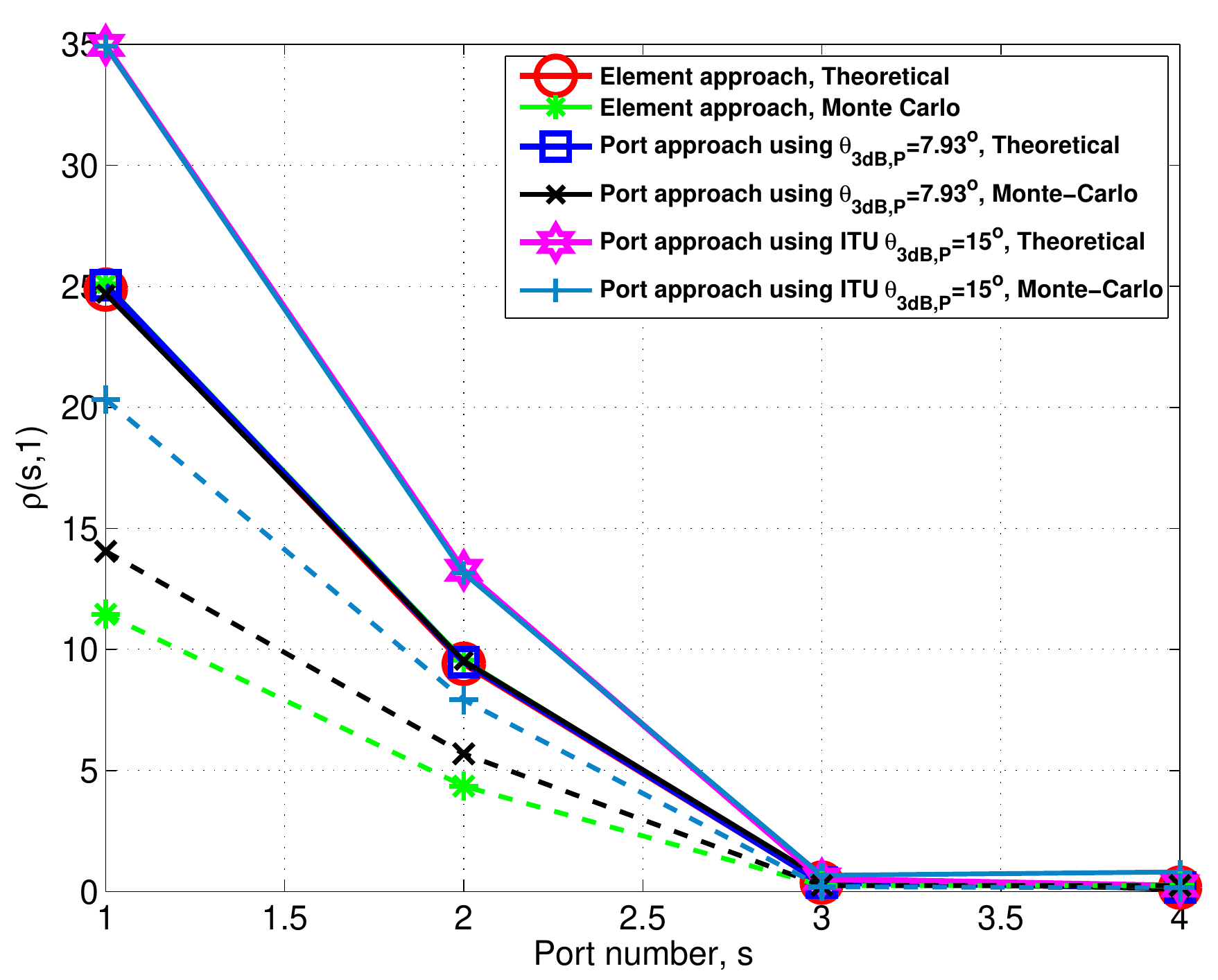}
\caption{Correlation between Tx antenna ports for the element approach in (\ref{corr_ports}) and \cite{ourworkTCOM} and port approach in (\ref{Tx_p}) and \cite{ourwork}. The dotted lines represent Monte-Carlo simulations for $\sigma_{l,t}=25^{o}$.}
\label{compare_corr}
\end{figure}

Both  the theoretical result in \cite{ourworkTCOM} and  the Monte-Carlo simulated result in (\ref{corr_ports}) using  (\ref{Tx}) are plotted in Fig. \ref{compare_corr} for the correlation between the antenna ports, based on the antenna element approach. The Monte Carlo simulations are performed over $10000$ realizations, with the weights calculated using (\ref{weight}) for $\theta_{tilt}=90^{o}$. Similarly, both the theoretical result in \cite{ourwork} and the Monte-Carlo simulated correlation in (\ref{Tx_p}) are plotted for the antenna port approach  for $\theta_{tilt}=90^{o}$ and $\theta_{3dB,P}=15^{o}$.  The antenna element gain and antenna port gain are incorporated into the simulations to study the effects on radiated power. The antennas are assumed to radiate isotropically in the azimuth. The theoretical results computed using the codes provided  in \cite{code} yield a perfect fit to the Monte Carlo simulated correlation. The developed SCFs in \cite{ourwork} and \cite{ourworkTCOM}  can therefore be efficiently utilized to generate correlation coefficients for FD-MIMO channels without relying on the time-consuming Monte-Carlo simulations. The SCFs only require the FS coefficients of the PAS and PES for the propagation scenario under study as inputs. 

%Note that the value of $\rho(s,1)$ at $s=1$ is the average energy transmitted by a port. 

We now compare the results for the element approach and the port approach in the same figure using the calculated values of $\theta_{3dB,P}$ and $G_{max,P}$ from Section IV-D. The values were calculated such that the $3$\rm{dB} beamwidth of the main lobes for both approaches were perfectly matched as shown in Fig. \ref{pattern_compere1}. This results in the correlation values for the port approach to be quite close to the values for the element approach as shown in Fig. \ref{compare_corr}. Since most of the radiated energy is concentrated in the main lobe of the radiation pattern so it seems reasonable to use the simpler antenna port approach for moderate values of angular spreads, as long as the correct values of $3 \rm{dB}$ beamwidth and maximum directional antenna port gain, $G_{max,P}$, are used. Using the $15^{o}$ value for the beamwidth as proposed in ITU \cite{ITU} directly results in significant overestimation of the radiated power due to the wider main lobe of the antenna radiation pattern. A wider lobe captures the energy of a higher number of propagation paths resulting in a higher radiated power as shown in Fig. \ref{compare_corr} by comparing $\rho(s,1)$  at $s=1$. 

%\begin{figure}[!t]
%\centering
%\includegraphics[width=2.5 in]{error.eps}
%\caption{Error in correlation caused by the antenna port approach compared to the antenna element approach.}
%\label{error}
%\end{figure}

When the same curves are plotted for a higher value of elevation angular spread, $\sigma_{l,t}=25^{o}$ (see dotted lines), the correlation values decrease due to richer propagation conditions. At the same time the radiated power from each antenna port also decreases because the energy from a higher number of propagation paths is not captured by the narrow beam transmitted from the antenna port. This is an illustration of how important it is to take into account the directional characteristics of antenna elements when evaluating the performance of FD-MIMO systems. Another important observation is that now the gap between the element approach and the port approach for the calculated values of  $\theta_{3dB,P}$ and $G_{max,P}$ is higher than that at $\sigma_{l,t}=8^{o}$. When the elevation spread is small and the user lies in the boresight of the antenna port, the energy of most of the propagation paths is captured by the Tx antenna pattern within its 3dB beamwidth. This beamwidth has been designed to be the same for both approaches utilizing (\ref{BW}) and (\ref{gain}), resulting in a smaller performance gap between the two approaches. On the other hand for higher values of elevation angular spreads, the approximate port approach towards modeling the vertical radiation pattern neglects the side lobes and the energy of many propagation paths (lying outside the 3dB beamwidth) is incorrectly scaled as seen in Fig. \ref{pattern_compere1}. This results in a mismatch in the correlation for the element approach and the port approach.

 %This effect is also confirmed in Fig. \ref{error}, which plots the difference between the correlation between adjacent antenna ports using both approaches. It can be seen that the error incurred by the port approach increases as the elevation spread increases. 

%The researchers interested in using the `antenna port approach' towards channel modeling should take into account the exact architecture of the port, the number of elements constituting it and the inter-element separation before proceeding to the analysis because the values proposed by ITU do not match the exact theoretical radiation pattern obtained by taking into account the construction of the port and individual patterns of the elements constituting it. 

Even though utilizing the calculated values of $\theta_{3dB,P}$ and $G_{max,P}$ using (\ref{BW}) and (\ref{gain}) respectively for the underlying structure of the port will minimize the gap between the performance of both approaches but it is still important to take the antenna element approach towards 3D channel modeling. The reason for this being the preferred channel representation is twofold; firstly it accounts for the side lobes in the radiation pattern, which are ignored in the antenna port approach; and secondly the linear dependence of the channel on the weights in the element approach as shown in (\ref{channel3}) allows for more flexibility in optimizing the downtilt weights to realize different elevation beamforming scenarios.

\section{Elevation Beamforming}

In conventional MIMO deployments, the BS antenna ports have a static vertical beam pattern and transmit at a fixed downtilt angle. The beams are only steered in the horizontal domain by feeding horizontally arranged antenna ports with  digitally preprocessed data symbols. As a result, the interference reduction and throughput optimization capabilities are also restricted to the horizontal plane. However several measurement campaigns have demonstrated that the angular spectrum of propagation channels has strong components in the elevation domain \cite{elevationcamp}, \cite{elevationcamp1}. As a consequence, beamforming in the azimuth plane alone can not fully exploit all the degrees of freedom offered by the channel. With the dynamic adaptation of the 3D radiation pattern of each antenna port in an AAA, the system performance can be further enhanced by using a variety of downtilting strategies that increase the desired signal power at an active user and/or suppress inter-user and inter-cell interference.  Encouraged by the initial implementations of this technology \cite{practicals, practicals1, practicals2, practicals3, trials}, the 3GPP is defining future mobile communication standards in the frame of its study items on 3D beamforming. 

In this section, we will first discuss the hybrid nature of beamforming allowed by FD-MIMO architectures and  the basic principles behind elevation beamforming. Later, we provide a mathematical framework for the design of elevation beamforming in a single-user MISO setting based on the element based channel representation. The objective is to  facilitate interested researchers in understanding how the aspects presented so far can be put together to devise elevation beamforming schemes. This is followed by an overview of related schemes in the literature designed for single-cell multi-user and multi-cell multi-user MISO systems. Finally, we compare the performance of different existing schemes using simulations.

\subsection{Hybrid Beamforming in FD-MIMO}

\begin{figure}[!t]
\centering
\includegraphics[width=2.5 in]{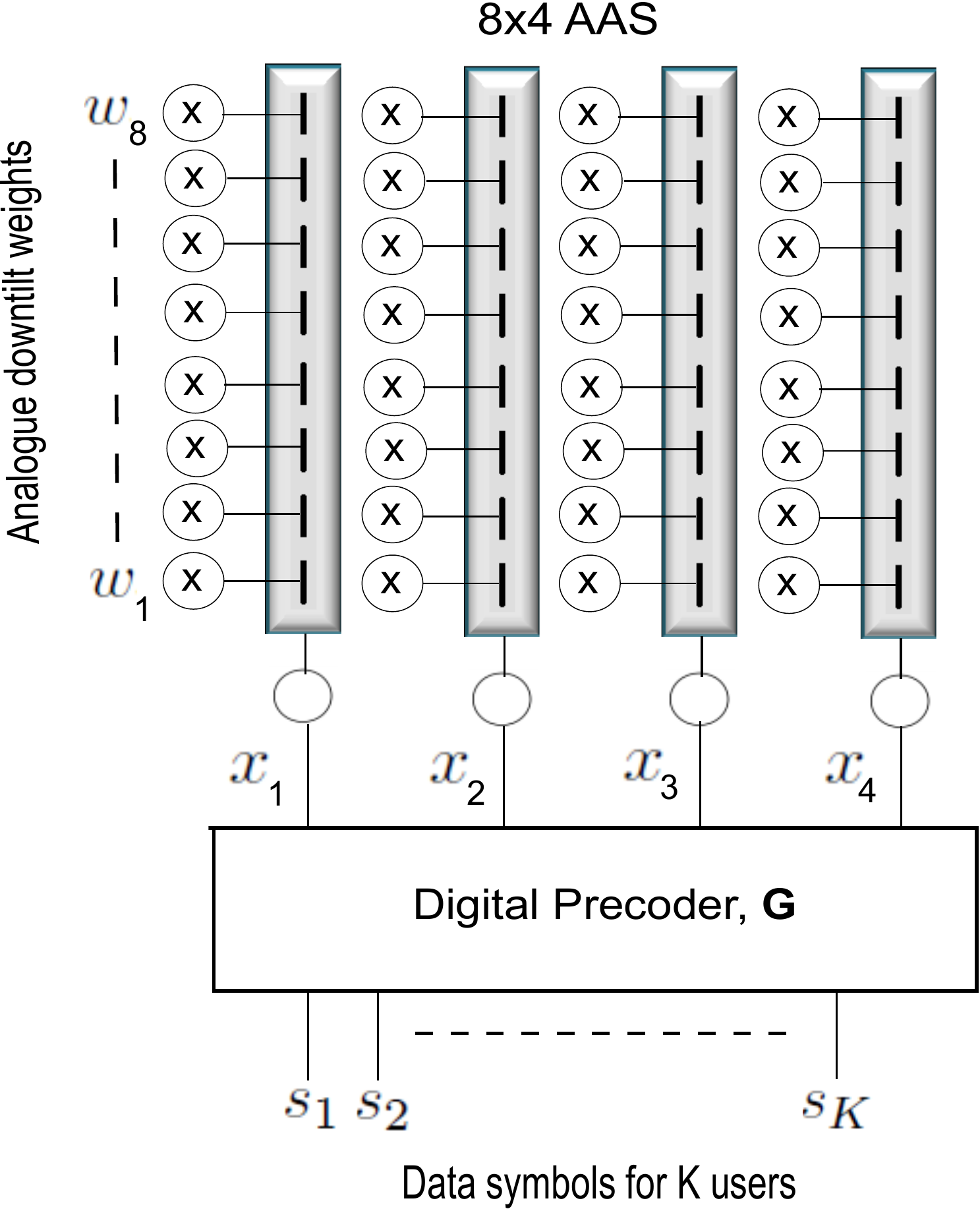}
\caption{Hybrid beamforming structure of an FD-MIMO AAA with $N=4$ and $M=8$.}
\label{hybrid}
\end{figure}

Multi-user MIMO precoding in the digital domain allows simultaneous transmission to several users by suppressing inter-user interference \cite{downlink_pre, downlink_pre1, downlink_pre2, downlink_pre3, downlink_pre4}. It has been shown that this interference can be handled in Massive MIMO settings with simple linear precoding techniques \cite{massive1, massive2}. Implementing these precoding techniques requires complete information of the channels, which is achieved through channel estimation. The canonical Massive MIMO systems operate a TDD protocol, where the uplink and downlink transmissions take place in the same frequency resource  but are separated in time. TDD systems benefit from channel reciprocity to acquire CSI in the uplink  and use it for downlink transmission. In every coherence interval, there is an uplink training phase where each user transmits a pilot sequence. The uplink estimation overhead is therefore proportional to the number of terminals and independent of the number of antennas, making it the preferred choice for Massive MIMO settings.

 %As the total number of orthogonal pilot sequences is limited by the channel coherence interval, so the pilot sequences have to be re-used in neighboring cells, contaminating the channel estimates. This effect is known as pilot contamination. There has been a significant amount of research on reducing the effect of pilot contamination in TDD systems \cite{PC1, PC2, PC3, PC4, PC5, PC6}. 

Unfortunately, the TDD systems require sophisticated calibration of the uplink and downlink signal chains to enable the exploitation of channel reciprocity through special hardware and design protocols \cite{cal2, cal1}.   Given that there is plenty of FDD spectrum currently in use so it is important to implement Massive MIMO in FDD as well, where uplink and downlink transmissions take place in separate bands and CSI is made available to the transmitter through explicit downlink training and uplink feedback. This downlink training and CSI feedback overhead is prohibitively high when implementing fully digital MU-MIMO precoding techniques in Massive MIMO settings.  

Generally, limited feedback methods have been proposed to reduce this feedback overhead \cite{LF, LF1}, which require high-resolution codebooks to minimize quantization errors. One way to reduce the feedback overhead is to compute the precoding vectors based on the users' channel covariance matrices which require knowledge of quasi-static large-scale channel parameters only instead of the full channels \cite{LF_SC}. However, this technique does not perform well in scenarios where the spatial correlation is low. Another approach is to employ opportunistic beamforming, where the BS transmits orthogonal beams with training signals in each time-slot and the users send back the SNR corresponding to the beams. The BS schedules the user with the highest SNR on each beam. This idea has been studied in several works \cite{hybrid1}, \cite{Opp_BF}.

Another popular technique employed in FDD based Massive MIMO systems is channel dimensionality reduction \cite{hybrid2, usergroups}. An example is the joint spatial division and multiplexing (JSDM) approach proposed in \cite{usergroups} where the authors split the downlink beamforming  into two stages - a pre-beamforming stage that  is implemented in the analog domain using quasi-static channel covariance matrices of the users and a MU-MIMO precoding stage that is implemented in the digital domain on the effective reduced-dimension channels formed by the pre-beamforming. The training dimension required to estimate the effective channels is significantly less than the total number of antenna elements, thanks to the dimensionality reduction of the analog pre-beamforming stage.

  The FD-MIMO AAAs realize this channel dimension reduction through the application of the downtilt antenna port weight vectors that are used to group antenna elements into a reduced number of antenna ports, as studied in Section III. These weights can be optimized to adapt the Tx beam dynamically in the vertical direction. This stage is implemented in the analog domain through the PAs connected directly to the antenna elements in each port and is referred to as elevation beamforming. The high spatial correlation in FD-MIMO arrays can be exploited to design these analog weights resulting in the implementation of the elevation beamforming stage using the slowly varying spatial covariance matrices of the users' channel vectors.  The digital precoding stage is implemented across the antenna ports instead of all the elements and requires the estimation of the effective channel of dimension $N\times 1$ instead of $NM\times 1$. 
	
As an example, consider a multi-user MISO system with $K$ users served by the AAA shown in Fig. \ref{hybrid}. First we apply analog beamforming across the antenna elements using the downtilt weights $w_{m}$, $m=1,\dots, M$. These weights group antenna elements into antenna ports and act as a way to reduce the channel dimension while providing the capability of controlling the beam pattern in the 3D space. To see this, note that the received signal at user $k$, $y_{k} \in \mathbb{C}$ is given as,
\begin{align}
&y_{k}=\textbf{h}_{k}^{H}\textbf{x} + {n}_{k}, 
\end{align}
where $\textbf{x}\in \mathbb{C}^{N\times 1}$ is the Tx signal vector from the BS, $n_{k} \sim \mathcal{CN}(0,\sigma_{n,k}^{2}$) is the additive white Gaussian noise (AWGN) with variance $\sigma_{n,k}^{2}$ at the user $k$ and  $\textbf{h}_{k}^{H} \in \mathbb{C}^{1\times N}$ is the channel vector from the BS to user $k$. If modeled using the Rayleigh correlated 3D channel model, it will be given as $\textbf{h}_{k}=\sqrt{10^{-(PL+\sigma_{SF})/10}} \textbf{R}_{BS_{k}}^{\frac{1}{2}}\textbf{z}_{k} \in \mathbb{C}^{N\times 1}$, as defined in (\ref{corrchannel}), where $\textbf{z}_{k}$ has i.i.d. zero mean, unit variance complex Gaussian entries and $\textbf{R}_{BS_{k}} \in \mathbb{C}^{N\times N}$ is the user's channel covariance matrix defined as $\textbf{R}_{BS_{k}} = \widetilde{\textbf{W}}^{H} \textbf{R}^{E}_{k} \widetilde{\textbf{W}}$, where $\widetilde{\textbf{W}}$ is $NM \times N$ block diagonal matrix of the weight vectors applied to the $N$ antenna ports given by (\ref{tilW}) and $\textbf{R}^{E}_{k}$ is the $NM \times NM$ correlation matrix for all the elements constituting the AAS with respect to user $k$ given by (\ref{corr_elements}). It is obvious that the analog weight vectors are playing the role of reducing the channel dimension from $NM \times 1$ to $N \times 1$.

%If the ray-tracing 3D SCM from section IV had been used, the channel vector would have entries given by (\ref{channel3}). Notice again that the downtilt weights $w_{m}$ reduce the channel dimension from $NM \times 1$ to $N \times 1$ in (\ref{channel3}). The optimization of these weights to yield desirable beam patterns in the elevation plane is referred to as elevation beamforming. 

The digital precoding is applied across the vector of data symbols of the $K$ users, denoted as $\textbf{s}=[s_{1}, s_{2}, \dots, s_{K}]^{T} \in \mathbb{C}^{K\times 1}$ resulting in the Tx signal vector $\textbf{x} \in \mathbb{C}^{N \times 1}$ written as,
\begin{align}
&\textbf{x}=\sum_{k=1}^{K} \sqrt{p_{k}} \textbf{g}_{k} s_{k},
\end{align}
where $\textbf{g}_{k} \in \mathbb{C}^{N\times 1}$ is the digital precoding vector for user $k$, and $p_{k} \geq 0$ and $s_{k} \sim \mathcal{CN}(0,1)$ are the signal power  and the data symbol for user $k$ respectively. The digital precoder, $\textbf{G} = [\textbf{g}_{1}, \dots, \textbf{g}_{K}] \in \mathbb{C}^{N\times K}$, satisfies the average total power constraint as,
\begin{align}
\label{p_cons}
& \mathbb{E}[||\textbf{x}||^{2}]=\mathbb{E}[\text{tr }(\textbf{P} \textbf{G}^{H} \textbf{G})] \leq \bar{P},
\end{align}
where $\bar{P}>0$ is the average total Tx power and $\textbf{P}=\text{diag}(p_{1}, p_{2}, \dots, p_{K}) \in \mathbb{R}^{K\times K}$. In order to form this digital precoder, the BS needs to estimate the $N\times 1$ channel vectors of the users instead of the $NM\times 1$ channel vectors thanks to the elevation beamforming stage, which significantly reduces the feedback overhead.

To summarize, an FD-MIMO array not only serves as a practical implementation of a Massive MIMO system but also lends itself to a hybrid digital analog implementation where the MU-MIMO precoding stage is implemented in the baseband across the antenna ports and the elevation beamforming stage is implemented in the analog domain across the antenna elements in each port. This hybrid implementation, illustrated in Fig. \ref{hybrid}, also makes FD-MIMO systems particularly suited to mm-wave bands, where the number of RF chains is limited. 

\subsection{Principle of Elevation Beamforming}

Elevation beamforming results from the adaptation of the vertical dimension of the antenna beam pattern. Passive antenna elements cannot adapt their directional radiation pattern dynamically since these elements are fed with a passive feeder network and can only transmit at a fixed downtilt angle. Active antenna elements, on the other hand, have individual PAs connected directly to each transmitting element as discussed in Section III. Arranging these elements in vertical columns allows the resulting 2D AAA to adapt its radiation pattern dynamically in the vertical and horizontal directions in each transmission interval. The AAAs support an electronic beam-tilt feature by providing control on the phase and amplitude weights applied to the individual antenna elements in each port \cite{AAS1, AAS2, AAS3}. As a consequence, the antenna pattern in the vertical direction and the MIMO precoding across the horizontal ports can be optimized simultaneously resulting in spatially separated transmissions to a large number of users in both the elevation and azimuth planes. The authors in \cite{beamforming2} analyze the average rate gain of AASs over passive antenna systems for a single user case. Active beamforming is shown to achieve an average rate gain equal to $3.986 (\sigma_{\theta}^{2}/\theta_{3dB}^{2})$ over passive beamforming in the high SNR regime, where $\sigma_{\theta}^{2}$ is the variance of the elevation angle distribution.

There are, however, some differences between the azimuth and elevation planes of the propagation channel that should be taken into account when designing elevation beamforming techniques. First, the coverage area in the azimuth plane is much wider that that in the elevation plane. Generally, a coverage of angles from $-60^{o}$ to $60^{o}$ relative to the BS boresight is required in the azimuth plane. However, the distribution of the users in the elevation is not as wide spread. This can be seen by noting that for a cell radius of 250m and BS and users heights equal to 30m and 1.5m respectively, the minimum downtilt angle at the edge of the cell is about $96^{o}$ and the maximum downtilt angle given the user can not be located within 30 m of the BS is about $130^{0}$. In fact it was shown in \cite{practicals} and \cite{portapp} that the downtilt angles required to cover 95\% of the cell area are between $90^{o}$ and $115^{o}$. Therefore the adjustment range for the downtilt angle of the vertical beam pattern in a typical 3D macro-cell environment is much less than that for the steering angle in the azimuth domain. Second difference is that since the users are located close to each other in the elevation plane, so the HPBW of the vertical beam has to be designed to be much smaller than the HPBW of the horizontal beam to allow for better separation of the users in the elevation. For example, ITU approximates the beamwidth of the main lobe of the vertical radiation pattern as $15^{o}$, whereas the beamwidth in the horizontal direction is set as $70^{o}$. This narrow beam in the elevation can be achieved by arranging a large number of antenna elements in the vertical direction to constitute an antenna port and feeding these elements with the same signal with appropriate weights to achieve the desired downtilt and a narrow beamwdith. The relation between the 3dB beamwidth and $M$ and $d_{V}$ is given in (\ref{BW}).

Therefore, the design goal of elevation beamforming schemes is to adapt the downtilt angle of the Tx beam dynamically within a small adjustment range  based on the users' channel information and/or statistics while the HPBW is kept as narrow as possible to allow for a better spatial separation of the users in the elevation plane. In the next section, we discuss a low channel feedback solution for the  downtilt weight vectors applied to the antenna ports in a single cell single user MISO system, incorporating the practical aspects of FD-MIMO presented so far.

\subsection{Elevation Beamforming in a Single User MISO System}

Consider the downlink of a single-cell where an  $M \times N$ AAS shown in Fig. \ref{antenna} serves a single user equipped with a single isotropic Rx antenna element. The received complex baseband signal $y \in \mathbb{C}$ at the user is given by,
\begin{align}
& y=\textbf{h}^{H} \textbf{x} + n,
\end{align}
where  $\textbf{h} \in \mathbb{C}^{N \times 1}$ is the 3D channel vector from the BS to the user,  $\textbf{x} \in \mathbb{C}^{N\times 1}$ is the Tx signal from the AAS, and $n \sim \mathcal{CN}(0,\sigma_{n}^{2}$) is the AWGN with variance $\sigma_{n}^{2}$ at the user.  Note that \textbf{h} is a function of $\theta_{tilt}$ directly in the port approach while it depends on $\theta_{tilt}$ through the downtilt weights in the element approach. 

The Tx signal vector is defined as $\textbf{x}=\textbf{g}s$, where $\textbf{g} \in \mathbb{C}^{N\times 1}$  denotes the Tx precoding vector with $||\textbf{g}||^{2}=P_{Tx}$, where $P_{Tx}$ is the Tx power of the BS, and $s$ denotes the Tx data symbol with unit variance. It is well known that MRT, i.e. $\textbf{g}=\sqrt{P_{Tx}}\textbf{h}/||\textbf{h}||$ is the optimal digital precoding scheme for single-user single-stream transmission when perfect CSI is available at the transmitter \cite{MRT_book}. Under this setting, the downlink signal-to-noise ratio (SNR) at the user is given by,
\begin{align}
\label{SNR_s}
&\gamma = \frac{P_{Tx}}{\sigma_{n}^{2}} \textbf{h}^{H}\textbf{h}.
\end{align}

%optimizing the downtilt weights dynamically using this model would impose high training and feedback overhead especially in FDD protocol based Massive MIMO settings. Also, 

The interest is in finding the optimal downtilt angle $\theta_{tilt}$ for the antenna ports that maximizes the SNR.  Preliminary notable works on single-user elevation beamforming, including \cite{practicals,beamforming2,3Dbeamforming}, considered the antenna port approach towards modeling the vertical radiation pattern of each port. In \cite{beamforming2}, the optimal downtilt angle was determined at the user end using training signals from the BS transmitted with different $\theta_{tilt}$. The index of the beam that maximized the downlink SNR was fed back to the BS and the corresponding downtilt angle was used for data transmission. It was shown in \cite{3Dbeamforming} that under active beamforming the optimal tilt angle corresponded to the elevation LoS angle of the user.

\begin{table*}
\centering
\caption{Popular Elevation beamforming schemes for a single-cell multi-user MISO system.}
\begin{tabular}{@{}l c c c c }
\hline
\rowcolor{LightCyan}
& \textbf{Small-scale channel} & \textbf{Radiation Pattern} & \textbf{Digital Precoding} & \textbf{Downtilting Strategy}  \\
\hline
& & & & \\
\textbf{Cell-Specific Tilting (CST)} & Any model & ITU port pattern (\ref{port_p}) & MRT/ZF/RZF & Common fixed tilt angle for \\
& & & & all antenna ports   \\
& & & & \\
\textbf{Switched Beam Tilting} & i.i.d. Rayleigh  & ITU port pattern  (\ref{port_p}) & ZF & Partitions the cell into vertical regions \\
\textbf{ (SBT)} \cite{beamforming3} & & & & and applies a pre-determined (tilt, HPBW)    \\
& & & &  pair when serving each region   \\

& & & & \\
\textbf{Multi-User Active}& i.i.d. Rayleigh & ITU port pattern  (\ref{port_p}) & ZF & Under high SNR, the optimal downtilt \\
\textbf{ Beamforming (MUAB)}  \cite{beamforming2} & & & & angle is the weighted arithmetic mean value  \\
& & & & of elevation LoS angles of all users   \\
& & & & \\
\textbf{Two Fixed}& LoS propagation & ITU port pattern with  & Not specified & The combination of near and far downtilts and  \\
\textbf{ Downtilts Scheme}  \cite{practicals3} & & measured values & &  the value of border between the two regions\\
& & & &  that maximize the SE is obtained numerically  \\
& & & & \\
\textbf{Antenna Tilt}& 3GPP TR36.873 & TR36.873 pattern (\ref{port})  & Not specified & Two users are scheduled simultaneously  \\
\textbf{Assignment}  \cite{downtilt1} & &  & &  using the proposed user scheduling algorithm \\
& & & &  and the downtilt angle of the port that serves one  \\
& & & &  user is set as that user's elevation LoS angle plus  \\
& & & &  an adjustment angle to avoid inter-port interference  \\
& & & & \\
\textbf{Single Downtilt Beamforming} & (\ref{corrchannel}), (\ref{corr_BS})-(\ref{corr_elements}) & TR36.873 pattern (\ref{port})  & MRT & Sub-optimal downtilt weight vectors  \\
\textbf{ (SDB)} \cite{ourworkTCOM} & & & & maximizing the minimum user SIR are\\
& & & & obtained through Dinkelbach's algorithm  \\
\hline
\end{tabular}
\label{multi_compare_table}
\end{table*}

These existing works on single-user elevation beamforming ignore the dependency of the vertical beam pattern on the construction of the port, i.e. the values of $M$, $d_{V}$, and the downtilt weights. We showed in \cite{ourworkTCOM} that the more recent antenna element approach can be used to yield an optimal closed-form solution for the downtilt antenna port weight vectors in an FD-MISO setting, when \textbf{h} is modeled using the correlated 3D channel model in (\ref{corrchannel}),  with the users' covariance matrices determined using (\ref{corr_BS})-(\ref{corr_elements}) developed for the element approach. Under this setting, the problem of single-user elevation beamforming is formulated as the maximization of the SNR with respect to the downtilt weight vectors applied to the $N$ antenna ports as follows.

\textit{Problem (P1):}
\begin{align}
& \underset{\textbf{w}}{\text{maximize }}  \gamma \\
\label{constraint}
& \text{subject to }  ||\textbf{w}^{s}||_{2}=1, \text{ for } s=1, \dots N.
\end{align}

The constraint in (\ref{constraint}) ensures that the total power of every antenna port is bounded with respect to the number of elements constituting it. This problem has a simple eigenvector solution in the large $(M, N)$ regime given as follows.

\textbf{Theorem 1 [\textbf{Theorem 2} \cite{ourworkTCOM}].} For a single user MISO system where the BS is equipped with an $M \times N$ AAS shown in Fig. \ref{antenna} and the channel is represented using the Rayleigh correlated  model in (\ref{corrchannel}), the optimal 3D beamforming weight vectors $\textbf{w}^{s}$, $s=1,\dots, N$ in the large $(N,M)$ regime can be computed as,
\begin{align}
\textbf{w}^{s}= \textbf{v}_{\lambda_{max}( \textbf{R}^{E}_{ss})}, \hspace{.3in} s=1, \dots N,
\end{align}
where $\textbf{v}_{\lambda_{max}( \textbf{R}^{E}_{ss})}$ is the eigenvector corresponding to the maximum eigenvalue $\lambda_{max}$ of $\textbf{R}^{E}_{ss}$, where $\textbf{R}^{E}_{ss}$ is a $M \times M$ matrix given by $\textbf{R}^{E}([(s-1)M+1:s M], [(s-1)M+1:s M])$, for $\textbf{R}^{E}$ defined in (\ref{corr_elements}).
%, such that,
%\begin{align}
%&{\textbf{R}^{E}_{ss}}^{\frac{1}{2}} \textbf{w}^{s}= \sqrt{\lambda_{max} (\textbf{R}^{E}_{ss})}  \textbf{w}^{s}.
%\end{align}

$\textbf{R}^{E}_{ss}$ basically refers to the correlation matrix formed by the elements of port $s$ and is therefore given by the $s^{th}$ $M \times M$ diagonal matrix from $\textbf{R}^{E}$ in (\ref{corr_elements}). Since all ports have an identical structure in the configuration considered in Fig. \ref{antenna} so $\textbf{w}=\textbf{w}^{s}$ $\forall s$.

The proof of \textbf{Theorem 1} follows similar steps as detailed in Appendix A of \cite{ourworkTCOM}. Note that the implementation of this solution imposes negligible feedback overhead since the channel correlation matrix for the user can be computed at the BS using knowledge of only the angular spread at the BS and the user’s LoS azimuth and elevation angles.

\subsection{Elevation Beamforming in Multi-User MISO Systems}

Elevation beamforming in single-cell and multi-cell MU-MISO settings has recently become a subject of interest, especially with the progress made in the 3GPP on various study items on FD-MIMO. As a consequence, several downtilting strategies have been proposed in literature while utilizing linear precoding schemes in the digital domain. In this section, we present these existing methods and provide some guidelines for the efficient design of elevation beamforming techniques under the element approach towards channel modeling. 

\subsubsection{Single-Cell Systems} Table \ref{multi_compare_table} summarizes some popular elevation beamforming schemes for a single-cell multi-user MISO system in terms of the small-scale channel model, radiation pattern, and horizontal and vertical beamforming techniques considered.

%In this section, we will present a few popular existing methods and compare them in terms of their computational cost and performance. 

Most of these schemes utilize linear precoding methods in the digital domain, which are generally asymptotically optimal in the large $(N,K)$ regime \cite{massive1} and robust to CSI imperfections \cite{massive2}. Zero-forcing (ZF) precoding completely cancels inter-user interference but also causes a reduction in the energy of the desired signal.  Moreover, ZF and the near-optimal regularized zeroforcing (RZF) precoding require inversion of the Gram matrix of the joint channel of all users, which has a complexity of $K^{2}N$ \cite{drabla2}, making the computational cost of these schemes prohibitively high in the large $(N,K)$ regime A notable exception that solves these issues is MRT \cite{MRT}, which is a popular scheme for large scale MIMO systems due to its low computational complexity, robustness, and high asymptotic performance \cite{massive1}.

In terms of the downtilting strategy, CST uses a common fixed downtilt at the BS for all antenna ports, determined initially using field trial evaluations. The value of this fixed tilt is used to determine the radiation pattern in (\ref{port_p}), that is required to model the channel in (\ref{channel1}) if the antenna port approach is adopted or used to compute the weights in (\ref{weight}) to model the channel in (\ref{channel3}) if the antenna element approach is adopted. The authors in \cite{beamforming3} proposed to partition the cell into vertical regions and apply one of the finite number of tilt and HPBW pairs at the BS when serving each region to increase the SNR over the coverage area. The transmission was scheduled to one of the vertical regions in each time slot, with the vertical region activity factor for region $s$ given by $\frac{|\mathcal{K}_{s}|}{K}$, where $\mathcal{K}_{s}$ was the set of the users in vertical region $s$.  The strategy was referred to as SBT. The authors in \cite{beamforming2} formulated the weighted sum rate maximization problem with ZF beamforming in the digital domain to completely cancel multi-user interference. The authors first proposed a joint optimization solution for the power allocation matrix of the $K$ users and the downtilt angle at the BS, which was prohibitively complex due to its iterative nature. To decrease the computational complexity, the authors proposed to separate the joint problem into two sub-problems. Under the high SNR approximation, the optimal downtilt angle for active beamforming was computed as the weighted arithmetic mean value of the elevation LoS angles of all users and the algorithm was referred to as MUAB.  The Two Fixed Downtilts scheme is similar to SBT, except here only the downtilt angles were selected - one for the near region and one for the far region. For each value of border between the two regions, a different combination of near and far downtilts provided the optimal value for the spectral efficiency. This combination was found numerically using simulations.

SBT and MUAB utilize the ITU approximation to model the  antenna port radiation pattern which assumes a constant gain outside the main lobe instead of considering the side lobes as explained in Section IV-D. The Two Fixed Downtilts scheme uses the ITU shape but the HPBW values are set as per measurements. In other words, these approaches directly optimize $\theta_{tilt}$ using (\ref{port_p}), without taking into account the effects of the number of elements in each port and the applied weights. In practice, elevation beamforming is performed by controlling the downtilt weights applied to the elements in the AAS, so  it is important to consider practical TXRU architectures outlined in the 3GPP for FD-MIMO systems and optimize the corresponding TXRU virtualization weight functions.

The Antenna Tilt Assignment scheme uses the element level 3D channel model outlined in TR36.873 but it does not optimize the weight functions directly but rather focuses on assigning the downtilt angles for the two simultaneously scheduled users (out of an arbitrary number of users)  such that the inter-user interference is minimized. The electrical tilt assignment is given as $\theta_{tilt} = \theta_{0} \pm \Delta \theta$ where $\Delta \theta$ is the adjustment angle whose optimal value is found so as to avoid interference between the antenna ports serving the two users in case the difference between the two user's elevation angles is less than the electrical tilt angular resolution. The example weight function from the 3GPP TR36.873 given in (\ref{weight}) is then used to compute the weights for the assigned tilt angles.

A preliminary contribution that deals with the downtilt weight vector optimization problem directly, for the TXRU architecture shown in Fig. \ref{antenna}, appeared in \cite{ourworkTCOM}. The authors utilized the  Rayleigh correlated channel model in (\ref{corrchannel}) with the covariance matrices determined using the correlation expressions  in (\ref{corr_BS})-(\ref{corr_elements}), developed for antenna element approach. The focus of \cite{ourworkTCOM} was on interference limited systems, so the performance metric employed was maximizing the minimum user signal to interference ratio (SIR), for which the asymptotic convergence result was derived under MRT precoding. The deterministic approximation of the SIR was then used to formulate a fractional optimization problem for the downtilt antenna port weight vectors, which was re-formulated using semi-definite relaxation and then solved using Dinkelbach's method  and Gaussian randomization technique. The algorithm was referred to as SDB. 

\begin{table*}
\centering
\caption{Popular elevation beamforming schemes for a multi-cell multi-user MISO system.}
\begin{tabular}{@{}l c c c c }
\hline
\rowcolor{LightCyan}
& \textbf{Small-scale channel} & \textbf{Radiation Pattern} & \textbf{Digital Precoding} & \textbf{Downtilting Strategy}  \\
\hline
& & & & \\
\textbf{Cell-Specific Tilting (CST)} & Any model & ITU port pattern (\ref{port_p}) & MRT/ZF/RZF & Common fixed tilt angle for \\
& & & & the antenna ports at each BS  \\
& & & & \\
\textbf{Coordinated User-Specific }  & i.i.d. Rayleigh  & ITU port pattern  (\ref{port_p}) & ICIC/MRT & Greedy tilt selection algorithm for
 \\
\textbf{Tilting (CUST)} \cite{portapp} & & & & coordinated tilt adaptation    \\
& & & & \\
\textbf{Coordinated 3D }& i.i.d. Rayleigh & ITU port pattern (\ref{port_p})  & MRT & DIviding RECTangle (DIRECT) method utilized  \\
\textbf{Beamforming} \cite{downtilt3} & &  & &  to select the tilt angles of all BSs for each \\
& & & &  realization of active users' locations, perfectly \\
& & & & known at the BS, to maximize the sum throughput \\
& & & & \\
\textbf{Adaptive Multicell}& i.i.d. Rayleigh & ITU port pattern (\ref{port_p})  & ZF & Partition the coverage area into disjoint vertical  \\
\textbf{3-D Beamforming}  \cite{7268913} & &  & &  regions and adapt the multicell cooperation \\
& & & &  strategy to serve the users in each vertical region \\
& & & & \\
\textbf{Utility Fair Tilt}& i.i.d. Rayleigh & ITU port pattern (\ref{port_p})  & Not specified & Jointly adjust antenna tilt angles  \\
\textbf{Optimization}  \cite{utility} & &  & &  within the cellular network so as to maximize \\
& & & &  user utility, subject to network constraints  \\
& & & & \\
\textbf{Direct Steering with}& 3GPP SCME (NLoS) & ITU port pattern (\ref{port_p})  & Not specified & Steer the beam directly to the scheduled  \\
\textbf{Downtilt Limitation}  \cite{practicals3} & &  & &   user under proportional fairness scheduling \\
& & & &  while limiting the minimal possible downtilt \\
& & & & \\
\textbf{Antenna tilt adaptation}& TR36.873 channel & TR36.873 pattern (\ref{port}) & MMSE detection & Gradient descent used to find the optimal \\
 \cite{tilt_letter} & & & & tilt in the uplink that maximizes the    \\
& & & & asymptotic SE under pilot contamination  \\
& & & & \\
\textbf{Multi-cell SDB} \cite{ourworkWCNC} & (\ref{corrchannel}), (\ref{corr_BS})-(\ref{corr_elements}) & TR36.873 pattern (\ref{port})  & MRT & Sub-optimal downtilt weight vectors   \\
& & & & maximizing the minimum user $\overline{SIR}$  \\
& & & &  under constraints on interference leakage \\
& & & & obtained through Dinkelbach's algorithm  \\
\hline
\end{tabular}
\label{multicell_compare_table}
\end{table*}

\subsubsection{Multi-cell systems}
The performance of multi-cell systems is generally limited by inter-cell interference. An efficient way to combat inter-cell interference is to use coordinated beamforming techniques, where  CSI is shared among all BSs to allow for the joint design of the beamforming matrix. The data for each user is only transmitted by the serving BS, making this approach practical when backhaul capacity is limited \cite{multicellref, multicellref1}. In the presence of perfect CSI, complete interference cancellation is possible by coordinated zero-forcing beamforming (ZFBF), but this also causes a loss in the energy of the desired signal and incurs a high computational cost in Massive MIMO settings.  Prior works have considered coordinated beamforming in the horizontal domain only, that enabled the cancellation of the interference in the azimuth plane only. However, in FD-MIMO architectures the vertical dimension of the antenna port radiation pattern can also be adapted for performance optimization through the downtilt weights applied to the elements in each port. This has motivated the design of coordinated beamforming schemes that mitigate inter-cell interference in both the azimuth and the elevation planes.

Initial works studied 3D beam steering strategies for the mitigation of inter-cell interference through system-level simulations \cite{beamforming1, practicals2, 3Dbeamformingmulticell}, without providing any theoretical framework or design guidelines for the selection of the downtilt angles. Later, some theoretical 3D coordinated beamforming strategies were proposed and have been in summarized in Table \ref{multicell_compare_table}. The authors in \cite{portapp} considered a multi-cell single-user system under CSI impairments that result from feedback delay and mobility of users. Possible extension to multi-user systems  was suggested. The authors proposed a  coordinated 3D beamforming strategy that jointly adapted beamforming in the horizontal and vertical dimensions depending on the location of the active user. For moderate delay/mobility values, the authors proposed to mitigate the effects of inter-cell interference using inter-cell interference cancellation (ICIC) in the digital domain and coordinated tilt adaptation in the analog domain. A greedy tilt selection algorithm was proposed for coordinated tilt adaptation and the algorithm was referred to as CUST.  The results of the proposed method seem promising in the propagation environment outlined in the work. 

In another work \cite{7268913}, the authors proposed a multi-cell cooperation strategy that would serve the users in either the cell-interior region or the cell-edge region in each time slot, with an appropriate transmission mode, i.e., conventional single-cell transmission or network MIMO transmission, and a corresponding appropriate tilt that would potentially achieve a tradeoff in maximizing all the performance metrics simultaneously. The authors denoted their proposed strategy as adaptive multicell 3-D beamforming. 

However, there are several limitations imposed on the propagation channel considered in both these works that make the results less general: 1) small-scale scattering is only assumed in the azimuth plane while a single LoS path is assumed in the elevation, which does not conform to the 3D channel model outlined in the 3GPP; 2) uncorrelated Rayleigh fading is assumed; 3) the scheme has been devised under the approximate antenna port radiation pattern expression given by (\ref{port_p}).

The authors in \cite{utility} formulated the problem of the adaptation of antenna tilt angles, subject to certain network constraints, as a utility fair optimization task. The tilts adjustments at all base stations were carried out jointly to manage interference in a coordinated manner. The resulting non-convex optimization problem was re-formulated as a convex optimization under certain conditions. Specifically the optimization was shown to be convex for any concave utility function in the high SNR regime, whereas for any SINR regime, the problem could be formulated in a convex manner for proportional fair rate allocation objective.

The authors in \cite{practicals3} studied the gains of 3D beamforming in a multi-cell multi-user setup with proportional fairness scheduling and interference coordination. Using, direct steering to the served user such that the minimal possible downtilt was limited in order to avoid severe interference in neighboring cells, a 50\% gain in cell edge throughput was achieved. 

Both these works again consider the ITU/3GPP TR36.814 radiation pattern  and thus do not take into account the actual geometry of the array. 

There have been two works so far that have considered the antenna element approach towards channel modeling in devising elevation beamforming strategies for multi-cell scenarios. The  antenna tilt adaptation scheme in \cite{tilt_letter} considered a TDD protocol and optimized the tilt angle in the uplink with MMSE detection, under the assumption that the elevation angles for each user were common for all propagation paths. The author took into account the effect of pilot contamination, and derived an asymptotic expression for the SE. It was shown that the asymptotic SE is a concave function of the antenna tilt, when some requirements on the SINR were met.  Under this case, the maximization of the SE resulted in one global solution, which was found using a gradient descent based approach. 

The authors in \cite{ourworkWCNC} considered a FD massive MISO setting, where the non-parametric channel model in (\ref{corrchannel}) formed using the element approach based correlation expressions was utilized and MRT precoding was used in the digital domain. Under this setting,  the authors derived a deterministic approximation for the SIR in the large $(M, N)$ regime and formulated an optimization task with the objective of maximizing the minimum user signal to intra-cell interference ratio in each cell, denoted by $\overline{SIR}$, subject to constraints on the interference leakage caused to the users in other cells \cite{ikram, leakage}. The problem was solved using semi-definite relaxation and Dinkelbach's method as done for the single-cell multi-user case in \cite{ourworkTCOM}. The authors referred to this algorithm as multi-cell SDB and it has been summarized in Algorithm 1 of \cite{ourworkWCNC}.

\subsubsection{Proposed Methodology}

Developing efficient algorithms for the downtilt weight vector optimization for different TXRU architectures under the element based channel representation is non-trivial. However, it is possible to solve these problems in the asymptotic regime where the number of BS antenna ports and users grow large - which corresponds to the Massive MIMO regime. The proposed methodology is to model the channel for each user using the Rayleigh correlated model in (\ref{corrchannel}), with the users' covariance matrices determined using (\ref{corr_BS})-(\ref{corr_elements}) developed for the element approach. The expressions for different performance metrics like the SINR and SIR will turn out to be quadratic forms in complex zero mean, unit variance Gaussian vectors, for which convenient convergence results exist in RMT that can be exploited to yield their deterministic approximations, that are tight in the asymptotic regime. These deterministic approximations are easier to optimize and yield solutions for the downtilt weight vectors for different TXRU architectures. The readers are referred to \cite{ourworkTCOM}  as an example implementation of the proposed method for the TXRU model in Fig. \ref{antenna}.

Note that the proposed method splits downlink beamforming into two linear stages: an elevation beamforming stage that depends only on the users' channel correlation matrices and a linear digital precoding stage for the effective channels with dimension reduced from $MN$ to $N$.  The feedback overhead for the implementation of the digital precoding stage is therefore significantly reduced, since the number of antenna ports is much less than the total number of antenna elements. 

The elevation beamforming stage imposes negligible feedback overhead since the channel correlation matrix for each user can be computed with knowledge of only the angular spread at the BS (locally estimated) and the user's LoS azimuth and elevation angles. The BS can estimate the location of each user in the uplink and compute the corresponding LoS angle. The users' channel correlation matrices are quasi-static and vary very slowly. Even for nomadic users, the correlation matrices evolve in time much more slowly than the actual Rayleigh fading process as discussed in \cite{usergroups}, and can be tracked using well-known existing algorithms.

\subsection{Performance Evaluation}

The performance gains realizable through elevation beamforming are now studied using simulations with parameter values set as given in Table \ref{parameters}, unless stated otherwise. All simulations use the Rayleigh correlated channel model in  (\ref{corrchannel}), where the correlation matrices are computed using (\ref{corr_BS}).

\subsubsection{Elevation Beamforming in the Single-User MISO Setting}

\begin{figure}[!t]
\centering
\includegraphics[width=2.65 in]{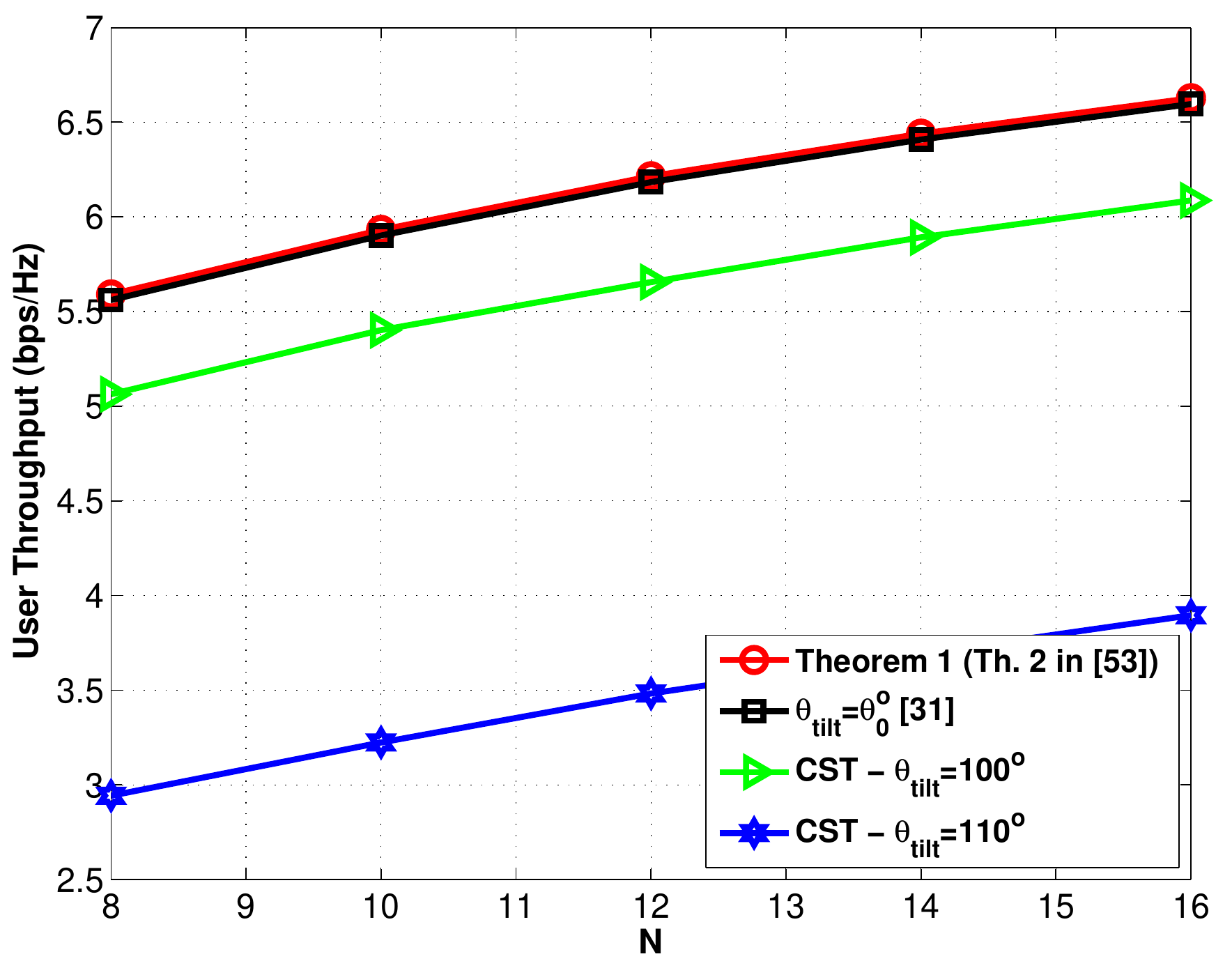}
\caption{Performance of a single-user MISO system.}
\label{single_user}
\end{figure}

The single-user MISO case is studied first using Monte-Carlo simulations of the SNR in (\ref{SNR_s}) for $P_{Tx}=56 \rm{dBm}$ and $\sigma_{n}^{2}=-100 \rm{dBm}$.  The BS equipped with an $8 \times N$ AAS serves an outdoor user located at the edge of a cell of radius $250m$, with the LoS angle $\theta_{0}$ computed to be $95.37^{o}$. The path loss and SF values are taken from the TR36.873 for the 3D urban macro scenario. \textbf{Theorem 1} is used to optimize the downtilt antenna port weight vector $\textbf{w}$ and the corresponding user throughput, averaged over $5000$ channel realizations, is plotted in red in Fig. \ref{single_user} along with the cases where the electrical downtilt angle is set to specific pre-defined values under CST. The solution given by \cite{3Dbeamforming}, where $\theta_{tilt}$ is set as the LoS angle of the user, i.e. $\theta_{0}=95.37^{o}$ is also plotted. To ensure fairness in comparison, all strategies are compared under the element based channel. The downtilt angles for CST and user specific tilting approach in \cite{3Dbeamforming} are used to compute the weight vectors utilizing the 3GPP relation in (\ref{weight}), which are then used to form the covariance matrices in (\ref{corr_BS}). User specific tilting in \cite{3Dbeamforming} and the eigenvector solution in  \cite{ourworkTCOM} yield somewhat similar performances, with the latter performing slightly better, while both perform much better than CST. 

\subsubsection{Elevation Beamforming in the Multi-User MISO Setting}

In this section, we study the multi-user MISO system with $K$ users placed randomly in a cell of radius $250 m$, at a minimum distance of $30m$ from the BS. We focus on an interference limited system as the value of thermal noise in the 3GPP is proposed to be very small. Monte-Carlo simulations are used to compare the achievable minimum user rate, averaged over $5000$ channel realizations, using the following downtilting strategies. 
\begin{itemize}
\item \textbf{SDB:} This strategy has been summarized in Table \ref{multi_compare_table} and deals with the direct optimization of the downtilt weight functions applied to the elements rather than the optimization of the overall downtilt angle of the port. These downtilt weights can be easily computed using Dinkelbach's method and Gaussian randomization technique, as explained in detail in Algorithm 1 of \cite{ourworkTCOM}. 
\item \textbf{Cell-specific tilting (CST):} A common fixed tilt of $90^{o}$ and $100^{o}$ is applied to all antenna ports with MRT precoding in the horizontal domain.
%\item  \textbf{SBT:} This tilting strategy proposed in \cite{beamforming3} has been summarized  in Table \ref{multi_compare_table}. The cell is partitioned into two vertical regions and one of the two ($\theta_{tilt}$, $\theta_{3\rm{dB}}$) pairs is applied at the BS when serving each region. The pairs are calculated as ($103.75^{o}, 21^{o}$) and ($96.51^{o},6.5^{o}$) for the inner and outer sectors respectively.  To allow for fair comparison with other strategies,  transmission using MRT precoding is scheduled to one of the two vertical regions in each time slot  and the strategy is denoted as MRT-SBT. 
\item \textbf{Center of Means (CoM):} The optimal downtilt angle is computed as the mean value of the elevation LoS angles of all the users at every transmission interval and MRT precoding is used in the digital domain. 
\end{itemize}
Note that CST and CoM are the most basic downtilting strategies that directly adapt the downtilt angle of the ports. However, to allow for a fair comparison with SDB, the obtained downtilt angle is used to compute the corresponding downtilt weight vector utilizing (\ref{weight}). The Rayleigh correlated channel model in (\ref{corrchannel}) is then simulated  with covariance matrices computed using $(\ref{corr_BS})$. SDB, on the other hand, deals with the direct optimization of the weight vectors and has been included in the simulations to provide a flavor of the performance gains that can be realized by optimizing the system at an element level.

\begin{figure}[!t]
\centering
\includegraphics[width=2.65 in]{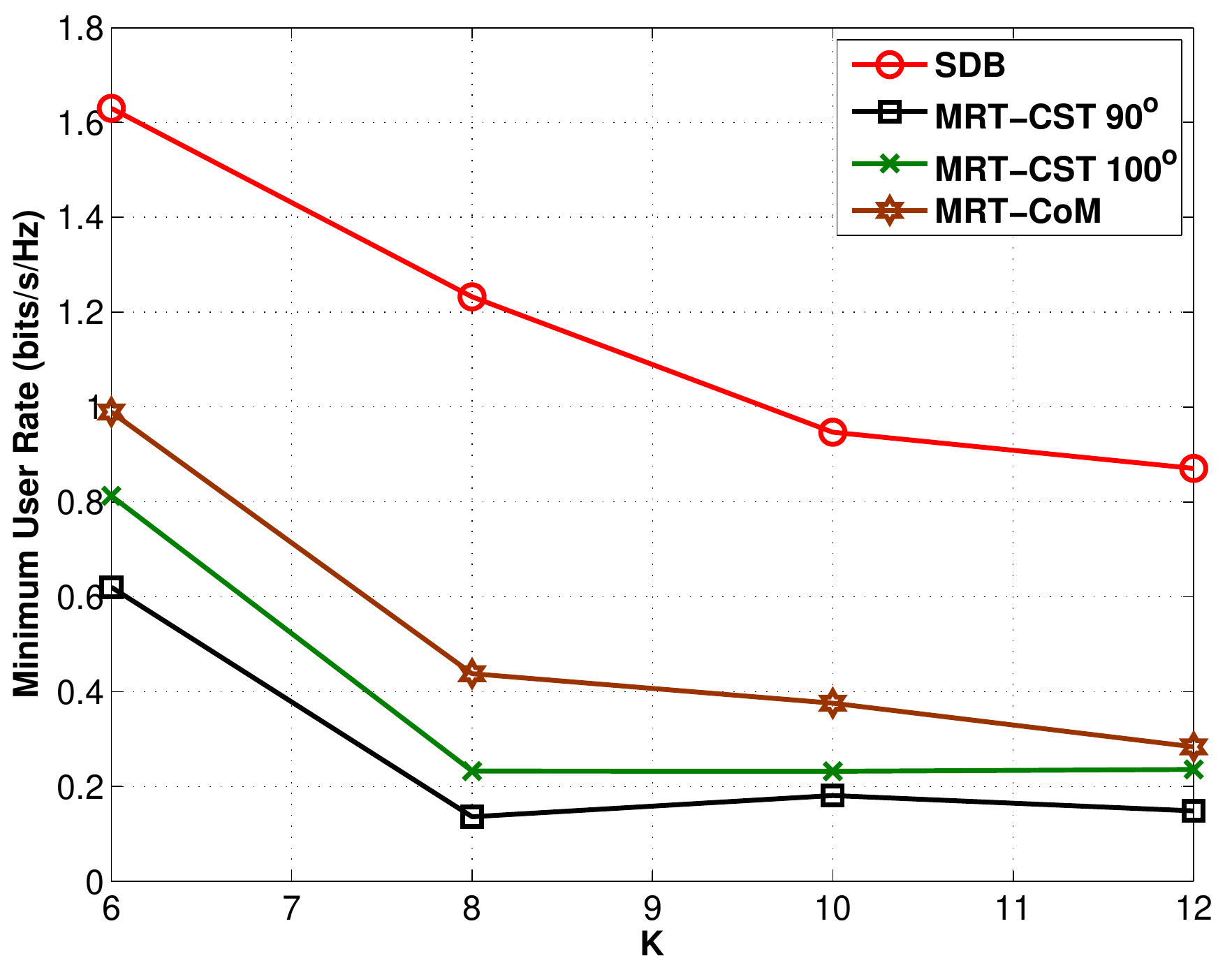}
\caption{Performance of a single-cell multi-user MISO system.}
\label{compare_perf1}
\end{figure}

Fig. \ref{compare_perf1} compares the Monte-Carlo simulated minimum user rate of the element approach based SDB strategy to the other tilting strategies outlined above for an $8 \times 12$ AAS. The SDB algorithm performs better than the other strategies. This is because the SDB strategy directly optimizes the weight vectors, thereby exploiting a higher number of degrees of freedom in generating the antenna radiation pattern to achieve the desired objective.

\subsubsection{Elevation Beamforming in the Multi-Cell Multi-User MISO Setting}

\begin{figure}[!t]
\centering
\includegraphics[width=2.65 in]{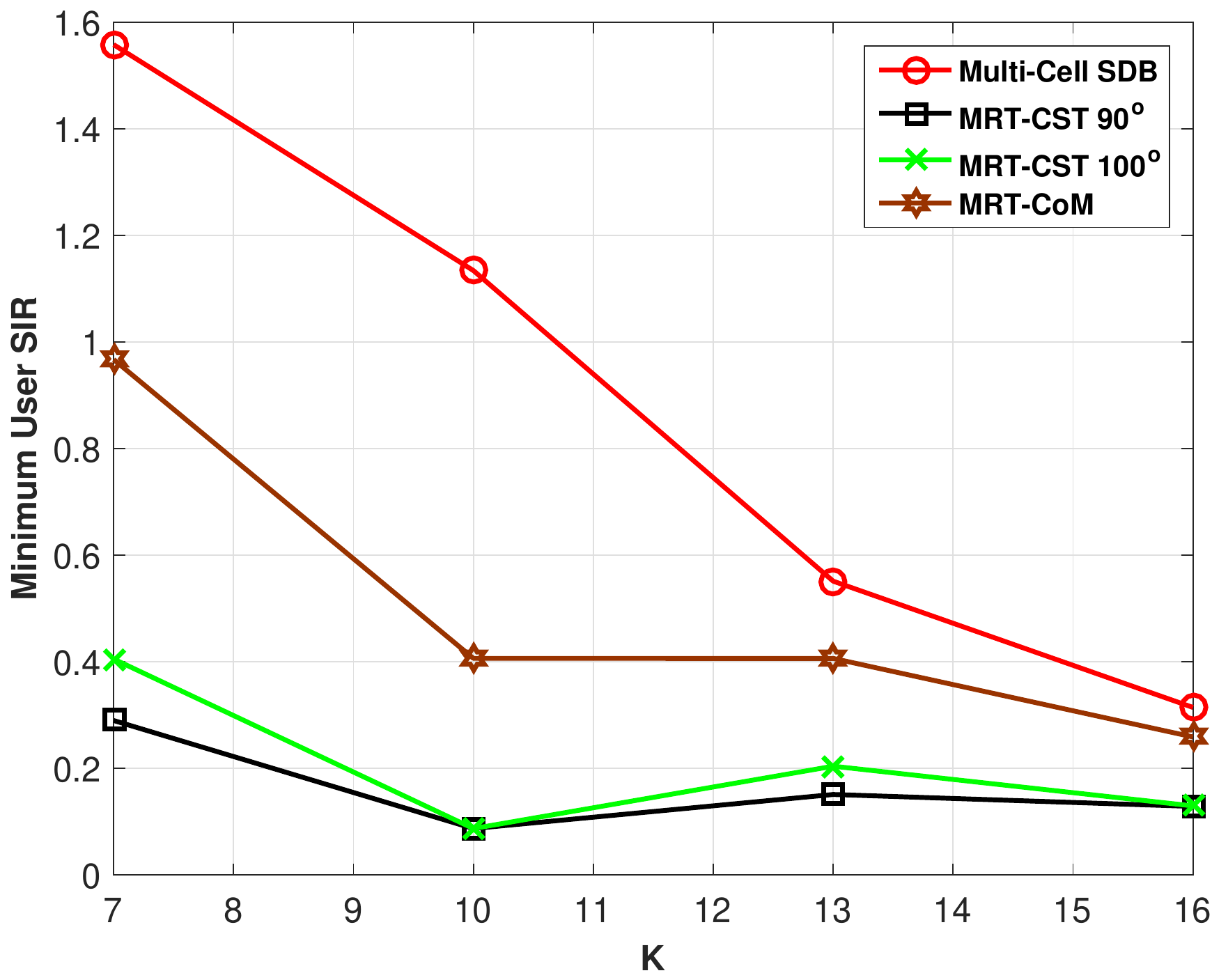}
\caption{Performance of a multi-cell multi-user MISO system.}
\label{multi_cell}
\end{figure}

The system comprises of three neighboring hexagonal cells with an equal number of users, $K$, placed randomly in each cell at a minimum distance of $30m$ from the BS. An $8 \times 18$ AAS is employed at the BS in each cell with Tx power of $5\rm{dB}$. The minimum user SIR performance of the following downtilting strategies is compared. 

\begin{enumerate}
\item \textbf{Multi-cell SDB}: The quasi-optimal weight vector $\textbf{w}_{i}^{*}$ for the antenna ports in cell $i$  is obtained using the multi-cell SDB Algorithm from \cite{ourworkWCNC}. This scheme is summarized in Table V and is important as it directly optimizes the downtilt weights applied to the elements in each port. The algorithm can be easily implemented by following the steps from Algorithm 1 in \cite{ourworkWCNC}.
\item \textbf{CST} -A common fixed tilt of $90^{o}$ and $100^{o}$ is applied to all antenna ports in all cells.
%\item \textbf{Coordinated user-specific tilting (CUST)} - This strategy, explained in Table \ref{multicell_compare_table}, uses the greedy tilt selection algorithm to adapt the downtilt angles in each cell, with max $\text{min}_{k}$ $SIR_{i,k}$ used as the objective function to allow for a fair comparison with multi-cell SDB. The modified strategy is referred to as Mod-CUST. 
\item \textbf{CoM} - The downtilt angle $\theta_{tilt,i}$ for the antenna ports in cell $i$ is computed as the mean of the elevation LoS angles of the users in cell $i$ at each transmission interval. 
\end{enumerate}
To ensure a fair comparison, the weight vectors for all cells are computed using (\ref{weight}), corresponding to the downtilt angles obtained as a result of the last two schemes. These weight vectors are then used to generate the Rayleigh correlated channel vectors for all users. All tilting strategies are simulated with MRT precoding in the horizontal domain under perfect CSI assumption. The multi-cell SDB scenario performs better in terms of the minimum user SIR as compared to other schemes as shown in Fig. \ref{multi_cell}. Note that the multi-cell SDB algorithm assumes the elements of the weight vectors to take arbitrary values such that the vectors have a unit norm, whereas approaches that adjust the tilt angle directly follow the 3GPP proposed weight vector expression in (\ref{weight}), which only allows the phase of the elements to vary.

These founding results provide a flavor of the performance gains realizable through the deployment of an AAS with a 2D planar array structure at the BS, where every antenna port is fed with a corresponding downtilt weight vector to realize spatially separated transmissions to a large number of users.  More sophisticated 3D beamforming techniques can be devised in the future that utilize other TXRU architectures presented in Section III, where every port can transmit at a different optimal downtilt angle. In fact, for  $N>>K$, spatially separated beams to almost all the users can be realized, which is why elevation beamforming can be highly advantageous when amalgamated with Massive MIMO techniques. 

\section{Further Research Directions}

This tutorial discussed the key features of FD-MIMO, focusing on 2D active antenna array design, transceiver architecture design, and 3D channel modeling using both ray-tracing and correlation based approaches, in  a way that was relevant to the design of beam adaptation and optimization strategies in the elevation plane. However, there are still many practical challenges down the road to the successful implementation of elevation beamforming techniques in actual FD Massive MIMO deployments, including pilot and feedback overhead reduction, advanced channel estimation, hardware limitations and wide beam generation for control signals. Some discussion on these issues is provided below.

\subsection{Channel Estimation and CSI Feedback in FD-MIMO}

Accurate downlink CSI at the BS is essential to reap the benefits of elevation beamforming in FD-MIMO systems, particularly where elevation beamforming strategies rely on knowledge of the instantaneous channels. Note that the current standards and dominant current cellular systems, e.g. 3GPP LTE and LTE-Advanced, are all based on FDD protocol. In order to support FDD FD-MIMO operation in LTE-A under time varying conditions, CSI feedback is needed to provide the serving BS with the quantized information about the downlink channels. To receive this information, the BS transmits RSs to the user. The CSI feedback mechanism then allows each user to report a recommended set of values, including the RI, PMI and the CQI, where the first two are used to assist the BS in performing beamforming  \cite{3GPP3D_channelfeed}. 

 Depending on the CSI-RS type used for CSI acquisition, two types of CSI feedback schemes are supported for FD-MIMO in LTE-A Pro: Class A and Class B CSI feedback \cite{FD3, feedback_FD}. In Class A, the user estimates the CSI values using measurements from non-precoded CSI-RS transmitted by the BS, and in Class B, the user estimates the CSI values using beamformed CSI-RS transmitted by the BS. Some details on the two schemes are now presented:

\textbf{Class-A CSI Feedback:} This scheme is usually implemented using the sub-array partition architectures shown in Fig. \ref{subarray1D} and \ref{subarray2D}. An identical weight vector is applied to all TXRUs. The TXRUs are fed with orthogonal CSI-RSs. Each user measures the CSI-RS during the training phase and then chooses the preferred codebook index $i^{*}$ maximizing the channel gain as,
\begin{align}
&i^{*}=\text{arg max}_{i} ||\bar{\textbf{h}}^{H}\textbf{g}^{i} ||_{2}^{2},
\end{align}
where $\bar{\textbf{h}}=\textbf{h}/||\textbf{h} ||_{2}$, where $\textbf{h} \in \mathbb{C}^{Q \times 1}$ is the channel vector and $\textbf{g}^{i} \in \mathbb{C}^{Q\times 1}$ is the $i^{th}$ precoder between the data channel and the antenna ports, where $Q$ is the total number of TXRUs/ports. Note that this scheme requires $Q$ CSI-RS resources  and the feedback overhead will also depend on the number of antennas and the resolution of the codebook. The CSI feedback mechanism relies on the use of a composite codebook $\textbf{G}_{P}$, which is divided into vertical and horizontal codebooks denoted as $\textbf{G}_{P,V}$ and $\textbf{G}_{P,H}$ respectively. Thus the channel information for the two dimensions can be separately sent to the BS. By combining the two codebooks, e.g. using the Kronecker product given as $\textbf{G}_{P}=\textbf{G}_{P,V} \otimes \textbf{G}_{P,H}$, the BS constructs the entire channel information. 
 
\textbf{Class-B CSI Feedback:} This scheme relies on beamformed CSI-RS transmission using the full connection TXRU architecture proposed in TR36.897. The BS transmits $N_{B}$ beamformed CSI-RSs using weight vectors $\textbf{w}^{i} \in \mathbb{C}^{MN \times 1}$, $i=1,\dots, N_{B}$, where $\textbf{w}^{i}$ is the TXRU virtualization weight vector for the $i^{th}$ beam. Note that in full connection architecture the CSI-RS is fed to all the elements using a weight vector of size $MN\times 1$. The user selects and feeds back the best beam index $i^{*}$ maximizing the received power as,
\begin{align}
&i^{*}=\text{arg max}_{i} ||\bar{\textbf{h}}^{H}\textbf{w}^{i}||_{2}^{2},
\end{align}
where $\bar{\textbf{h}}=\textbf{h}/||\textbf{h} ||_{2}$, with $\textbf{h} \in \mathbb{C}^{MN \times 1}$ being the channel vector with respect to all the elements in the AAA. This will enable the BS to acquire knowledge of the spatial angle between the BS and each user. The CSI-RS overhead scales with $N_{B}$ in this case where $N_{B}$ is generally much less than the total number of antennas. The feedback overhead also depends on the number of beams. This renders this scheme more suitable for Massive MIMO implementations as compared to Class-A feedback scheme. It is important to note that the conventional codebook for Class A FD-MIMO proposed in the 3GPP cannot be used to measure the CSI using the beamformed RS so a new channel feedback mechanism supporting the beamformed CSI transmission is required.  For this purpose, beam index feedback has been proposed, where  the BS only needs to feedback the beam index that maximizes the channel gain and corresponding CQI. When the BS receives this information, it uses the weight vector corresponding to the selected beam for data transmission.

Accurate CSI at the BS is essential for attaining the gains offered by elevation beamforming methods in FD-MIMO settings. Existing works on elevation beamforming  employ TDD protocol to exploit the  channel reciprocity and often assume the availability of perfect CSI at the BS.  However, majority of the existing cellular deployments employ FDD protocol which makes it important to explore backwards compatible FD-MIMO upgrades of such systems. Therefore,  elevation beamforming schemes should be studied in light of the channel estimation and feedback schemes proposed for FDD FD-MIMO systems in the 3GPP and academia, and the effect of channel estimation errors on the performance of these schemes should be analyzed.  

\subsection{Hardware Limitations}

%. One way to compensate the residual error of calibration is to use T-junctions in the calibration circuits with equal lengths on opposite branches to introduce pair wise symmetry 
In practice, low-cost imperfect hardware components are employed to implement the AAS. The performance of analog beamforming implemented using phase shifters and PAs, connected to the antenna elements, degrades due to  phase errors and non-linear behaviour of the PAs. Also, TDD systems face phase and amplitude mismatch between different transmitter and receiver paths due to construction variation and temperature drift of the analog components. If this mismatch is not corrected through calibration circuits, the 3D channel estimated from the uplink pilot symbols is not aligned with the downlink channel. Any phase inaccuracy in the calibration circuit will result in significant residual errors and degrade the performance of elevation beamforming algorithms in practice \cite{FD4}. 

 The imperfections of non-ideal hardware, such as the non-linear behaviour of the PAs, analog-to-digital and digital-to-analog non-linearities, phase errors and hardware calibration errors \cite{assumptions}  should be taken into account in the design and performance evaluation of elevation beamforming schemes. 

\subsection{Wide Beam Design for Control Signals}

Multi-user elevation beamforming allows for the formation of multiple narrow beams that serve several users in the same time-frequency resource. The design of narrow beams for user-specific data transmission is desirable as it reduces inter-user interference. However, besides the user-specific data, FD-MIMO BSs also need to broadcast data intended for all users  simultaneously, such as the RSs. Such signals need to be transmitted via a wide beam that covers the entire sector, instead of a narrow beam.  Two methods can be utilized to generate a wide beam: first is to transmit only from one antenna element and second is to transmit from multiple elements with certain amplitude weights. However, the first method transmits very little power since only one element is excited, which limits the range. The second method suffers from a similar problem, since it uses a beam pattern that heavily excites a few antenna elements  while the others are excited with very little power. To generate a wide beam with all antenna elements fully excited is challenging, because a fully excited antenna array generates a narrow radiation pattern. To ensure a wide coverage, the downtilt weight vectors need to be intelligently designed for control symbols so that their transmission from the BS has a wide beam width \cite{FD4, FD7}, which is non-trivial in FD-MIMO settings that involve a much higher number of antenna elements as compared to conventional MIMO systems.

\subsection{FD-MIMO in mm-Wave bands}
The large bandwidth in mm-Wave spectrum has made MIMO communication in mm-Wave bands a promising candidate for future cellular networks. The small values of wavelength allow a large number of antenna elements to be packed within feasible BS form factors, which has led to the Massive MIMO concept for mm-Wave bands. However, channel measurements have shown severe path and penetration loss and rain fading in mm-Wave bands as compared to current cellular bands, resulting in short propagation distances \cite{mmwave_sur1}.  In order to attain both high data rates and sufficient coverage, dense deployment of small cells is proposed for next generation mm-Wave communication systems \cite{mmwave2}. 

Moreover, it is not practical to implement conventional fully digital precoding schemes in mm-Wave Massive MIMO systems, as it would require one dedicated RF chain per antenna element, which is impractical from both cost and power consumption perspectives in mm-Wave bands. To overcome this challenge, hybrid beamforming schemes have been proposed \cite{hybrid2, hybrid_sur, usergroups} which concatenate an analog beamformer with a low-dimension digital beamformer. These hybrid schemes are particularly suited for FD-MIMO settings. Some light on this has already been shed in Section VI-A. 

Although the design of mm-Wave Massive MIMO systems has received considerable attention but there are still many uncertain practical issues  related to the implementation of FD-MIMO technology in mm-Wave bands such as the hardware design of the AAS, interference control using elevation beamforming in small cell deployments and resource management, that need attention.

\section{Conclusion}
In this article, we presented a tutorial on elevation beamforming using FD-MIMO architectures for evolution towards 5G cellular systems. We first presented the  2D AAS  and corresponding TXRU architectures that support elevation beamforming. In order to devise beam adaptation and optimization strategies and evaluate their performance, we outlined and compared two 3D ray-tracing spatial channel models that capture the characteristics of the wireless channel in both the azimuth and elevation planes. We stressed on the importance of utilizing the new and hardly used antenna element approach towards 3D channel modeling  to allow for the better optimization of elevation beamforming techniques in practice.  Then, we presented and compared the spatial correlation analysis for both channel models motivated by the fact that the high spatial correlation in FD-MIMO channels can be exploited not only to facilitate the theoretical analysis of elevation beamforming techniques but also to reduce the CSI feedback overhead incurred in the implementation of these techniques. All these aspects were put together to provide a mathematical framework for the design of elevation beamforming schemes. Simulation examples associated with comparisons and discussions were also presented. Finally, we pointed out future research directions for elevation beamforming in the context of 5G networks.

\bibliographystyle{IEEEtran}
\bibliography{bib}
\end{document}